\documentclass[12pt]{report}
\usepackage{amsfonts}
\usepackage{amsmath}
\usepackage{graphicx}
\usepackage{subfig}
\usepackage{caption}

\def\theenumi{\arabic{enumi}}

\def\theenumii{\alph{enumii}}
\def\p@enumii{\theenumi.}

\def\theenumiii{\arabic{enumiii}}
\def\p@enumiii{(\theenumi)(\theenumii)}

\def\p@enumiv{\p@enumiii.\theenumiii}
\newenvironment{proof}[1][Proof]{\noindent\textbf{#1.} }{\ \rule{0.5em}{0.5em}}
\input{tcilatex}
\begin{document}

\begin{center}
{\Large A Bivariate Compound Dynamic Contagion Process for Cyber Insurance}
\end{center}

\bigskip

\begin{center}
\textbf{Jiwook Jang}
\end{center}

\begin{quote}
{\small Department of Actuarial Studies \& Business Analytics, Macquarie
Business School, Macquarie University, Sydney NSW 2109, Australia, E-mail:
jiwook.jang@mq.edu.au}
\end{quote}

\begin{center}
\textbf{Rosy Oh}
\end{center}

\begin{quote}
{\small Institute of Mathematical Sciences, Ewha Womans University, Seoul, 03760,
Korea, E-mail: rosy.oh5@gmail.com}
\end{quote}

\bigskip

\textbf{Abstract \ }As corporates and governments become more digital, they
become vulnerable to various forms of cyber attack. Cyber insurance products
have been used as risk management tools, yet their pricing does not reflect
actual risk, including that of multiple, catastrophic and contagious losses.
\ For the modelling of aggregate losses from cyber events, in this paper we
introduce a bivariate \textit{compound} dynamic contagion process, where the
bivariate dynamic contagion process is a point process that includes both
externally excited joint jumps, which are distributed according to a shot
noise Cox process and two separate self-excited jumps, which are distributed
according to the branching structure of a Hawkes process with an exponential
fertility rate, respectively. \ We analyse the theoretical distributional
properties for these processes systematically, based on the piecewise
deterministic Markov process developed by Davis (1984) and the univariate
dynamic contagion process theory developed by Dassios and Zhao (2011). \ The
analytic expression of the Laplace transform of the compound process and its
moments are presented, which have the potential to be applicable to a
variety of problems in credit, insurance, market and other operational
risks. \ As an application of this process, we provide insurance premium
calculations based on its moments. \ Numerical examples show that this
compound process can be used for the modelling of aggregate losses from
cyber events. \ We also provide the simulation algorithm for statistical
analysis, further business applications and research.

\bigskip

\textbf{Keywords}: \ Aggregate losses from cyber events; Contagion risk;
Bivariate compound dynamic contagion process; Hawkes process; Piecewise
deterministic Markov process; Martingale methodology; Insurance premium

\bigskip

\textbf{1 } \textbf{Introduction}

\bigskip

Due to the digitalisation of business and economic activities via the
Internet of Things (IoT), cloud computing, mobile and other innovative
technologies, cyber risk is inherent and extreme. Cyber risks refer to any
risk of financial loss, disruption to operations, or damage to the
reputation of an organisation due to failure of its information technology
(IT) systems, as defined by the Institute of Risk Management (IRM).
Financial losses from malicious cyber activities result from IT
security/data/digital assets recovery, liability in respect of identity
theft and data breaches, reputation/brand damage, legal liability, cyber
extortion, regulatory defence and penalties coverage and business
interruption.

The frequency of malicious cyber activities is rapidly increasing, with the
scope and nature dependent on an organisation's industry, size and location.
According to a 2016 Allianz survey, cyber risk is the top long-term risk to
business and currently a top-three global business risk. It is therefore
critical that corporations and governments focus on IT and network security
enhancement. Unless public and private sector organisations have effective
cyber security plans and strategies in place, and tools to manage and
mitigate losses from cyber risks, cyber events have the potential to affect
their business significantly, possibly damaging hard-earned reputations
irreparably.

Insurance has served to mitigate liability since the 17th century, after the
Great Fire of London in 1666. As part of a cyber risk mitigation strategy,
cyber insurance can be purchased by organisations to cover economic and
financial losses occurring from cyber incidents. Since the widespread Y2K
concerns raised the profile of the possible security vulnerabilities of
digitalisation, the cyber insurance industry has grown to a total annual
premium of \$2.5 billion, and the market is expected to reach \$20 billion
by 2025 globally. However, due to the complexity of cyber incidents, i.e.
multiple, catastrophic and contagious losses, it is difficult for insurers
to price cyber insurance products accurately. Inaccurate pricing could have
severe market effects in the event of a significant claim.

To date however there has been little theoretical work done on developing
acceptable cyber insurance pricing models. Also due to the complexity of
cyber risks, the previous studies (Mukhopadhyay et al. 2006; Herath and
Herath 2011and Xu and Hua 2017) do not provide a suitable framework to
measure cyber risks as they have not accounted for future cyber attacks
dynamically. \ Also traditionally insurance claim modelling has used
homogeneous/non-homogeneous Poisson processes as a claim arrival process.
However, for cyber events, the assumption that resulting claims occur in
terms of the Poisson process is inadequate due to its deterministic
intensity. Therefore, an alternative point process needs to be used to
predict claim arrivals from cyber incidents.

To this effect, we introduce a bivariate \textit{compound} dynamic contagion
process (BCDCP) for the modelling of aggregate losses from cyber events,
where the bivariate dynamic contagion process (BDCP) is a point process
which has both externally excited joint jumps, which are distributed
according to a shot noise Cox process and two separate self-excited jumps,
which are Hawkes processes. Since Hawkes (1971a, 1971b) and Hawkes and Oakes
(1974) introduced a self-exciting point process, the applications and
modelling of Hawkes processes in finance and insurance can be found in
Chavez-Demoulin et al. (2005), McNeil et al. (2005), Bauwens and Hautsch
(2009), Bowsher (2007), Errais et al. (2010), Stabile and Torrisi (2010),
Embrechts et al. (2011), Giesecke and Kim (2011) and A\"{\i}t-Sahalia et al.
(2014, 2015).

Dassios and Zhao (2011) introduced a dynamic contagion process, which is a
generalisation of the externally excited Cox process with shot noise
intensity and the self-excited Hawkes process applying to credit risk.
Dassios and Zhao (2012) also examined infinite horizon ruin probability with
its Monte Carlo simulation using this process as the claim arrival process.
Dassios and Zhao (2017a) extended this process with diffusion component to
calculate the default probability and to price defaultable zero-coupon
bonds. \ We have found dynamic contagion processes to be flexible and
realistic in modelling claims with contagion.

These aforementioned papers are neither the bivariate dynamic contagion
models nor the compound models. \ In contrast we extend it further to
quantify aggregate losses from cyber events using a bivariate \textit{%
compound} dynamic contagion process as they are multiple, catastrophic and
contagious losses. \ Biener et al. (2015) emphasised that one of
characteristics of cyber risk is highly interrelated losses, and modelling
cyber risk would be a great deal of promise to test them when enough cyber
loss data become available.

Bivariate modelling with self-exciting Hawkes processes can be noticed in
Jang and Dassios (2013), where they introduced a bivariate shot noise
self-exciting process that can be used for the modelling of catastrophic
losses. Dong (2014) examined the stationarity of bivariate dynamic contagion
processes including the cross-exciting contagion effect in his doctoral
thesis. \ Applications and modelling of multivariate Hawkes process in
high-frequency limit order book data can be found in Rombaldi et al. (2017)
and Lu and Abergel (2018). \ Yang et al. (2018) investigated the
interactions between market return events and investor sentiment using a
multivariate Hawkes process.

Compound modelling with univariate self-exciting Hawkes processes can be
noticed in Dassios and Zhao (2017b), where they developed the algorithms for
a generalised self-exciting point process with CIR-type intensities. \ Gao
et al. (2018) applied the joint Laplace transform of the classical Hawkes
process and its compound process in dark pool trading, which do not display
bid and ask quotes to the public.

This project develops a new model for pricing cyber risk using a BCDCP,
which accommodate the interdependence dynamics of IT system and the
frequency and impact of cyber events. \ Our research offers a new framework
to enable insurance companies to price cyber insurance policies
accommodating clustering of losses.

This paper is structured as follows. In Section 2, we provide a mathematical
definition of the BCDCP and the BDCP, respectively via the stochastic
intensity representation adopted the one used by Dassios and Zhao (2011) and
the algorithm for simulating these processes in Section 5. In Section 3, we
analyse these processes systematically for their theoretical distributional
properties, based on the piecewise deterministic Markov process theory
developed by Davis (1984), and the martingale methodology used by Dassios
and Jang (2003). The joint moment of two processes, its covariance and
linear correlation are derived in Section 4, where for simplicity, we use
the case for the stationary distribution of the intensity processes. As an
application of this process, we provide cyber insurance premium calculations
based on these quantities in Section 5. Section 6 concludes the paper.

\bigskip

\textbf{2 } \textbf{Definition}

\bigskip

In this section, we have a mathematical definition for the BCDCP in
Definition 2.2. \ Before that, let us have a mathematical definition for the
BDCP in Definition 2.1 via the stochastic intensity representation adopted
the one used by Dassios and Zhao (2017). \ For an alternative definition for
this process, we refer you Dassios and Zhao (2011), Jang and Dassios (2013)
and Dong (2014), where they gave as a cluster process representation for the
univariate dynamic contagion process, the bivariate shot noise self-exciting
process and the bivariate dynamic contagion process, respectively.

\bigskip

\textbf{Definition 2.1 (Bivariate dynamic contagion process). }Bivariate
dynamic contagion process is a point process $\left( 
\begin{array}{c}
N_{t}^{\left( 1\right) } \\ 
N_{t}^{\left( 2\right) }%
\end{array}%
\right) _{t>0}=\left( 
\begin{array}{c}
\sum\limits_{j\geq 1}\mathbb{I}\left( T_{2,j}\leq t\right) _{j=1,2,\cdots }
\\ 
\sum\limits_{k\geq 1}\mathbb{I}\left( T_{2,k}\leq t\right) _{k=1,2,\cdots }%
\end{array}%
\right) $ with the non-negative $\Im _{t}-$stochastic bivariate intensity
process $\left( 
\begin{tabular}{l}
$\lambda _{t}^{\left( 1\right) }$ \\ 
$\lambda _{t}^{\left( 2\right) }$%
\end{tabular}%
\right) $, i.e.%
\begin{eqnarray}
\lambda _{t}^{\left( 1\right) } &=&a^{\left( 1\right) }+\left( \lambda
_{0}^{\left( 1\right) }-a^{\left( 1\right) }\right) e^{-\delta ^{\left(
1\right) }t}+\sum\limits_{i\geq 1}X_{i}^{\left( 1\right) }e^{-\delta
^{\left( 1\right) }\left( t-T_{1,i}\right) }\mathbb{I}\left( T_{1,i}\leq
t\right)  \notag \\
&&+\sum\limits_{j\geq 1}Y_{j}e^{-\delta ^{\left( 1\right) }\left(
t-T_{2,j}\right) }\mathbb{I}\left( T_{2,j}\leq t\right) ,  \notag \\
\lambda _{t}^{\left( 2\right) } &=&a^{\left( 2\right) }+\left( \lambda
_{0}^{\left( 2\right) }-a^{\left( 2\right) }\right) e^{-\delta ^{\left(
2\right) }t}+\sum\limits_{i\geq 1}X_{i}^{\left( 2\right) }e^{-\delta
^{\left( 2\right) }\left( t-T_{1,i}\right) }\mathbb{I}\left( T_{1,i}\leq
t\right)  \notag \\
&&+\sum\limits_{k\geq 1}Z_{k}e^{-\delta ^{\left( 2\right) }\left(
t-T_{2,k}\right) }\mathbb{I}\left( T_{2,k}\leq t\right) ,  \TCItag{2.1}
\end{eqnarray}

where

\begin{quote}
\textbullet\ $\left\{ \Im _{t}\right\} _{t\geq 0}$ is a history of the joint
process$\left( 
\begin{tabular}{l}
$N_{t}^{\left( 1\right) }$ \\ 
$N_{t}^{\left( 2\right) }$%
\end{tabular}%
\right) ,$ with respect to which $\left\{ 
\begin{tabular}{l}
$\lambda _{t}^{\left( 1\right) }$ \\ 
$\lambda _{t}^{\left( 2\right) }$%
\end{tabular}%
\right\} _{t\geq 0}$ is adapted;

\textbullet\ $\lambda _{0}^{\left( d\right) }$ $>0$ is the initial intensity
at time $t=0$, where $d=1,2$;

\textbullet\ $a^{\left( d\right) }$ $\geq 0$ is the constant mean-reverting
level;

\textbullet\ $\delta ^{\left( d\right) }$ $>0$ is the constant
mean-reverting rate;

\textbullet\ $\left\{ X_{i}^{\left( 1\right) },\text{ }X_{i}^{\left(
2\right) }\right\} _{i=1,2,\cdots }$ is a sequence of \textit{i}.\textit{i}.%
\textit{d}. positive externally-excited \textit{joint} jumps with
distribution $F(x^{\left( 1\right) },x^{\left( 2\right) }),$ $x^{\left(
1\right) }>0,$ $x^{\left( 2\right) }>0$, where margins are $F_{X^{\left(
1\right) }}$ and $F_{X^{\left( 2\right) }}$ at the corresponding random
times $\left\{ T_{1,i}\right\} _{i=1,2,\cdots }$ following a Poisson process 
$M_{t}$ with constant rate $\rho >0$, and $\mathbb{I}$ is the indicator
function.

\textbullet\ $\left\{ Y_{j}\right\} _{j=1,2,\cdots }$ is a sequence of 
\textit{i}.\textit{i}.\textit{d}. positive self-excited jumps with
distribution function $G(y)$, $y>0$, at the corresponding random times $%
\left\{ T_{2,j}\right\} _{j=1,2,\cdots }$.

\textbullet\ $\left\{ Z_{k}\right\} _{k=1,2,\cdots }$ is another sequence of 
\textit{i}.\textit{i}.\textit{d}. positive self-excited jumps with
distribution function $H(z)$, $z>0$, at the corresponding random times $%
\left\{ T_{2,k}\right\} _{k=1,2,\cdots }$.

\textbullet\ $\left\{ X_{i}^{\left( 1\right) },X_{i}^{\left( 2\right)
}\right\} _{i=1,2,\cdots }$, $\left\{ Y_{j}\right\} _{j=1,2,\cdots }$, $%
\left\{ Z_{k}\right\} _{k=1,2,\cdots }$, $\left\{ T_{1,i}\right\}
_{i=1,2,\cdots }$, $\left\{ T_{2,j}\right\} _{j=1,2,\cdots }$ and $\left\{
T_{2,k}\right\} _{k=1,2,\cdots }$are assumed to be independent of each other.
\end{quote}

\bigskip

The bivariate \textit{compound} model we consider has the following
structure:

\begin{eqnarray}
L_{t}^{(1)} &=&\sum\limits_{j\geq 1}\Xi _{j}^{\left( 1\right) }\mathbb{I}%
\left( T_{2,j}\leq t\right) ,\text{ }  \notag \\
L_{t}^{(2)} &=&\sum\limits_{k\geq 1}\Xi _{k}^{\left( 2\right) }\mathbb{I}%
\left( T_{2,k}\leq t\right) ,  \TCItag{2.2}
\end{eqnarray}%
where $L_{t}^{(d)}$ is the total amount of claims/losses arising from risk
type $d=1,2$ and $N_{t}^{\left( d\right) }$ is the number of points (i.e.
claims/losses) up to time $t$. \ The random variables $\Xi _{j}^{\left(
1\right) }$ and $\Xi _{k}^{\left( 2\right) }$ denote the individual
claim/loss amounts, where we assume that they are independent identically
distributed with distributions $J_{Y^{\left( 1\right) }}$ and $K_{Y^{\left(
2\right) }}$, respectively. \ Our intensity processes for $N_{t}^{\left(
1\right) }$ and $N_{t}^{\left( 2\right) }$ are modelled by jump processes,
which are in the form of (2.1).

\bigskip

\textbf{Definition 2.2 (Bivariate compound dynamic contagion process). \ }%
Bivariate compound dynamic contagion process is a \textit{compound} point
process $\left( 
\begin{array}{c}
L_{t}^{\left( 1\right) } \\ 
L_{t}^{\left( 2\right) }%
\end{array}%
\right) _{t>0}=\left( 
\begin{array}{c}
\sum\limits_{j\geq 1}\Xi _{j}^{\left( 1\right) }\mathbb{I}\left( T_{2,j}\leq
t\right) _{j=1,2,\cdots } \\ 
\sum\limits_{k\geq 1}\Xi _{k}^{\left( 2\right) }\mathbb{I}\left( T_{2,k}\leq
t\right) _{k=1,2,\cdots }%
\end{array}%
\right) $ with the non-negative $\Im _{t}-$stochastic bivariate intensity
process $\left( 
\begin{tabular}{l}
$\lambda _{t}^{\left( 1\right) }$ \\ 
$\lambda _{t}^{\left( 2\right) }$%
\end{tabular}%
\right) $ which is in the form of (2.1), where

\begin{quote}
\textbullet\ $\left\{ \Xi _{j}^{\left( 1\right) }\right\} _{j=1,2,\cdots }$
is a sequence of \textit{i}.\textit{i}.\textit{d}. positive individual
claim/loss amounts from risk type $d=1$ with distribution function $J(\xi
^{\left( 1\right) })$, $\xi ^{\left( 1\right) }>0$, at the corresponding
random times $\left\{ T_{2,j}\right\} _{j=1,2,\cdots }$.

\textbullet\ $\left\{ \Xi _{k}^{\left( 2\right) }\right\} _{k=1,2,\cdots }$
is another sequence of \textit{i}.\textit{i}.\textit{d}. positive individual
claim/loss amounts from risk type $d=2$ with distribution function $K(\xi
^{\left( 2\right) })$, $\xi ^{\left( 2\right) }>0$, at the corresponding
random times $\left\{ T_{2,k}\right\} _{k=1,2,\cdots }$.

\textbullet\ $\left\{ X_{i}^{\left( 1\right) },X_{i}^{\left( 2\right)
}\right\} _{i=1,2,\cdots }$, $\left\{ Y_{j}\right\} _{j=1,2,\cdots }$, $%
\left\{ Z_{k}\right\} _{k=1,2,\cdots }$, $\left\{ \Xi _{j}^{\left( 1\right)
}\right\} _{j=1,2,\cdots }$, $\left\{ \Xi _{k}^{\left( 1\right) }\right\}
_{k=1,2,\cdots }$, $\left\{ T_{1,i}\right\} _{i=1,2,\cdots }$ $\left\{
T_{2,j}\right\} _{j=1,2,\cdots }$ and $\left\{ T_{2,k}\right\}
_{k=1,2,\cdots }$are assumed to be independent of each other.
\end{quote}

\bigskip

The joint process of $\left\{ \left( 
\begin{tabular}{l}
$\lambda _{t}^{\left( 1\right) }$ \\ 
$\lambda _{t}^{\left( 2\right) }$%
\end{tabular}%
\right) ,\left( 
\begin{array}{c}
N_{t}^{\left( 1\right) } \\ 
N_{t}^{\left( 2\right) }%
\end{array}%
\right) ,\left( 
\begin{array}{c}
L_{t}^{\left( 1\right) } \\ 
L_{t}^{\left( 2\right) }%
\end{array}%
\right) \right\} _{t\geq 0}$ is a Markov process in the state space $\mathbb{%
R}^{+}\times \mathbb{N}_{0}\times \mathbb{R}_{0}^{+}$. \ With the aid of
piecewise deterministic Markov process theory and using the results in Davis
(1984), the infinitesimal generator of the bivariate compound dynamic
contagion process $\left( \lambda _{t}^{\left( 1\right) },N_{t}^{\left(
1\right) },L_{t}^{(1)},\lambda _{t}^{\left( 2\right) },N_{t}^{\left(
2\right) },L_{t}^{(2)},t\right) $ acting on a function $f\left( \lambda
^{\left( 1\right) },n^{\left( 1\right) },l^{(1)},\lambda ^{\left( 2\right)
},n^{\left( 2\right) },l^{(2)},t\right) $ within its domain $\mathcal{D}%
\left( \mathcal{A}\right) $ is given by

\begin{eqnarray}
&&\mathcal{A}\text{ }f\left( \lambda ^{\left( 1\right) },n^{\left( 1\right)
},l^{(1)},\lambda ^{\left( 2\right) },n^{\left( 2\right) },l^{(2)},t\right) 
\notag \\
&=&\frac{\partial f}{\partial t}+\delta ^{\left( 1\right) }\left( a^{\left(
1\right) }-\lambda ^{\left( 1\right) }\right) \frac{\partial f}{\partial
\lambda ^{\left( 1\right) }}+\delta ^{\left( 2\right) }\left( a^{\left(
2\right) }-\lambda ^{\left( 2\right) }\right) \frac{\partial f}{\partial
\lambda ^{\left( 2\right) }}  \notag \\
&&+\lambda ^{\left( 1\right) }\left[ 
\begin{array}{c}
\int\limits_{0}^{\infty }\int\limits_{0}^{\infty }f\left( \lambda ^{\left(
1\right) }+y,n^{\left( 1\right) }+1,l^{(1)}+\xi ^{\left( 1\right) },\lambda
^{\left( 2\right) },n^{\left( 2\right) },l^{(2)},t\right) dG(y)dJ(\xi
^{\left( 1\right) }) \\ 
-f\left( \lambda ^{\left( 1\right) },n^{\left( 1\right) },l^{(1)},\lambda
^{\left( 2\right) },n^{\left( 2\right) },l^{(2)},t\right) 
\end{array}%
\right]   \notag \\
&&+\lambda ^{\left( 2\right) }\left[ 
\begin{array}{c}
\int\limits_{0}^{\infty }\int\limits_{0}^{\infty }f\left( \lambda ^{\left(
1\right) },n^{\left( 1\right) },l^{(1)},\lambda ^{\left( 2\right)
}+z,n^{\left( 2\right) }+1,l^{(2)}+\xi ^{\left( 2\right) },t\right)
dH(z)dK(\xi ^{\left( 2\right) }) \\ 
-f\left( \lambda ^{\left( 1\right) },n^{\left( 1\right) },l^{(1)},\lambda
^{\left( 2\right) },n^{\left( 2\right) },l^{(2)},t\right) 
\end{array}%
\right]   \notag \\
&&+\rho \left[ 
\begin{array}{c}
\int\limits_{0}^{\infty }\int\limits_{0}^{\infty }f\left( \lambda ^{\left(
1\right) }+x^{\left( 1\right) },n^{\left( 1\right) },l^{(1)},\lambda
^{\left( 2\right) }+x^{\left( 2\right) },n^{\left( 2\right)
},l^{(2)},t\right) dF_{X^{\left( 1\right) },X^{\left( 2\right) }}(x^{\left(
1\right) }\text{, }x^{\left( 2\right) }) \\ 
-f\left( \lambda ^{\left( 1\right) },n^{\left( 1\right) },l^{(1)},\lambda
^{\left( 2\right) },n^{\left( 2\right) },l^{(2)},t\right) 
\end{array}%
\right] ,  \notag \\
&&  \TCItag{2.3}
\end{eqnarray}%
where $\mathcal{D}\left( \mathcal{A}\right) $ is the domain of the generator 
$\mathcal{A}$ such that $f\left( \lambda ^{\left( 1\right) },n^{\left(
1\right) },l^{(1)},\lambda ^{\left( 2\right) },n^{\left( 2\right)
},l^{(2)},t\right) $ is differentiable with respect to $\lambda ^{\left(
1\right) },\lambda ^{\left( 2\right) }$ and $t$ for all $\lambda ^{\left(
1\right) },\lambda ^{\left( 2\right) }$ and $t,$ and

\begin{equation*}
\left\vert \int\limits_{0}^{\infty }\int\limits_{0}^{\infty }f\left( \cdot
,\lambda ^{\left( 1\right) }+y,n^{\left( 1\right) }+1,l^{(1)}+\xi ^{\left(
1\right) },\cdot \right) dG(y)dJ(\xi ^{\left( 1\right) })-f\left( \cdot
,\lambda ^{\left( 1\right) },n^{\left( 1\right) },l^{(1)},\cdot \right)
\right\vert <\infty \text{,}
\end{equation*}

\begin{equation*}
\left\vert \int\limits_{0}^{\infty }\int\limits_{0}^{\infty }f\left( \cdot
,\lambda ^{\left( 2\right) }+z,n^{\left( 2\right) }+1,l^{(2)}+\xi ^{\left(
2\right) },\cdot \right) dH(z)dK(\xi ^{\left( 2\right) })-f\left( \cdot
,\lambda ^{\left( 2\right) },n^{\left( 2\right) },l^{(2)},\cdot \right)
\right\vert <\infty ,
\end{equation*}

\begin{equation*}
\left\vert \int\limits_{0}^{\infty }\int\limits_{0}^{\infty }f\left( \cdot
,\lambda ^{\left( 1\right) }+x^{\left( 1\right) },\lambda ^{\left( 2\right)
}+x^{\left( 2\right) },\cdot \right) dF\left( x^{\left( 1\right) },x^{\left(
2\right) }\right) -f\left( \cdot ,\lambda ^{\left( 1\right) }+x^{\left(
1\right) },\lambda ^{\left( 2\right) }+x^{\left( 2\right) },\cdot \right)
\right\vert <\infty \text{.}
\end{equation*}

\bigskip

\textbf{3. }\ \textbf{Bivariate Compound Dynamic Contagion Process}

\bigskip

In this section, we derive the joint Laplace transform of the process $%
(L_{T}^{\left( 1\right) },$ $L_{T}^{\left( 2\right) })$ in Theorem 3.4, for
which we start with Theorem 3.1. \ Theorem 3.1 leads to the key results of
the paper as we also derive the joint probability generating function of the
process $(N_{T}^{\left( 1\right) },$ $N_{T}^{\left( 2\right) })$ in Theorem
3.2. The joint Laplace transform of the process $(\lambda _{T}^{\left(
1\right) },$ $\lambda _{T}^{\left( 2\right) })$ can be also derived using
this theorem as presented in Jang and Dassios (2013).

\bigskip

\textbf{3.1. \ Joint Laplace Transform - Probability Generating Function of} 
$(\lambda _{t}^{\left( 1\right) },$ $\lambda _{t}^{\left( 2\right) },$ $%
N_{t}^{\left( 1\right) },$ $N_{t}^{\left( 2\right) },$ $L_{t}^{\left(
1\right) },$ $L_{t}^{\left( 2\right) })$

\bigskip

\textbf{Theorem 3.1 \ }\textit{Considering the constants}, $0\leq \theta
\leq 1,$ $0\leq \eta \leq 1,$ $\nu \geq 0,$ $\zeta \geq 0,$ $\upsilon \geq
0, $ $\gamma \geq 0$ \textit{and time} $0\leq t\leq T,$ \textit{we have the
conditional joint Laplace transform}, \textit{probability generating
function of the process} $(\lambda _{T}^{\left( 1\right) },$ $\lambda
_{T}^{\left( 2\right) })$, \textit{the} \textit{point process} ($%
N_{T}^{\left( 1\right) },$ $N_{T}^{\left( 2\right) })$ \textit{and the
compound point process} $(L_{T}^{\left( 1\right) },$ $L_{T}^{\left( 2\right)
})$ \textit{is given by}

\begin{eqnarray}
&&E\left[ \theta ^{\left\{ N_{T}^{\left( 1\right) }-N_{t}^{\left( 1\right)
}\right\} }\eta ^{\left\{ N_{T}^{\left( 2\right) }-N_{t}^{\left( 2\right)
}\right\} }e^{-\nu \left\{ L_{T}^{\left( 1\right) }-L_{t}^{\left( 1\right)
}\right\} }e^{-\zeta \left\{ L_{T}^{\left( 2\right) }-L_{t}^{\left( 2\right)
}\right\} }\times e^{-\upsilon \lambda _{T}^{\left( 1\right) }}e^{-\gamma
\lambda _{T}^{\left( 2\right) }}\mid \Im _{t}\right]  \notag \\
&=&e^{-B_{1}(t)\lambda _{t}^{\left( 1\right) }}e^{-B_{2}(t)\lambda
_{t}^{\left( 2\right) }}e^{-\left\{ C(T)-C(t)\right\} },  \TCItag{3.1}
\end{eqnarray}%
\textit{where} $B_{1}(t)$ \textit{and} $B_{2}(t)$\ \textit{are determined by
two non-linear ordinary differential equations }(\textit{ODEs})

\begin{eqnarray}
-B_{1}^{\prime }\left( t\right) +\delta ^{\left( 1\right) }B_{1}\left(
t\right) +\theta \text{ }\overset{\wedge }{g}\left\{ B_{1}\left( t\right)
\right\} \text{ }\overset{\wedge }{j}\left( \nu \right) -1 &=&0, 
\TCItag{3.2} \\
-B_{2}^{\prime }\left( t\right) +\delta ^{\left( 2\right) }B_{2}\left(
t\right) +\eta \text{ }\overset{\wedge }{h}\left\{ B_{2}\left( t\right)
\right\} \text{ }\overset{\wedge }{k}\left( \zeta \right) -1 &=&0, 
\TCItag{3.3}
\end{eqnarray}%
\textit{with the boundary condition }$B_{1}(T)=\upsilon $ \textit{and }$%
B_{2}(T)=\gamma $\textit{, respectively, where}

\begin{eqnarray}
\overset{\wedge }{g}\left( \varepsilon \right) &=&\int\limits_{0}^{\infty
}e^{-\varepsilon y}\text{ }dG(y)\text{, }\overset{\wedge }{h}\left(
\varepsilon \right) =\int\limits_{0}^{\infty }e^{-\varepsilon z}\text{ }%
dH(z),\text{\ }\overset{\wedge }{j}\left( \kappa \right)
=\int\limits_{0}^{\infty }e^{-\kappa \zeta ^{\left( 1\right) }}dJ(\zeta
^{\left( 1\right) })\text{ \ }  \notag \\
\text{\textit{and} \ }\overset{\wedge }{k}\left( \kappa \right)
&=&\int\limits_{0}^{\infty }\text{ }e^{-\kappa \zeta ^{\left( 2\right)
}}dK(\zeta ^{\left( 2\right) }).  \TCItag{3.4}
\end{eqnarray}%
$C(t)$ \textit{is determined by}

\begin{equation}
C(t)=\rho \int\limits_{0}^{t}\left[ 1-\overset{\wedge }{f}\left\{
B_{1}\left( s\right) ,B_{2}\left( s\right) \right\} \right] ds+a^{\left(
1\right) }\delta ^{\left( 1\right) }\int\limits_{0}^{t}B_{1}\left( s\right)
ds+a^{\left( 2\right) }\delta ^{\left( 2\right)
}\int\limits_{0}^{t}B_{2}\left( s\right) ds,  \tag{3.5}
\end{equation}%
\textit{where}%
\begin{equation}
\overset{\wedge }{f}\left( \varepsilon ,\kappa \right)
=\int\limits_{0}^{\infty }\int\limits_{0}^{\infty }e^{-\varepsilon x^{\left(
1\right) }}e^{-\kappa x^{\left( 2\right) }}dF\left( x^{\left( 1\right)
},x^{\left( 2\right) }\right) .  \tag{3.6}
\end{equation}%
\textit{It is assumed that the Laplace transforms of above}, \textit{i}.%
\textit{e}. $\overset{\wedge }{g}\left( \varepsilon \right) ,$ $\overset{%
\wedge }{h}\left( \varepsilon \right) ,$ $\overset{\wedge }{j}\left( \kappa
\right) ,$ $\overset{\wedge }{k}\left( \kappa \right) $ \textit{and the
joint Laplace transform}, $\overset{\wedge }{f}\left( \varepsilon ,\kappa
\right) $ \textit{are finite}.

\begin{proof}
Consider a function $f\left( \lambda ^{\left( 1\right) },n^{\left( 1\right)
},l^{(1)},\lambda ^{\left( 2\right) },n^{\left( 2\right) },l^{(2)},t\right) $
with an exponential affine form%
\begin{eqnarray*}
&&f\left( \lambda ^{\left( 1\right) },n^{\left( 1\right) },l^{(1)},\lambda
^{\left( 2\right) },n^{\left( 2\right) },l^{(2)},t\right)  \\
&=&\theta ^{n^{\left( 1\right) }}\eta ^{n^{\left( 2\right) }}e^{-\nu
l^{\left( 1\right) }}e^{-\zeta l^{\left( 2\right) }}e^{-B_{1}\left( t\right)
\lambda ^{\left( 1\right) }}e^{-B_{2}\left( t\right) \lambda ^{\left(
2\right) }}e^{C(t)},
\end{eqnarray*}%
substitute into $\mathcal{A}$ $f=0$ in (2.3), we have%
\begin{eqnarray*}
&&-\lambda ^{\left( 1\right) }B_{1}^{\prime }\left( t\right) -\lambda
^{\left( 2\right) }B_{2}^{\prime }\left( t\right) +C^{\prime }\left(
t\right)  \\
&&+\lambda ^{\left( 1\right) }\left[ \theta \text{ }\overset{\wedge }{g}%
\left\{ B_{1}\left( t\right) \right\} \text{ }\overset{\wedge }{j}\left( \nu
\right) \right] +\lambda ^{\left( 2\right) }\left[ \eta \text{ }\overset{%
\wedge }{h}\left\{ B_{2}\left( t\right) \right\} \text{ }\overset{\wedge }{k}%
\left( \zeta \right) \right]  \\
&&+\delta ^{\left( 1\right) }\left( a^{\left( 1\right) }-\lambda ^{\left(
1\right) }\right) \left\{ -B_{1}\left( t\right) \right\} +\delta ^{\left(
2\right) }\left( a^{\left( 2\right) }-\lambda ^{\left( 2\right) }\right)
\left\{ -B_{2}\left( t\right) \right\}  \\
&&+\rho \left[ \overset{\wedge }{f}\left\{ B_{1}\left( t\right) ,B_{2}\left(
t\right) \right\} -1\right]  \\
&=&0.
\end{eqnarray*}

\begin{eqnarray}
&&\left[ -B_{1}^{\prime }\left( t\right) +\delta ^{\left( 1\right)
}B_{1}\left( t\right) +\theta \text{ }\overset{\wedge }{g}\left\{
B_{1}\left( t\right) \right\} \text{ }\overset{\wedge }{j}\left( \nu \right)
-1\right] \lambda ^{\left( 1\right) }  \notag \\
&&\left[ -B_{2}^{\prime }\left( t\right) +\delta ^{\left( 2\right)
}B_{2}\left( t\right) +\eta \text{ }\overset{\wedge }{h}\left\{ B_{2}\left(
t\right) \right\} \text{ }\overset{\wedge }{k}\left( \zeta \right) -1\right]
\lambda ^{\left( 2\right) }  \notag \\
&&+\left[ C^{\prime }\left( t\right) +\rho \text{ }\overset{\wedge }{f}%
\left\{ B_{1}\left( t\right) ,B_{2}\left( t\right) \right\} -\rho -\delta
^{\left( 1\right) }a^{\left( 1\right) }B_{1}\left( t\right) -\delta ^{\left(
2\right) }a^{\left( 2\right) }B_{2}\left( t\right) \right]  \notag \\
&=&0.  \TCItag{3.7}
\end{eqnarray}%
where%
\begin{eqnarray*}
\overset{\wedge }{g}\left( \varepsilon \right) \overset{\wedge }{j}\left(
\kappa \right) &=&\int\limits_{0}^{\infty }\int\limits_{0}^{\infty
}e^{-\varepsilon y}\text{ }e^{-\kappa \zeta ^{\left( 1\right)
}}dG(y)dJ(\zeta ^{\left( 1\right) }), \\
\overset{\wedge }{h}\left( \varepsilon \right) \overset{\wedge }{k}\left(
\kappa \right) &=&\int\limits_{0}^{\infty }\int\limits_{0}^{\infty
}e^{-\varepsilon z}\text{ }e^{-\kappa \zeta ^{\left( 2\right)
}}dH(z)dK(\zeta ^{\left( 2\right) }), \\
\overset{\wedge }{f}\left( \varepsilon ,\kappa \right)
&=&\int\limits_{0}^{\infty }\int\limits_{0}^{\infty }e^{-\varepsilon
x^{\left( 1\right) }}e^{-\kappa x^{\left( 2\right) }}dF\left( x^{\left(
1\right) },x^{\left( 2\right) }\right) .
\end{eqnarray*}

\bigskip

Since this equation holds for any $l^{(1)},l^{(2)},n^{\left( 1\right) },$ $%
n^{\left( 2\right) },$ $\lambda ^{\left( 1\right) }$ and $\lambda ^{\left(
2\right) }$, it is equivalent to solving three separated equations, i.e.%
\begin{eqnarray}
-B_{1}^{\prime }\left( t\right) +\delta ^{\left( 1\right) }B_{1}\left(
t\right) +\theta \text{ }\overset{\wedge }{g}\left\{ B_{1}\left( t\right)
\right\} \text{ }\overset{\wedge }{j}\left( \nu \right) -1 &=&0, 
\TCItag{3.8.1} \\
-B_{2}^{\prime }\left( t\right) +\delta ^{\left( 2\right) }B_{2}\left(
t\right) +\eta \text{ }\overset{\wedge }{h}\left\{ B_{2}\left( t\right)
\right\} \text{ }\overset{\wedge }{k}\left( \zeta \right) -1 &=&0, 
\TCItag{3.8.2}
\end{eqnarray}%
\begin{equation}
C^{\prime }\left( t\right) +\rho \text{ }\overset{\wedge }{f}\left\{
B_{1}\left( t\right) ,B_{2}\left( t\right) \right\} -\rho -\delta ^{\left(
1\right) }a^{\left( 1\right) }B_{1}\left( t\right) -\delta ^{\left( 2\right)
}a^{\left( 2\right) }B_{2}\left( t\right) =0.  \tag{3.8.3}
\end{equation}

\bigskip

We have two ODEs of (3.8.1) and (3.8.2) with the boundary condition $%
B_{1}\left( T\right) =\upsilon $ and $B_{2}\left( T\right) =\gamma $,
respectively. \ By (3.8.3) with boundary condition $C\left( 0\right) =0,$
the integration of (3.5) follows. \ Since $\theta ^{N_{t}^{\left( 1\right)
}}\eta ^{N_{t}^{\left( 2\right) }}e^{-\nu L_{t}^{\left( 1\right) }}e^{-\zeta
L_{t}^{\left( 2\right) }}e^{-B_{1}\left( t\right) \lambda _{t}^{\left(
1\right) }}e^{-B_{2}\left( t\right) \lambda _{t}^{\left( 2\right) }}e^{C(t)}$%
\ is a $\Im $-martingale by the property of the infinitesimal generator, we
have%
\begin{eqnarray}
&&E\left[ \theta ^{N_{T}^{\left( 1\right) }}\eta ^{N_{T}^{\left( 2\right)
}}e^{-\nu L_{T}^{\left( 1\right) }}e^{-\zeta L_{T}^{\left( 2\right)
}}e^{-B_{1}(T)\lambda _{T}^{\left( 1\right) }}e^{-B_{1}(T)\lambda
_{T}^{\left( 2\right) }}e^{C(T)}\mid \Im _{t}\right]  \notag \\
&=&\theta ^{N_{t}^{\left( 1\right) }}\eta ^{N_{t}^{\left( 2\right) }}e^{-\nu
L_{t}^{\left( 1\right) }}e^{-\zeta L_{t}^{\left( 2\right)
}}e^{-B_{1}(t)\lambda _{t}^{\left( 1\right) }}e^{-B_{2}(t)\lambda
_{t}^{\left( 2\right) }}e^{C(t)}.  \TCItag{3.9}
\end{eqnarray}

Then, by the boundary condition $B_{1}\left( T\right) =\upsilon $ and $%
B_{2}\left( T\right) =\gamma ,$ (3.1) follows.
\end{proof}

\bigskip

\textbf{3.2. \ Joint Laplace Transform of} $(\lambda _{T}^{\left( 1\right)
}, $ $\lambda _{T}^{\left( 2\right) })$

\bigskip

Based on (3.1), we can easily derive the joint Laplace transform for the
process $(\lambda _{T}^{\left( 1\right) },$ $\lambda _{T}^{\left( 2\right)
}) $ setting $\theta =1,$ $\eta =1,$ $\nu =0,$ $\zeta =0.$ \ As it has
already presented in Jang and Dassios (2013), we state two propositions
adopted from them in this section. \ $\mathcal{G}_{\upsilon ,1}^{-1}(T)$ and 
$\mathcal{H}_{\gamma ,1}^{-1}(T)$ in the proposition will become apparent in
Theorem 3.3.

\bigskip

\textbf{Proposition 3.1.} \ \textit{The conditional joint Laplace transform
for the process} $\left( \lambda _{T}^{\left( 1\right) },\lambda
_{T}^{\left( 2\right) }\right) $ \textit{given }$\lambda _{0}^{\left(
1\right) }$\textit{\ and }$\lambda _{0}^{\left( 2\right) }$\textit{\ at time}
$t=0$ \textit{is given by}

\begin{eqnarray}
&&E\left[ e^{-\upsilon \lambda _{T}^{\left( 1\right) }}e^{-\gamma \lambda
_{T}^{\left( 2\right) }}\mid \lambda _{0}^{\left( 1\right) },\lambda
_{0}^{\left( 2\right) }\right]  \notag \\
&=&\exp \left\{ -\mathcal{G}_{\upsilon ,1}^{-1}(T)\text{ }\lambda
_{0}^{\left( 1\right) }\right\} \exp \left\{ -\mathcal{H}_{\gamma ,1}^{-1}(T)%
\text{ }\lambda _{0}^{\left( 2\right) }\right\}  \notag \\
&&\times \exp \left[ -\rho \int\limits_{0}^{T}\left[ 1-\overset{\wedge }{f}%
\left\{ \mathcal{G}_{\upsilon ,1}^{-1}(\tau ),\mathcal{H}_{\gamma
,1}^{-1}(\tau )\right\} \right] d\tau \right]  \notag \\
&&\times \exp \left[ -\int\limits_{\mathcal{G}_{\upsilon
,1}^{-1}(T)}^{\upsilon }\left\{ \frac{a^{\left( 1\right) }\delta ^{\left(
1\right) }\text{ }u}{\delta ^{\left( 1\right) }\text{ }u+\overset{\wedge }{g}%
\left( u\right) -1}\right\} du\right]  \notag \\
&&\times \exp \left[ -\int\limits_{\mathcal{H}_{\gamma ,1}^{-1}(T)}^{\gamma
}\left\{ \frac{a^{\left( 2\right) }\delta ^{\left( 2\right) }\text{ }u}{%
\delta ^{\left( 2\right) }\text{ }u+\overset{\wedge }{h}\left( u\right) -1}%
\right\} du\right] ,  \TCItag{3.10}
\end{eqnarray}

\textit{where}%
\begin{equation*}
\mu _{1_{G}}=\int\limits_{0}^{\infty }\text{ }ydG(y)\text{, \ \ }\mathcal{G}%
_{\upsilon ,1}(\Psi _{1})=:\int\limits_{\Psi _{1}}^{\upsilon }\left[ \frac{1%
}{\delta ^{\left( 1\right) }\text{ }u+\overset{\wedge }{g}\left( u\right) -1}%
\right] du\text{,}
\end{equation*}

\begin{equation*}
\mu _{1_{H}}=\int\limits_{0}^{\infty }zdH(z),\ \ \mathcal{H}_{\gamma
,1}(\Psi _{2})=:\int\limits_{\Psi _{2}}^{\gamma }\left[ \frac{1}{\delta
^{\left( 2\right) }\text{ }u+\overset{\wedge }{h}\left( u\right) -1}\right]
du,
\end{equation*}%
\begin{equation*}
\delta ^{\left( 1\right) }>\mu _{1_{G}}\text{ \ \textit{and \ }}\delta
^{\left( 2\right) }>\mu _{1_{H}}\text{.}
\end{equation*}

\bigskip

\textbf{Remark 1}. \ (3.10) is the conditional joint Laplace transform of
the process $\left( \lambda _{T}^{\left( 1\right) },\lambda _{T}^{\left(
2\right) }\right) $ given $\lambda _{0}^{\left( 1\right) }$\ and $\lambda
_{0}^{\left( 2\right) }$\ at time $t=0,$ where the jumps $X^{\left( 1\right)
}$ and $X^{\left( 2\right) }$ with distribution function $F\left( x^{\left(
1\right) },x^{\left( 2\right) }\right) $, occur simultaneously/collaterally
with constant intensity $\rho $. \ Because of these two dependences in the
process, this conditional joint Laplace transform is not the product of
conditional Laplace transform of $\lambda _{T}^{(1)}$ given $\lambda
_{0}^{\left( 1\right) }$ and the Laplace transform of $\lambda _{T}^{(2)}$
given $\lambda _{0}^{\left( 2\right) },$ i.e.

\begin{equation}
E\left[ e^{-\upsilon \lambda _{T}^{\left( 1\right) }}e^{-\gamma \lambda
_{T}^{\left( 2\right) }}\mid \lambda _{0}^{\left( 1\right) },\lambda
_{0}^{\left( 2\right) }\right] \neq E\left[ e^{-\upsilon \lambda
_{T}^{\left( 1\right) }}\mid \lambda _{0}^{\left( 1\right) }\right] \text{ }E%
\left[ e^{-\gamma \lambda _{T}^{\left( 2\right) }}\mid \lambda _{0}^{\left(
2\right) }\right] .  \tag{3.11}
\end{equation}

\bigskip

\textbf{Proposition 3.2.} \ \textit{The joint Laplace transform of the
asymptotic distribution of }$\left( \lambda _{T}^{\left( 1\right) },\lambda
_{T}^{\left( 2\right) }\right) $ \textit{is} \textit{given by}

\begin{eqnarray}
\underset{T\rightarrow \infty }{\lim }E\left[ e^{-\upsilon \lambda
_{T}^{\left( 1\right) }}e^{-\gamma \lambda _{T}^{\left( 2\right) }}\mid
\lambda _{0}^{\left( 1\right) },\lambda _{0}^{\left( 2\right) }\right]
&=&\exp \left[ -\rho \int\limits_{0}^{\infty }\left[ 1-\overset{\wedge }{f}%
\left\{ \mathcal{G}_{\upsilon ,1}^{-1}(\tau ),\mathcal{H}_{\gamma
,1}^{-1}(\tau )\right\} \right] d\tau \right]  \notag \\
&&\times \exp \left[ -\int\limits_{0}^{\upsilon }\left\{ \frac{a^{\left(
1\right) }\delta ^{\left( 1\right) }\text{ }u}{\delta ^{\left( 1\right) }%
\text{ }u+\overset{\wedge }{g}\left( u\right) -1}\right\} du\right]  \notag
\\
&&\times \exp \left[ -\int\limits_{0}^{\gamma }\left\{ \frac{a^{\left(
2\right) }\delta ^{\left( 2\right) }\text{ }u}{\delta ^{\left( 2\right) }%
\text{ }u+\overset{\wedge }{h}\left( u\right) -1}\right\} du\right] , 
\TCItag{3.12}
\end{eqnarray}

\textit{where }$\delta ^{\left( 1\right) }>\mu _{1_{G}}$ \ \textit{and \ }$%
\delta ^{\left( 2\right) }>\mu _{1_{H}}$.

\bigskip

\textbf{Remark 2}. We can easily derive the Laplace transform of $\lambda
_{T}^{\left( 1\right) }$ and $\lambda _{T}^{\left( 2\right) }$ for a fixed
time $T$, respectively using (3.10). \ This can also be found in Theorem 3.2
in Dassios and Zhao (2011). \ Setting $\rho =0$, we can obtain\textit{\ }the
conditional Laplace transform of $\lambda _{T}^{\left( d\right) }$ $(d=1,2)$%
\ given $\lambda _{0}^{\left( d\right) }$\ at time $t=0$ for the
self-exciting process with exponential decay. \ These processes can be
considered in modelling the bivariate intensity process only when
self-excited jumps are involved eliminating the effect of the externally
excited jumps, or to see the contribution of \textquotedblleft after-cyber
attacks\textquotedblright\ to the intensity eliminating the contribution of
\textquotedblleft initial-cyber attacks\textquotedblright\ to the intensity
in cyber insurance context.

\bigskip

\textbf{3.3} \ \textbf{Joint Probability Generating Function of} $%
(N_{T}^{\left( 1\right) },$ $N_{T}^{\left( 2\right) })$

\bigskip

We derive the\textit{\ }joint probability generating function for the
process $(N_{T}^{\left( 1\right) },$ $N_{T}^{\left( 2\right) })$ for a fixed
time $T$ in Theorem 3.2 using the result in Theorem 3.1.

\bigskip

\textbf{Theorem 3.2. \ }\textit{The conditional joint probability generating
function for the process} $(N_{T}^{\left( 1\right) },$ $N_{T}^{\left(
2\right) })$ \textit{given }$\lambda _{0}^{\left( 1\right) }$\textit{\ and }$%
\lambda _{0}^{\left( 2\right) },$\textit{\ and }$N_{0}^{\left( 1\right) }=0$ 
\textit{and} $N_{0}^{\left( 2\right) }=0$ \textit{at time} $t=0$ \textit{is
given by}

\begin{eqnarray}
&&E\left[ \theta ^{N_{T}^{\left( 1\right) }}\eta ^{N_{T}^{\left( 2\right)
}}\mid \lambda _{0}^{\left( 1\right) },\text{ }\lambda _{0}^{\left( 2\right)
}\right]  \notag \\
&=&\exp \left\{ -\mathcal{G}_{0,\theta }^{-1}(T)\text{ }\lambda _{0}^{\left(
1\right) }\right\} \exp \left\{ -\mathcal{H}_{0,\eta }^{-1}(T)\lambda
_{0}^{\left( 2\right) }\right\}  \notag \\
&&\times \exp \left[ -\rho \int\limits_{0}^{T}\left[ 1-\overset{\wedge }{f}%
\left\{ \mathcal{G}_{0,\theta }^{-1}(\tau ),\text{ }\mathcal{H}_{0,\eta
}^{-1}(\tau )\right\} \right] d\tau \right]  \notag \\
&&\times \exp \left[ -\int\limits_{0}^{\mathcal{G}_{0,\theta
}^{-1}(T)}\left\{ \frac{a^{\left( 1\right) }\delta ^{\left( 1\right) }\text{ 
}u}{1-\delta ^{\left( 1\right) }\text{ }u-\theta \overset{\wedge }{g}\left(
u\right) }\right\} du\right]  \notag \\
&&\times \exp \left[ -\int\limits_{0}^{\mathcal{H}_{0,\eta }^{-1}(T)}\left\{ 
\frac{a^{\left( 2\right) }\delta ^{\left( 2\right) }u}{1-\delta ^{\left(
2\right) }\text{ }u-\eta \overset{\wedge }{h}\left( u\right) }\right\} du%
\right] .  \TCItag{3.13}
\end{eqnarray}

\begin{proof}
By setting $t=0,$ $\nu =0,$ $\zeta =0,$ $\upsilon =0$ and $\gamma =0$ in
(3.1) with the assumption that $N_{0}^{\left( 1\right) }=0$ and $%
N_{0}^{\left( 2\right) }=0$, we have%
\begin{equation}
E\left[ \theta ^{N_{T}^{\left( 1\right) }}\eta ^{N_{T}^{\left( 2\right)
}}\mid \Im _{0}\right] =e^{-B_{1}(0)\lambda _{0}^{\left( 1\right)
}}e^{-B_{2}(0)\lambda _{0}^{\left( 2\right) }}e^{-C(T)},  \tag{3.14}
\end{equation}

where $B_{1}(0)$ is uniquely determined by the \textbf{non}-linear ordinary
differential equation (ODE)%
\begin{equation}
-B_{1}^{\prime }(t)+\delta ^{\left( 1\right) }B_{1}(t)+\theta \text{ }%
\overset{\wedge }{g}\left\{ B_{1}(t)\right\} -1=0  \tag{3.15}
\end{equation}%
with boundary condition $B_{1}(T)=0$ and similarly, $B_{2}(0)$ is uniquely
determined by the \textbf{non}-linear ODE%
\begin{equation}
-B_{2}^{\prime }(t)+\delta ^{\left( 2\right) }B_{2}(t)+\eta \text{ }\hat{h}%
\left\{ B_{2}(t)\right\} -1=0  \tag{3.16}
\end{equation}%
with boundary condition $B_{2}(T)=0.$

\bigskip

(3.15) can be solved, under the condition $\delta ^{\left( 1\right) }>\mu
_{1_{G}}$, by the following steps (1)-(7).

\bigskip

(1) \ Set $B_{1}(t)=\Psi _{1}(T-t)=\Psi _{1}(\tau ).$ \ Then it becomes%
\begin{equation}
\frac{d\Psi _{1}(\tau )}{d\tau }=1-\delta ^{\left( 1\right) }B_{1}(t)-\theta 
\overset{\wedge }{g}\left\{ B_{1}(t)\right\} =1-\delta _{1}^{\left( 1\right)
}\Psi _{1}(\tau )-\theta \overset{\wedge }{g}\left\{ \Psi _{1}(\tau
)\right\} =:f_{1}(\Psi _{1}),\text{ \ \ }0\leq \theta \leq 1  \tag{3.17}
\end{equation}%
with initial condition $\Psi _{1}(0)=0;$ \ we define the right-hand side as
the function, $f_{1}(\Psi _{1})$.

\bigskip

(2) \ There is only one positive singular point, denoted by $\upsilon ^{\ast
}>0,$ which can be obtained by solving the equation%
\begin{equation}
1-\delta ^{\left( 1\right) }u-\theta \overset{\wedge }{g}\left( u\right) =0,
\tag{3.18}
\end{equation}%
at which the uniqueness of the solution of equation (3.18) is violated. \
This is because, for the case $0<\theta <1,$ $f_{1}(\Psi _{1})=0$ is
equivalent to%
\begin{equation}
\overset{\wedge }{g}\left( u\right) =\frac{1}{\theta }\left( 1-\delta
^{\left( 1\right) }\text{ }u\right) ,\text{ \ }0<\theta <1\text{.} 
\tag{3.19}
\end{equation}%
Note that the left-hand side of (3.19) is a convex function, hence it is
clear that there is only one positive solution to $f_{1}(\Psi _{1})$. \ For
the case that $\theta =0$, there is only one singular point 
\begin{equation*}
\upsilon ^{\ast }=\frac{1}{\delta ^{\left( 1\right) }}>0.
\end{equation*}

For both cases, we have%
\begin{equation*}
\upsilon ^{\ast }=\frac{1-\theta \overset{\wedge }{g}\left( \upsilon ^{\ast
}\right) }{\delta ^{\left( 1\right) }}\geq \frac{1-\theta }{\delta ^{\left(
1\right) }}>0,
\end{equation*}%
hence, we have $f_{1}(\Psi _{1})>0$ for $0\leq \Psi _{1}<\upsilon ^{\ast }$
\ and \ $f_{1}(\Psi _{1})<0$ for $\Psi _{1}>\upsilon ^{\ast }$.

\bigskip

(3) \ (3.17) can be written as%
\begin{equation*}
\frac{d\Psi _{1}(\tau )}{1-\delta _{1}^{\left( 1\right) }\Psi _{1}(\tau
)-\theta \overset{\wedge }{g}\left\{ \Psi _{1}(\tau )\right\} }=d\tau .
\end{equation*}%
Integrate both sides from time $0$ to $\tau ,$ then we have%
\begin{equation*}
\int\limits_{0}^{\Psi _{1}(\tau )}\left[ \frac{1}{1-\delta ^{\left( 1\right)
}u-\theta \overset{\wedge }{g}\left( u\right) }\right] du=\tau ,
\end{equation*}%
where $0\leq \Psi _{1}(\tau )<\upsilon ^{\ast }$. \ Now we define the
left-hand side as the function%
\begin{equation*}
\mathcal{G}_{0,\theta }(\Psi _{1})=:\int\limits_{0}^{\Psi _{1}(\tau )}\left[ 
\frac{1}{1-\delta ^{\left( 1\right) }u-\theta \overset{\wedge }{g}\left(
u\right) }\right] du.
\end{equation*}%
Then we have%
\begin{equation*}
\mathcal{G}_{0,\theta }(\Psi _{1})=\tau \text{ \ }(=T-t),
\end{equation*}%
which is the time difference between $T$ and $t$, and it is obvious that $%
\Psi _{1}(\tau )\rightarrow 0$ when $\tau $ $\rightarrow 0$ \ and $\Psi
_{1}(\tau )\rightarrow \upsilon ^{\ast }$ when $\tau $ $\rightarrow \infty $%
. \ The integrand is positive in the domain $u\in (0,\upsilon ^{\ast }]$ and
for $\Psi _{1}(\tau )\geq 0$, $\mathcal{G}_{0,\theta }(\Psi _{1})$ is a
strictly \textbf{increasing} function. \ Therefore%
\begin{equation*}
\mathcal{G}_{0,\theta }(\Psi _{1})=\tau :[0,\upsilon ^{\ast })\rightarrow
\lbrack 0,\infty )
\end{equation*}%
is a well defined function and it inverse function%
\begin{equation*}
\mathcal{G}_{0,\theta }^{-1}(\tau )=\Psi _{1}:[0,\infty )\rightarrow \lbrack
0,\upsilon ^{\ast })
\end{equation*}%
exists.

\bigskip

(4) \ The unique solution is found by%
\begin{equation*}
\Psi _{1}\left( \tau \right) =\Psi _{1}\left( T-t\right) =B_{1}(t)=\mathcal{G%
}_{0,\theta }^{-1}(\tau )=\mathcal{G}_{0,\theta }^{-1}(T-t)
\end{equation*}%
and hence $B_{1}(0)$ is obtained,%
\begin{equation*}
B_{1}(0)=\Psi _{1}\left( T\right) =\mathcal{G}_{0,\theta }^{-1}(T).
\end{equation*}%
(5) \ Similar to solving (3.15), under the condition $\delta ^{\left(
2\right) }>\mu _{1_{H}},$ the unique solution for (3.16) is given by%
\begin{equation*}
\Psi _{2}\left( \tau \right) =\Psi _{2}\left( T-t\right) =B_{2}(t)=\mathcal{H%
}_{0,\eta }^{-1}(\tau )=\mathcal{H}_{0,\eta }^{-1}(T-t)
\end{equation*}%
and hence $B_{2}(0)$ is obtained,%
\begin{equation*}
B_{2}(0)=\Psi _{2}\left( T\right) =\mathcal{H}_{0,\eta }^{-1}(T),
\end{equation*}%
where%
\begin{equation*}
\mathcal{H}_{0,\eta }(\Psi _{2})=\int\limits_{0}^{\Psi _{2}(\tau )}\left[ 
\frac{1}{1-\delta ^{\left( 2\right) }\text{ }u-\eta \overset{\wedge }{h}%
\left( u\right) }\right] du
\end{equation*}%
is also a strictly \textbf{increasing} function: the integrand is positive
in the domain $u\in (0,\gamma ^{\ast }]$ and for $\Psi _{2}(\tau )\geq 0$ and%
\begin{equation*}
\mathcal{H}_{0,\eta }(\Psi _{2})=\tau :[0,\gamma ^{\ast })\rightarrow
\lbrack 0,\infty )
\end{equation*}%
is a well defined function and it inverse function%
\begin{equation*}
\mathcal{H}_{0,\eta }^{-1}(\tau )=\Psi _{2}:[0,\infty )\rightarrow \lbrack
0,\gamma ^{\ast })
\end{equation*}%
exists.

\bigskip

(6) \ $C(T)$ is determined by%
\begin{equation*}
C(T)=\rho \int\limits_{0}^{T}\left[ 1-\overset{\wedge }{f}\left\{ \mathcal{G}%
_{0,\theta }^{-1}(\tau ),\text{ }\mathcal{H}_{0,\eta }^{-1}(\tau )\right\} %
\right] d\tau +\delta ^{\left( 1\right) }a^{\left( 1\right)
}\int\limits_{0}^{T}\mathcal{G}_{0,\theta }^{-1}(\tau )d\tau +\delta
^{\left( 2\right) }a^{\left( 2\right) }\int\limits_{0}^{T}\mathcal{H}%
_{0,\eta }^{-1}(\tau )d\tau ,
\end{equation*}%
and by the change of variable $\mathcal{G}_{0,\theta }^{-1}(\tau )=u,$ we
have $\tau =\mathcal{G}_{0,\theta }(u)$ ($\rightarrow $ $d\tau =\frac{%
\partial \mathcal{G}_{0,\theta }(u)}{\partial u}du$), and%
\begin{equation*}
\int\limits_{0}^{T}\mathcal{G}_{0,\theta }^{-1}(\tau )d\tau
=\int\limits_{0}^{\mathcal{G}_{0,\theta }^{-1}(T)}\frac{u}{1-\delta ^{\left(
1\right) }\text{ }u-\theta \overset{\wedge }{g}\left( u\right) }du
\end{equation*}%
and similarly, $\mathcal{H}_{0,\eta }^{-1}(\tau )=u,$ we have $\tau =%
\mathcal{H}_{0,\eta }(u)$ ($\rightarrow $ $d\tau =\frac{\partial \mathcal{H}%
_{0,\eta }(u)}{\partial u}du$), and%
\begin{equation*}
\int\limits_{0}^{T}\mathcal{H}_{0,\eta }^{-1}(\tau )d\tau =\int\limits_{0}^{%
\mathcal{H}_{0,\eta }^{-1}(T)}\frac{u}{1-\delta ^{\left( 2\right) }\text{ }%
u-\eta \overset{\wedge }{h}\left( u\right) }du
\end{equation*}

\bigskip

(7) \ Finally, substitute $B_{1}(0),$ $B_{2}(0)$ and $C(T)$ into (3.14) and
the result follows.
\end{proof}

\bigskip

\textbf{Remark 3}. We can easily derive the Laplace transform of $%
N_{T}^{\left( 1\right) }$ and $N_{T}^{\left( 2\right) }$ for a fixed time $T$%
, respectively, using (3.13). \ This can also be found in Theorem 3.4 in
Dassios and Zhao (2011). \ Setting $\rho =0$, we can obtain\textit{\ }the
conditional Laplace transform of $N_{T}^{\left( d\right) }$ $(d=1,2)$\ given 
$\lambda _{0}^{\left( d\right) }$\ at time $t=0$ for the self-exciting
process with exponential decay. \ These processes can be considered in
modelling the bivariate point process only when self-excited jumps are
involved in the bivariate intensity process eliminating the effect of the
externally excited jumps, or to see the number of losses from the
contribution of \textquotedblleft after-cyber attacks\textquotedblright\ to
the intensity eliminating the contribution of \textquotedblleft
initial-cyber attacks\textquotedblright\ to the intensity in cyber insurance
context.

\bigskip

\textbf{3.4. \ Joint Laplace Transform of} $(L_{T}^{\left( 1\right) },$ $%
L_{T}^{\left( 2\right) })$

To derive the joint Laplace transform of the process $(L_{T}^{\left(
1\right) },$ $L_{T}^{\left( 2\right) })$ for a fixed time $T$, we start with
deriving the conditional joint Laplace transform, probability generating
function of the process $(\lambda _{T}^{\left( 1\right) },$ $\lambda
_{T}^{\left( 2\right) })$ and the compound point process $(L_{T}^{\left(
1\right) },$ $L_{T}^{\left( 2\right) })$ in Theorem 3.3.

\bigskip

\textbf{Theorem 3.3 \ }\textit{The conditional joint Laplace transform}, 
\textit{probability generating function of the process} $(\lambda
_{T}^{\left( 1\right) },$ $\lambda _{T}^{\left( 2\right) })$ \textit{and the
compound point process} $(L_{T}^{\left( 1\right) },$ $L_{T}^{\left( 2\right)
})$ \textit{given }$\lambda _{0}^{\left( 1\right) }$\textit{\ and }$\lambda
_{0}^{\left( 2\right) }$,\textit{\ and }$L_{0}^{\left( 1\right) }=0$ \textit{%
and} $L_{0}^{\left( 2\right) }=0$ \textit{at time} $t=0$ \textit{is given by}

\begin{eqnarray}
&&E\left[ e^{-\nu L_{T}^{\left( 1\right) }}e^{-\zeta L_{T}^{\left( 2\right)
}}\times e^{-\upsilon \lambda _{T}^{\left( 1\right) }}e^{-\gamma \lambda
_{T}^{\left( 2\right) }}\mid \lambda _{0}^{\left( 1\right) },\text{ }\lambda
_{0}^{\left( 2\right) }\right]  \notag \\
&=&\exp \left\{ -\mathcal{G}_{\upsilon ,\nu }^{-1}(T)\text{ }\lambda
_{0}^{\left( 1\right) }\right\} \exp \left\{ -\mathcal{H}_{\gamma ,\zeta
}^{-1}(T)\text{ }\lambda _{0}^{\left( 2\right) }\right\}  \notag \\
&&\times \exp \left[ -\rho \int\limits_{0}^{T}\left[ 1-\overset{\wedge }{f}%
\left\{ \mathcal{G}_{\upsilon ,\nu }^{-1}(\tau ),\mathcal{H}_{\gamma ,\zeta
}^{-1}(\tau )\right\} \right] d\tau \right]  \notag \\
&&\times \exp \left[ -\int\limits_{\mathcal{G}_{\upsilon ,\nu
}^{-1}(T)}^{\upsilon }\left\{ \frac{a^{\left( 1\right) }\delta ^{\left(
1\right) }\text{ }u}{\delta ^{\left( 1\right) }\text{ }u+\text{ }\overset{%
\wedge }{j}\left( \nu \right) \overset{\wedge }{g}\left( u\right) -1}%
\right\} du\right]  \notag \\
&&\times \exp \left[ -\int\limits_{\mathcal{H}_{\gamma ,\zeta
}^{-1}(T)}^{\gamma }\left\{ \frac{a^{\left( 2\right) }\delta ^{\left(
2\right) }\text{ }u}{\delta ^{\left( 2\right) }\text{ }u+\text{ }\overset{%
\wedge }{k}\left( \xi \right) \overset{\wedge }{h}\left( u\right) -1}%
\right\} du\right] ,  \notag \\
&&  \TCItag{3.20}
\end{eqnarray}

where

\begin{equation*}
\mu _{1_{G}}=\int\limits_{0}^{\infty }\text{ }ydG(y)\text{, \ \ }\mathcal{G}%
_{\upsilon ,\nu }(\Psi _{1})=\int\limits_{\Psi _{1}}^{\upsilon }\left[ \frac{%
1}{\delta ^{\left( 1\right) }\text{ }u+\text{ }\overset{\wedge }{j}\left(
\nu \right) \overset{\wedge }{g}\left( u\right) -1}\right] du\text{,}
\end{equation*}

\begin{equation*}
\mu _{1_{H}}=\int\limits_{0}^{\infty }\text{ }zdH(z)\text{, \ \ }\mathcal{H}%
_{\gamma ,\zeta }(\Psi _{2})=\int\limits_{\Psi _{2}}^{\gamma }\left[ \frac{1%
}{\delta ^{\left( 2\right) }\text{ }u+\text{ }\overset{\wedge }{k}\left(
\zeta \right) \overset{\wedge }{h}\left( u\right) -1}\right] du,
\end{equation*}%
\begin{equation*}
\delta ^{\left( 1\right) }>\text{ }\overset{\wedge }{j}\left( \nu \right)
\mu _{1_{G}}\text{ \ \textit{and \ }}\delta ^{\left( 2\right) }>\text{ }%
\overset{\wedge }{k}\left( \xi \right) \mu _{1_{H}}\text{.}
\end{equation*}

\begin{proof}
By setting $t=0,$ $\theta =1,$ and $\eta =1,$ in (3.1), we have%
\begin{equation}
E\left[ e^{-\nu L_{T}^{\left( 1\right) }}e^{-\zeta L_{T}^{\left( 2\right) }}%
\text{ }e^{-\upsilon \lambda _{T}^{\left( 1\right) }}e^{-\gamma \lambda
_{T}^{\left( 2\right) }}\mid \Im _{0}\right] =e^{-B_{1}(0)\lambda
_{0}^{\left( 1\right) }}e^{-B_{2}(0)\lambda _{0}^{\left( 2\right)
}}e^{-C(T)},  \tag{3.21}
\end{equation}%
where $B_{1}(0)$ is uniquely determined by the \textbf{non}-linear ordinary
differential equation (ODE)%
\begin{equation}
-B_{1}^{\prime }\left( t\right) +\delta ^{\left( 1\right) }B_{1}\left(
t\right) +\overset{\wedge }{g}\left\{ B_{1}\left( t\right) \right\} \text{ }%
\overset{\wedge }{j}\left( \nu \right) -1=0  \tag{3.22}
\end{equation}%
with boundary condition $B_{1}\left( T\right) =\upsilon $, and similarly $%
B_{2}(0)$ is uniquely determined by the \textbf{non}-linear ODE%
\begin{equation}
-B_{2}^{\prime }\left( t\right) +\delta ^{\left( 2\right) }B_{2}\left(
t\right) +\overset{\wedge }{h}\left\{ B_{2}\left( t\right) \right\} \text{ }%
\overset{\wedge }{k}\left( \zeta \right) -1=0  \tag{3.23}
\end{equation}%
with boundary condition $B_{2}(T)=\gamma .$

\bigskip

(3.22) can be solved, under the condition $\delta ^{\left( 1\right) }>$ $%
\overset{\wedge }{j}\left( \nu \right) $ $\mu _{1_{G}}$, by the following
steps (1)-(8):

\bigskip

(1) \ Let us set $B_{1}(t)=\Psi _{1}(T-t)=\Psi _{1}(\tau ).$ \ Then it
becomes%
\begin{equation}
\frac{d\Psi _{1}(\tau )}{d\tau }=1-\delta ^{\left( 1\right) }B_{1}(t)-%
\overset{\wedge }{g}\left\{ B_{1}(t)\right\} \overset{\wedge }{j}\left( \nu
\right) =1-\delta ^{\left( 1\right) }\Psi _{1}(\tau )-\overset{\wedge }{g}%
\left\{ \Psi _{1}(\tau )\right\} \overset{\wedge }{j}\left( \nu \right)
=:f_{2}(\Psi _{1})  \tag{3.24}
\end{equation}%
with initial condition $\Psi _{1}(0)=\upsilon ;$ \ we define the right-hand
side as the function, $f_{2}(\Psi _{1})$.

\bigskip

(2) \ For $\nu =0$, we have%
\begin{equation*}
f_{2}(\Psi _{1})=1-\delta ^{\left( 1\right) }\Psi _{1}(\tau )-\overset{%
\wedge }{g}\left\{ \Psi _{1}(\tau )\right\}
\end{equation*}%
and its unique solution is found by $\Psi _{1}(\tau )=\mathcal{G}_{\upsilon
,1}^{-1}(\tau ),$ that has been shown in Proposition 3.1.

\bigskip

Under the condition of $\delta ^{\left( 1\right) }>$ $\overset{\wedge }{j}%
\left( \nu \right) $ $\mu _{1_{G}}$, we have%
\begin{equation*}
\frac{\partial f_{2}(\Psi _{1})}{\partial \Psi _{1}}=\overset{\wedge }{j}%
\left( \nu \right) \int\limits_{0}^{\infty }ye^{-\Psi _{1}\text{ }y\text{ }%
}dG(y)-\delta ^{\left( 1\right) }\text{ }\leq \text{ }\overset{\wedge }{j}%
\left( \nu \right) \int\limits_{0}^{\infty }\text{ }ydG(y)-\delta ^{\left(
1\right) }=\text{ }\overset{\wedge }{j}\left( \nu \right) \mu
_{1_{G}}-\delta ^{\left( 1\right) }<0,\text{ \ for }\Psi _{1}\geq 0,
\end{equation*}%
then $f_{2}(\Psi _{1})<0$ for $\Psi _{1}>0$.

\bigskip

(3) \ (3.24) can be written as%
\begin{equation*}
\frac{d\Psi _{1}(\tau )}{\delta ^{\left( 1\right) }\Psi _{1}(\tau )-\text{ }%
\overset{\wedge }{j}\left( \nu \right) \overset{\wedge }{g}\left\{ \Psi
_{1}(\tau )\right\} -1}=-d\tau .
\end{equation*}%
Integrate both sides from time 0 to $\tau $ with initial condition $\Psi
_{1}(0)=\upsilon >0,$ then we have%
\begin{equation*}
\int\limits_{\Psi _{1}}^{\upsilon }\left[ \frac{1}{\delta ^{\left( 1\right) }%
\text{ }u+\text{ }\overset{\wedge }{j}\left( \nu \right) \overset{\wedge }{g}%
\left( u\right) -1}\right] du=\tau ,
\end{equation*}%
where $\Psi _{1}\geq 0.$ \ Now we define the left-hand side as the function%
\begin{equation*}
\mathcal{G}_{\upsilon ,\nu }(\Psi _{1})=:\int\limits_{\Psi _{1}}^{\upsilon }%
\left[ \frac{1}{\delta ^{\left( 1\right) }\text{ }u+\text{ }\overset{\wedge }%
{j}\left( \nu \right) \overset{\wedge }{g}\left( u\right) -1}\right] du.
\end{equation*}%
Then we have%
\begin{equation*}
\mathcal{G}_{\upsilon ,\nu }(\Psi _{1})=\tau \text{ }(=T-t),
\end{equation*}%
which is the time difference between $T$ and $t$ and it is obvious that $%
\Psi _{1}\rightarrow \upsilon $ when $\tau $ $(=T-t)\rightarrow 0.$

\bigskip

(4) \ As $\delta ^{\left( 1\right) }-$ $\overset{\wedge }{j}\left( \nu
\right) \mu _{1_{G}}>0$ by convergence test, we have

\begin{equation*}
\int\limits_{0}^{\upsilon }\left[ \frac{1}{\delta ^{\left( 1\right) }\text{ }%
u+\text{ }\overset{\wedge }{j}\left( \nu \right) \overset{\wedge }{g}\left(
u\right) -1}\right] du=\infty
\end{equation*}

so $\Psi _{1}\rightarrow 0$ when $\tau \rightarrow \infty .$ \ The integrand
is positive in the domain $u\in (0,\upsilon ]$ and for $\Psi _{1}\leq
\upsilon $, $\mathcal{G}_{\upsilon ,\nu }(\Psi _{1})$ is a strictly \textbf{%
decreasing} function. \ Therefore%
\begin{equation*}
\mathcal{G}_{\upsilon ,\nu }(\Psi _{1})=\tau :(0,\upsilon ]\rightarrow
\lbrack 0,\infty )
\end{equation*}

is a well defined (monotone) function and its inverse function%
\begin{equation*}
\mathcal{G}_{\upsilon ,\nu }^{-1}(\tau )=\Psi _{1}:[0,\infty )\rightarrow
(0,\upsilon ]
\end{equation*}

exists.

\bigskip

(5) \ The unique solution is found by%
\begin{equation*}
\Psi _{1}\left( \tau \right) =\Psi _{1}\left( T-t\right) =B_{1}(t)=\mathcal{G%
}_{\upsilon ,\nu }^{-1}(\tau )=\mathcal{G}_{\upsilon ,\nu }^{-1}(T-t)
\end{equation*}

and hence $B_{1}(0)$ is obtained,%
\begin{equation*}
B_{1}(0)=\Psi _{1}\left( T\right) =\mathcal{G}_{\upsilon ,\nu }^{-1}(T).
\end{equation*}

\bigskip

(6) \ Similar to solving (3.22), under the condition $\delta ^{\left(
2\right) }>$ $\overset{\wedge }{k}\left( \zeta \right) \mu _{1_{H}}$, the
unique solution for (3.23) is found by%
\begin{equation*}
\Psi _{2}\left( \tau \right) =\Psi _{2}\left( T-t\right) =B_{2}(t)=\mathcal{H%
}_{\gamma ,\zeta }^{-1}(\tau )=\mathcal{H}_{\gamma ,\zeta }^{-1}(T-t)
\end{equation*}

and hence $B_{2}(0)$ is obtained,%
\begin{equation*}
B_{2}(0)=\Psi _{2}\left( T\right) =\mathcal{H}_{\gamma ,\zeta }^{-1}(T).
\end{equation*}%
Hence%
\begin{equation*}
\mathcal{H}_{\gamma ,\zeta }(\Psi _{2})=:\int\limits_{\Psi _{2}}^{\gamma }%
\left[ \frac{1}{\delta ^{\left( 2\right) }\text{ }u+\text{ }\overset{\wedge }%
{k}\left( \zeta \right) \overset{\wedge }{h}\left( u\right) -1}\right] du
\end{equation*}%
is a strictly \textbf{decreasing} function, where the integrand is positive
in the domain $u\in (0,\gamma ]$ and for $\Psi _{2}\leq \gamma $, $\mathcal{H%
}_{\gamma ,\zeta }(\Psi _{2})$ is a strictly \textbf{decreasing} function. \
Therefore%
\begin{equation*}
\mathcal{H}_{\gamma ,\zeta }(\Psi _{2})=\tau :(0,\gamma ]\rightarrow \lbrack
0,\infty )
\end{equation*}

is a well defined (monotone) function and its inverse function%
\begin{equation*}
\mathcal{H}_{\gamma ,\zeta }^{-1}(\tau )=\Psi _{2}:[0,\infty )\rightarrow
(0,\gamma ]
\end{equation*}

exists.

\bigskip

(7) \ Now $C(T)$ is determined by%
\begin{equation*}
C(T)=\rho \int\limits_{0}^{T}\left[ 1-\overset{\wedge }{f}\left\{ \mathcal{G}%
_{\upsilon ,\nu }^{-1}(\tau ),\mathcal{H}_{\gamma ,\zeta }^{-1}(\tau
)\right\} \right] d\tau +a^{\left( 1\right) }\delta ^{\left( 1\right)
}\int\limits_{0}^{T}\mathcal{G}_{\upsilon ,\nu }^{-1}(\tau )d\tau +a^{\left(
2\right) }\delta ^{\left( 2\right) }\int\limits_{0}^{T}\mathcal{H}_{\gamma
,\zeta }^{-1}(\tau )d\tau .
\end{equation*}

By the change of variable $\mathcal{G}_{\upsilon ,\nu }^{-1}(\tau )=u,$ we
have $\tau =\mathcal{G}_{\upsilon ,\nu }^{-1}(u)$, and%
\begin{equation*}
\int\limits_{0}^{T}\mathcal{G}_{\upsilon ,\nu }^{-1}(\tau )d\tau
=\int\limits_{\mathcal{G}_{\upsilon ,\nu }^{-1}(0)}^{\mathcal{G}_{\upsilon
,\nu }^{-1}(T)}u\frac{\partial \tau }{\partial u}du=\int\limits_{\mathcal{G}%
_{\upsilon ,\nu }^{-1}(T)}^{\upsilon }\left\{ \frac{u}{\delta ^{\left(
1\right) }\text{ }u+\text{ }\overset{\wedge }{j}\left( \nu \right) \overset{%
\wedge }{g}\left( u\right) -1}\right\} du.
\end{equation*}

Similarly, we have%
\begin{equation*}
\int\limits_{0}^{T}\mathcal{H}_{\gamma ,\zeta }^{-1}(\tau )d\tau
=\int\limits_{\mathcal{H}_{\gamma ,\zeta }^{-1}(0)}^{\mathcal{H}_{\gamma
,\zeta }^{-1}(T)}u\frac{\partial \tau }{\partial u}du=\int\limits_{\mathcal{H%
}_{\gamma ,\zeta }^{-1}(\tau )(T)}^{\gamma }\left\{ \frac{u}{\delta ^{\left(
2\right) }\text{ }u+\text{ }\overset{\wedge }{k}\left( \xi \right) \overset{%
\wedge }{h}\left( u\right) -1}\right\} du.
\end{equation*}

\bigskip

(8) \ Finally, substitute $B_{1}(0),$ $B_{2}(0)$ and $C(T)$ into (3.21) and
the result follows.
\end{proof}

\bigskip

Now let us derive the joint Laplace transform of the process $(L_{T}^{\left(
1\right) },$ $L_{T}^{\left( 2\right) })$ for a fixed time $T$ in Theorem 3.4.

\bigskip

\textbf{Theorem 3.4. \ }\textit{The conditional joint Laplace transform of
the process} $(L_{T}^{\left( 1\right) },$ $L_{T}^{\left( 2\right) })$ 
\textit{given }$\lambda _{0}^{\left( 1\right) }$\textit{\ and }$\lambda
_{0}^{\left( 2\right) }$,\textit{\ and }$L_{0}^{\left( 1\right) }=0$ \textit{%
and} $L_{0}^{\left( 2\right) }=0$ \textit{at time} $t=0$ \textit{is given by}

\begin{eqnarray}
&&E\left[ e^{-\nu L_{T}^{\left( 1\right) }}e^{-\zeta L_{T}^{\left( 2\right)
}}\mid \lambda _{0}^{\left( 1\right) },\text{ }\lambda _{0}^{\left( 2\right)
}\right]  \notag \\
&=&\exp \left\{ -\mathcal{G}_{0,\nu }^{-1}(T)\text{ }\lambda _{0}^{\left(
1\right) }\right\} \exp \left\{ -\mathcal{H}_{0,\zeta }^{-1}(T)\text{ }%
\lambda _{0}^{\left( 2\right) }\right\}  \notag \\
&&\times \exp \left[ -\rho \int\limits_{0}^{T}\left[ 1-\overset{\wedge }{f}%
\left\{ \mathcal{G}_{0,\nu }^{-1}(\tau ),\mathcal{H}_{0,\zeta }^{-1}(\tau
)\right\} \right] d\tau \right]  \notag \\
&&\times \exp \left[ -\int\limits_{\mathcal{G}_{0,\nu }^{-1}(T)}^{0}\left\{ 
\frac{a^{\left( 1\right) }\delta ^{\left( 1\right) }\text{ }u}{\delta
^{\left( 1\right) }\text{ }u+\text{ }\overset{\wedge }{j}\left( \nu \right) 
\overset{\wedge }{g}\left( u\right) -1}\right\} du\right]  \notag \\
&&\times \exp \left[ -\int\limits_{\mathcal{H}_{0,\zeta
}^{-1}(T)}^{0}\left\{ \frac{a^{\left( 2\right) }\delta ^{\left( 2\right) }%
\text{ }u}{\delta ^{\left( 2\right) }\text{ }u+\text{ }\overset{\wedge }{k}%
\left( \xi \right) \overset{\wedge }{h}\left( u\right) -1}\right\} du\right]
.  \notag \\
&&  \TCItag{3.25}
\end{eqnarray}

\begin{proof}
Set $\upsilon =0$, and $\gamma =0$ in (3.20), then the result follows
immediately.
\end{proof}

\textbf{Remark 4}. We can easily derive the Laplace transform of $%
L_{T}^{\left( 1\right) }$ and $L_{T}^{\left( 2\right) }$ for a fixed time $T$%
, respectively, using (3.25). \ Setting $\rho =0$, we can obtain\textit{\ }%
the conditional Laplace transform of $L_{T}^{\left( d\right) }$ $(d=1,2)$\
given $\lambda _{0}^{\left( d\right) }$\ at time $t=0$ for the self-exciting
process with exponential decay. \ These processes can be considered in
modelling the bivariate \textit{compound} point process only when
self-excited jumps are involved in the bivariate intensity process
eliminating the effect of the externally excited jumps, or to see the
aggregate losses from the contribution of \textquotedblleft after-cyber
attacks\textquotedblright\ to the intensity eliminating the contribution of
\textquotedblleft initial-cyber attacks\textquotedblright\ to the intensity
in cyber insurance context.

\bigskip

\textbf{4.} \ \textbf{Moments,} \textbf{covariance and linear correlation }

\bigskip

In this section, we derive the expectation of $L_{t}^{\left( i\right) }$ ($%
i=1,2$) and the joint expectation of $L_{t}^{\left( 1\right) }$ and $%
L_{t}^{\left( 2\right) }$, which is another key result of this paper, for
which we need the expectations of $\lambda _{t}^{\left( 1\right) }$ and $%
\lambda _{t}^{\left( 2\right) }$, respectively and the joint expectation of $%
\lambda _{t}^{\left( 1\right) }$ and $\lambda _{t}^{\left( 2\right) }$. \ So
let us start with stating three propositions adopted from Dassios and Zhao
(2011) and Jang and Dassios (2013).

\textbf{Proposition 4.1.} \ \textit{The conditional expectation of the
process} $\lambda _{t}^{\left( 1\right) }$ \textit{given }$\lambda
_{0}^{\left( 1\right) }$\textit{\ at time} $t=0$, \textit{is} \textit{given
by}

\begin{eqnarray}
E\left( \lambda _{t}^{\left( 1\right) }\mid \lambda _{0}^{\left( 1\right)
}\right) &=&\lambda _{0}^{\left( 1\right) }e^{-\left( \delta ^{\left(
1\right) }-\mu _{1_{G}}\right) t}+\frac{\mu _{1_{F_{1}}}\text{ }\rho
+a^{\left( 1\right) }\delta ^{\left( 1\right) }}{\delta ^{\left( 1\right)
}-\mu _{1_{G}}}\left( 1-e^{-\left( \delta ^{\left( 1\right) }-\mu
_{1_{G}}\right) t}\right) ,\text{ \textit{for} }\delta ^{\left( 1\right)
}\neq \mu _{1_{G}},  \notag \\
&&  \TCItag{4.1} \\
E\left( \lambda _{t}^{\left( 1\right) }\mid \lambda _{0}^{\left( 1\right)
}\right) &=&\lambda _{0}^{\left( 1\right) }+\left( \mu _{1_{F_{1}}}\text{ }%
\rho +a^{\left( 1\right) }\delta ^{\left( 1\right) }\right) t,\text{ \textit{%
for} }\delta ^{\left( 1\right) }=\mu _{1_{G}},  \TCItag{4.2}
\end{eqnarray}

\textit{where}%
\begin{equation*}
\mu _{1_{F_{1}}}=\int\limits_{0}^{\infty }x^{\left( 1\right) }dF\left(
x^{\left( 1\right) }\right)
\end{equation*}

\textit{and }$F\left( x^{\left( 1\right) }\right) $\textit{\ is the marginal
distribution function for }$\left\{ X_{i}^{\left( 1\right) }\right\}
_{i=1,2,\cdots }$\textit{.}

\bigskip

\textit{The conditional expectation of the process} $\lambda _{t}^{\left(
2\right) }$ \textit{given }$\lambda _{0}^{\left( 2\right) }$\textit{\ at time%
} $t=0$, \textit{is} \textit{given by}

\begin{eqnarray}
E\left( \lambda _{t}^{\left( 2\right) }\mid \lambda _{0}^{\left( 2\right)
}\right) &=&\lambda _{0}^{\left( 2\right) }e^{-\left( \delta ^{\left(
2\right) }-\mu _{1_{H}}\right) t}+\frac{\mu _{1_{F_{2}}}\text{ }\rho
+a^{\left( 2\right) }\delta ^{\left( 2\right) }}{\delta ^{\left( 2\right)
}-\mu _{1_{H}}}\left( 1-e^{-\left( \delta ^{\left( 2\right) }-\mu
_{1_{H}}\right) t}\right) ,\text{ \textit{for}\ }\delta ^{\left( 2\right)
}\neq \mu _{1_{H}},  \notag \\
&&  \TCItag{4.3} \\
E\left( \lambda _{t}^{\left( 2\right) }\mid \lambda _{0}^{\left( 2\right)
}\right) &=&\lambda _{0}^{\left( 2\right) }+\left( \mu _{1_{F_{2}}}\text{ }%
\rho +a^{\left( 2\right) }\delta ^{\left( 2\right) }\right) t,\text{ \textit{%
for} }\delta ^{\left( 2\right) }=\mu _{1_{H}},  \TCItag{4.4}
\end{eqnarray}

\textit{where}%
\begin{equation*}
\mu _{1_{F_{2}}}=\int\limits_{0}^{\infty }x^{\left( 2\right) }dF\left(
x^{\left( 2\right) }\right)
\end{equation*}

\textit{and }$F\left( x^{\left( 2\right) }\right) $\textit{\ is the marginal
distribution function for }$\left\{ X_{i}^{\left( 2\right) }\right\}
_{i=1,2,\cdots }$\textit{.}

\bigskip

\textit{Assuming that }$\delta ^{\left( 1\right) }>\mu _{1_{G}}$\textit{\
and }$\delta ^{\left( 2\right) }>\mu _{1_{H}}$,\textit{\ } \textit{and
setting time }$t$ $\rightarrow \infty $ \textit{in} (4.1) \textit{and} (4.3) 
\textit{respectively}, \textit{the expectations of the stationary
distribution of the process} $\lambda _{t}^{\left( i\right) }$ ($i=1,2$) 
\textit{are given by}

\begin{equation}
E\left( \lambda _{t}^{\left( 1\right) }\right) =\frac{\mu _{1_{F_{1}}}\text{ 
}\rho +a^{\left( 1\right) }\delta ^{\left( 1\right) }}{\delta ^{\left(
1\right) }-\mu _{1_{G}}}  \tag{4.5}
\end{equation}

\textit{and}

\begin{equation}
E\left( \lambda _{t}^{\left( 2\right) }\right) =\frac{\mu _{1_{F_{2}}}\text{ 
}\rho +a^{\left( 2\right) }\delta ^{\left( 2\right) }}{\delta ^{\left(
2\right) }-\mu _{1_{H}}}.  \tag{4.6}
\end{equation}

\bigskip

\textbf{Proposition 4.2.} \ \textit{The conditional joint expectation of }$%
\lambda _{t}^{\left( 1\right) }$\textit{\ and }$\lambda _{t}^{\left(
2\right) }$ \textit{given }$\lambda _{0}^{\left( 1\right) }$\textit{\ and }$%
\lambda _{0}^{\left( 2\right) }$\textit{\ at time }$t=0$, \textit{is} 
\textit{given by}

\begin{equation*}
E\left( \lambda _{t}^{\left( 1\right) }\lambda _{t}^{\left( 2\right) }\mid
\lambda _{0}^{\left( 1\right) },\lambda _{0}^{\left( 2\right) }\right)
=\lambda _{0}^{\left( 1\right) }\lambda _{0}^{\left( 2\right) }e^{-\left\{
\left( \delta ^{\left( 1\right) }-\mu _{1_{G}}\right) +\left( \delta
^{\left( 2\right) }-\mu _{1_{H}}\right) \right\} t}
\end{equation*}

\begin{equation*}
+\left( a^{\left( 2\right) }\delta ^{\left( 2\right) }+\mu _{1_{F_{2}}}\text{
}\rho \right) \left[ 
\begin{array}{c}
\left( \lambda _{0}^{\left( 1\right) }-\frac{\mu _{1_{F_{1}}}\rho +a^{\left(
1\right) }\delta ^{\left( 1\right) }}{\delta ^{\left( 1\right) }-\mu _{1_{G}}%
}\right) \left\{ \frac{e^{-\left( \delta ^{\left( 1\right) }-\mu
_{1_{G}}\right) t}-e^{-\left\{ \left( \delta ^{\left( 1\right) }-\mu
_{1_{G}}\right) +\left( \delta ^{\left( 2\right) }-\mu _{1_{H}}\right)
\right\} t}}{\delta ^{\left( 2\right) }-\mu _{1_{H}}}\right\} \\ 
+\left( \frac{\mu _{1_{F_{1}}}\rho +a^{\left( 1\right) }\delta ^{\left(
1\right) }}{\delta ^{\left( 1\right) }-\mu _{1_{G}}}\right) \left\{ \frac{%
1-e^{-\left\{ \left( \delta ^{\left( 1\right) }-\mu _{1_{G}}\right) +\left(
\delta ^{\left( 2\right) }-\mu _{1_{H}}\right) \right\} t}}{\left( \delta
^{\left( 1\right) }-\mu _{1_{G}}\right) +\left( \delta ^{\left( 2\right)
}-\mu _{1_{H}}\right) }\right\}%
\end{array}%
\right]
\end{equation*}

\begin{equation*}
+\left( a^{\left( 1\right) }\delta ^{\left( 1\right) }+\mu _{1_{F_{1}}}\text{
}\rho \right) \left[ 
\begin{array}{c}
\left( \lambda _{0}^{\left( 2\right) }-\frac{\mu _{1_{F_{2}}}\rho +a^{\left(
2\right) }\delta ^{\left( 2\right) }}{\delta ^{\left( 2\right) }-\mu _{1_{H}}%
}\right) \left\{ \frac{e^{-\left( \delta ^{\left( 2\right) }-\mu
_{1_{H}}\right) t}-e^{-\left\{ \left( \delta ^{\left( 1\right) }-\mu
_{1_{G}}\right) +\left( \delta ^{\left( 2\right) }-\mu _{1_{H}}\right)
\right\} t}}{\delta ^{\left( 1\right) }-\mu _{1_{G}}}\right\} \\ 
+\left( \frac{\mu _{1_{F_{2}}}\rho +a^{\left( 2\right) }\delta ^{\left(
2\right) }}{\delta ^{\left( 2\right) }-\mu _{1_{H}}}\right) \left\{ \frac{%
1-e^{-\left\{ \left( \delta ^{\left( 1\right) }-\mu _{1_{G}}\right) +\left(
\delta ^{\left( 2\right) }-\mu _{1_{H}}\right) \right\} t}}{\left( \delta
^{\left( 1\right) }-\mu _{1_{G}}\right) +\left( \delta ^{\left( 2\right)
}-\mu _{1_{H}}\right) }\right\}%
\end{array}%
\right]
\end{equation*}

\begin{equation}
+\mu _{1_{F_{1,2}}}\rho \left\{ \frac{1-e^{-\left\{ \left( \delta ^{\left(
1\right) }-\mu _{1_{G}}\right) +\left( \delta ^{\left( 2\right) }-\mu
_{1_{H}}\right) \right\} t}}{\left( \delta ^{\left( 1\right) }-\mu
_{1_{G}}\right) +\left( \delta ^{\left( 2\right) }-\mu _{1_{H}}\right) }%
\right\} ,\text{ \ \ \textit{for} }\delta ^{\left( 1\right) }\neq \mu
_{1_{G}}\text{\ \textit{and} \ }\delta ^{\left( 2\right) }\neq \mu _{1_{H}}.
\tag{4.7}
\end{equation}

\bigskip

\begin{equation*}
E\left( \lambda _{t}^{\left( 1\right) }\lambda _{t}^{\left( 2\right) }\mid
\lambda _{0}^{\left( 1\right) },\lambda _{0}^{\left( 2\right) }\right)
=\lambda _{0}^{\left( 1\right) }\lambda _{0}^{\left( 2\right) }
\end{equation*}

\begin{equation*}
+\left( a^{\left( 2\right) }\delta ^{\left( 2\right) }+\mu _{1_{F_{2}}}\rho
\right) \left[ \lambda _{0}^{\left( 1\right) }t+\left( \frac{\mu
_{1_{F_{1}}}\rho +a^{\left( 1\right) }\delta ^{\left( 1\right) }}{2}\right)
t^{2}\right]
\end{equation*}

\begin{equation*}
+\left( a^{\left( 1\right) }\delta ^{\left( 1\right) }+\mu _{1_{F_{1}}}\text{
}\rho \right) \left[ \lambda _{0}^{\left( 2\right) }t+\left( \frac{\mu
_{1_{F_{2}}}\text{ }\rho +a^{\left( 2\right) }\delta ^{\left( 2\right) }}{2}%
\right) t^{2}\right]
\end{equation*}

\begin{equation}
+\mu _{1_{F_{1,2}}}\rho t,\text{ \ \ \ \ \textit{for} }\delta ^{\left(
1\right) }=\mu _{1_{G}}\text{ \ \ \textit{and} \ \ }\delta ^{\left( 2\right)
}=\mu _{1_{H}}.  \tag{4.8}
\end{equation}

\textit{where }$\mu _{1_{F_{1,2}}}=\int\limits_{0}^{\infty
}\int\limits_{0}^{\infty }x^{\left( 1\right) }x^{\left( 2\right) }dF\left(
x^{\left( 1\right) },x^{\left( 2\right) }\right) $.

\bigskip

\textit{Assuming that }$\delta ^{\left( 1\right) }>\mu _{1_{G}}$\textit{\
and }$\delta ^{\left( 2\right) }>\mu _{1_{H}}$,\textit{\ } \textit{and
setting time }$t$ $\rightarrow \infty $ \textit{in} (4.7), \textit{the joint
expectation of the stationary distribution of the process} $\lambda
_{t}^{\left( i\right) }$ ($i=1,2$) \textit{is given by}

\begin{equation}
E\left( \lambda _{t}^{\left( 1\right) }\lambda _{t}^{\left( 2\right)
}\right) =\left( \frac{1}{\left( \delta ^{\left( 1\right) }-\mu
_{1_{G}}\right) +\left( \delta ^{\left( 2\right) }-\mu _{1_{H}}\right) }%
\right) \times \left\{ 
\begin{array}{c}
\left( a^{\left( 2\right) }\delta ^{\left( 2\right) }+\mu _{1_{F_{2}}}\text{ 
}\rho \right) \left( \frac{\mu _{1_{F_{1}}}\rho +a^{\left( 1\right) }\delta
^{\left( 1\right) }}{\delta ^{\left( 1\right) }-\mu _{1_{G}}}\right) \\ 
+\left( a^{\left( 1\right) }\delta ^{\left( 1\right) }+\mu _{1_{F_{1}}}\text{
}\rho \right) \left( \frac{\mu _{1_{F_{2}}}\rho +a^{\left( 2\right) }\delta
^{\left( 2\right) }}{\delta ^{\left( 2\right) }-\mu _{1_{H}}}\right) \\ 
+\mu _{1_{F_{1,2}}}\rho%
\end{array}%
\right\}  \tag{4.9}
\end{equation}

\bigskip

\textbf{Proposition 4.3.} \ \textit{The second moment of the process} $%
\lambda _{t}^{\left( 1\right) }$ \textit{given }$\lambda _{0}^{\left(
1\right) }$\textit{\ at time} $t=0$, \textit{is} \textit{given by}

\begin{eqnarray*}
&&E\left[ \left\{ \lambda _{t}^{\left( 1\right) }\right\} ^{2}\mid \lambda
_{0}^{\left( 1\right) }\right] \\
&=&\left( \lambda _{0}^{\left( 1\right) }\right) ^{2}e^{-2\left( \delta
^{\left( 1\right) }-\mu _{1_{G}}\right) t}+\frac{2\left( \mu _{1_{F_{1}}}%
\text{ }\rho +a^{\left( 1\right) }\delta ^{\left( 1\right) }\right) +\mu
_{2_{G}}}{\delta ^{\left( 1\right) }-\mu _{1_{G}}} \\
&&\times \left( \lambda _{0}^{\left( 1\right) }-\frac{\mu _{1_{F_{1}}}\text{ 
}\rho +a^{\left( 1\right) }\delta ^{\left( 1\right) }}{\delta ^{\left(
1\right) }-\mu _{1_{G}}}\right) \left( e^{-\left( \delta ^{\left( 1\right)
}-\mu _{1_{G}}\right) t}-e^{-2\left( \delta ^{\left( 1\right) }-\mu
_{1_{G}}\right) t}\right)
\end{eqnarray*}

\begin{equation*}
+\left[ \frac{\left\{ 2\left( \mu _{1_{F_{1}}}\text{ }\rho +a^{\left(
1\right) }\delta ^{\left( 1\right) }\right) +\mu _{2_{G}}\right\} \left( \mu
_{1_{F_{1}}}\text{ }\rho +a^{\left( 1\right) }\delta ^{\left( 1\right)
}\right) }{2\left( \delta ^{\left( 1\right) }-\mu _{1_{G}}\right) ^{2}}+%
\frac{\mu _{2_{F_{1}}}\text{ }\rho }{2\left( \delta ^{\left( 1\right) }-\mu
_{1_{G}}\right) }\right] \left( 1-e^{-2\left( \delta ^{\left( 1\right) }-\mu
_{1_{G}}\right) t}\right) ,
\end{equation*}

\begin{equation}
\text{\textit{for} }\delta ^{\left( 1\right) }\neq \mu _{1_{G}},  \tag{4.10}
\end{equation}

\bigskip

\begin{eqnarray*}
&&E\left[ \left\{ \lambda _{t}^{\left( 1\right) }\right\} ^{2}\mid \lambda
_{0}^{\left( 1\right) }\right] \\
&=&\left( \lambda _{0}^{\left( 1\right) }\right) ^{2}+\left\{ 2\left( \mu
_{1_{F_{1}}}\text{ }\rho +a^{\left( 1\right) }\delta ^{\left( 1\right)
}\right) +\mu _{2_{G}}\right\} \left\{ \lambda _{0}^{\left( 1\right) }t+%
\frac{1}{2}\left( \mu _{1_{F_{1}}}\text{ }\rho +a^{\left( 1\right) }\delta
^{\left( 1\right) }\right) t^{2}\right\} +\mu _{2_{F_{1}}}\text{ }\rho t,
\end{eqnarray*}

\begin{equation}
\text{\textit{for} }\delta ^{\left( 1\right) }=\mu _{1_{G}},  \tag{4.11}
\end{equation}

\textit{where}%
\begin{equation*}
\mu _{1_{F_{1}}}=\int\limits_{0}^{\infty }x^{\left( 1\right) }dF\left(
x^{\left( 1\right) }\right) ,\text{ \ \textit{and} \ }\mu
_{2_{F_{1}}}=\int\limits_{0}^{\infty }\left\{ x^{\left( 1\right) }\right\}
^{2}dF\left( x^{\left( 1\right) }\right)
\end{equation*}

\textit{and }$F\left( x^{\left( 1\right) }\right) $\textit{\ is the marginal
distribution function for }$\left\{ X_{i}^{\left( 1\right) }\right\}
_{i=1,2,\cdots }$\textit{.}

\bigskip

\textit{The second moment of the process} $\lambda _{t}^{\left( 2\right) }$ 
\textit{given }$\lambda _{0}^{\left( 2\right) }$\textit{\ at time} $t=0$, 
\textit{is} \textit{given by}

\begin{eqnarray*}
&&E\left[ \left\{ \lambda _{t}^{\left( 2\right) }\right\} ^{2}\mid \lambda
_{0}^{\left( 2\right) }\right] \\
&=&\left( \lambda _{0}^{\left( 2\right) }\right) ^{2}e^{-2\left( \delta
^{\left( 2\right) }-\mu _{1_{H}}\right) t}+\frac{2\left( \mu _{1_{F_{2}}}%
\text{ }\rho +a^{\left( 2\right) }\delta ^{\left( 2\right) }\right) +\mu
_{2_{H}}}{\delta ^{\left( 2\right) }-\mu _{1_{H}}} \\
&&\times \left( \lambda _{0}^{\left( 2\right) }-\frac{\mu _{1_{F_{2}}}\text{ 
}\rho +a^{\left( 2\right) }\delta ^{\left( 2\right) }}{\delta ^{\left(
2\right) }-\mu _{1_{H}}}\right) \left( e^{-\left( \delta ^{\left( 2\right)
}-\mu _{1_{H}}\right) t}-e^{-2\left( \delta ^{\left( 2\right) }-\mu
_{1_{H}}\right) t}\right)
\end{eqnarray*}

\begin{equation*}
+\left[ \frac{\left\{ 2\left( \mu _{1_{F_{2}}}\text{ }\rho +a^{\left(
2\right) }\delta ^{\left( 2\right) }\right) +\mu _{2_{H}}\right\} \left( \mu
_{1_{F_{2}}}\text{ }\rho +a^{\left( 2\right) }\delta ^{\left( 2\right)
}\right) }{2\left( \delta ^{\left( 2\right) }-\mu _{1_{H}}\right) ^{2}}+%
\frac{\mu _{2_{F_{2}}}\text{ }\rho }{2\left( \delta ^{\left( 2\right) }-\mu
_{1_{H}}\right) }\right] \left( 1-e^{-2\left( \delta ^{\left( 2\right) }-\mu
_{1_{H}}\right) t}\right) ,
\end{equation*}

\begin{equation}
\text{ \textit{for} }\delta ^{\left( 2\right) }\neq \mu _{1_{H}},  \tag{4.12}
\end{equation}

\bigskip

\begin{eqnarray*}
&&E\left[ \left\{ \lambda _{t}^{\left( 2\right) }\right\} ^{2}\mid \lambda
_{0}^{\left( 2\right) }\right] \\
&=&\left( \lambda _{0}^{\left( 2\right) }\right) ^{2}+\left\{ 2\left( \mu
_{1_{F_{2}}}\text{ }\rho +a^{\left( 2\right) }\delta ^{\left( 2\right)
}\right) +\mu _{2_{H}}\right\} \left\{ \lambda _{0}^{\left( 2\right) }t+%
\frac{1}{2}\left( \mu _{1_{F_{2}}}\text{ }\rho +a^{\left( 2\right) }\delta
^{\left( 2\right) }\right) t^{2}\right\} +\mu _{2_{F_{2}}}\text{ }\rho t,%
\text{ }
\end{eqnarray*}

\begin{equation}
\text{\textit{for} }\delta ^{\left( 2\right) }=\mu _{1_{H}},  \tag{4.13}
\end{equation}

\textit{where}%
\begin{equation*}
\mu _{1_{F_{2}}}=\int\limits_{0}^{\infty }x^{\left( 2\right) }dF\left(
x^{\left( 2\right) }\right) \text{\textit{and} \ }\mu
_{2_{F_{2}}}=\int\limits_{0}^{\infty }\left\{ x^{\left( 2\right) }\right\}
^{2}dF\left( x^{\left( 2\right) }\right)
\end{equation*}

\textit{and }$F\left( x^{\left( 2\right) }\right) $\textit{\ is the marginal
distribution function for }$\left\{ X_{i}^{\left( 2\right) }\right\}
_{i=1,2,\cdots }$\textit{.}

\bigskip

\textit{Assuming that }$\delta ^{\left( 1\right) }>\mu _{1_{G}}$\textit{\
and }$\delta ^{\left( 2\right) }>\mu _{1_{H}}$,\textit{\ } \textit{and
setting time }$t$ $\rightarrow \infty $ \textit{in} (4.10) \textit{and}
(4.12) \textit{respectively}, \textit{the second moments of the stationary
distribution of the process} $\lambda _{t}^{\left( i\right) }$ ($i=1,2$) 
\textit{are given by}

\begin{equation}
E\left[ \left\{ \lambda _{t}^{\left( 1\right) }\right\} ^{2}\right] =\frac{%
\left\{ 2\left( \mu _{1_{F_{1}}}\text{ }\rho +a^{\left( 1\right) }\delta
^{\left( 1\right) }\right) +\mu _{2_{G}}\right\} \left( \mu _{1_{F_{1}}}%
\text{ }\rho +a^{\left( 1\right) }\delta ^{\left( 1\right) }\right) }{%
2\left( \delta ^{\left( 1\right) }-\mu _{1_{G}}\right) ^{2}}+\frac{\mu
_{2_{F_{1}}}\text{ }\rho }{2\left( \delta ^{\left( 1\right) }-\mu
_{1_{G}}\right) }  \tag{4.14}
\end{equation}

\textit{and}

\begin{equation}
E\left[ \left\{ \lambda _{t}^{\left( 2\right) }\right\} ^{2}\right] =\frac{%
\left\{ 2\left( \mu _{1_{F_{2}}}\text{ }\rho +a^{\left( 2\right) }\delta
^{\left( 2\right) }\right) +\mu _{2_{H}}\right\} \left( \mu _{1_{F_{2}}}%
\text{ }\rho +a^{\left( 2\right) }\delta ^{\left( 2\right) }\right) }{%
2\left( \delta ^{\left( 2\right) }-\mu _{1_{H}}\right) ^{2}}+\frac{\mu
_{2_{F_{2}}}\text{ }\rho }{2\left( \delta ^{\left( 2\right) }-\mu
_{1_{H}}\right) }.  \tag{4.15}
\end{equation}

\bigskip

Using Proposition 4.1, we now derive the expectation of $L_{T}^{\left(
i\right) }$ ($i=1,2$) directly solving an ODE in Theorem 4.1. \ We can
derive them by differentiating the Laplace transform of $L_{T}^{\left(
i\right) }$ ($i=1,2$) with respect to $\nu $ and $\xi $, and then setting $%
\nu =0$ and $\xi =0$, respectively. \ However solving the ODE directly is
easier to generalise to derive higher moments beyond the conditions $\delta
^{\left( 1\right) }>\mu _{1_{G}}$ and $\delta ^{\left( 2\right) }>\mu
_{1_{H}}$ , if necessary.

The moments of $N_{t}$ can also be derived directly solving relevant ODEs,
for which we refer you Dassios and Zhao (2011, 2017).

\bigskip

\textbf{Theorem 4.1.} \ \textit{The conditional expectation of the process} $%
L_{t}^{\left( 1\right) }$ \textit{given }$\lambda _{0}^{\left( 1\right) }$%
\textit{\ at time} $t=0$, \textit{is} \textit{given by}

\begin{eqnarray}
E\left( L_{t}^{\left( 1\right) }\mid \lambda _{0}^{\left( 1\right) }\right)
&=&L_{0}^{\left( 1\right) }+\mu _{1_{J}}\left\{ 
\begin{array}{c}
\left( \lambda _{0}^{\left( 1\right) }-\frac{\mu _{1_{F_{1}}}\text{ }\rho
+a^{\left( 1\right) }\delta ^{\left( 1\right) }}{\delta ^{\left( 1\right)
}-\mu _{1_{G}}}\right) \left( \frac{1-e^{-\left( \delta ^{\left( 1\right)
}-\mu _{1_{G}}\right) t}}{\delta ^{\left( 1\right) }-\mu _{1_{G}}}\right) \\ 
+\left( \frac{\mu _{1_{F_{1}}}\text{ }\rho +a^{\left( 1\right) }\delta
^{\left( 1\right) }}{\delta ^{\left( 1\right) }-\mu _{1_{G}}}\right) t%
\end{array}%
\right\} ,\text{ \textit{for} }\delta ^{\left( 1\right) }\neq \mu _{1_{G}}, 
\notag \\
&&  \TCItag{4.16} \\
&&  \notag \\
E\left( L_{t}^{\left( 1\right) }\mid \lambda _{0}^{\left( 1\right) }\right)
&=&L_{0}^{\left( 1\right) }+\mu _{1_{J}}\left\{ \lambda _{0}^{\left(
1\right) }t+\left( \frac{\mu _{1_{F_{1}}}\text{ }\rho +a^{\left( 1\right)
}\delta ^{\left( 1\right) }}{2}\right) t^{2}\right\} ,\text{ \textit{for} }%
\delta ^{\left( 1\right) }=\mu _{1_{G}},  \TCItag{4.17}
\end{eqnarray}

\textit{where}%
\begin{equation*}
\mu _{1_{J}}=\int\limits_{0}^{\infty }\zeta ^{\left( 1\right) }dJ(\zeta
^{\left( 1\right) })\text{.}
\end{equation*}

\bigskip

\textit{The conditional expectation of the process} $L_{t}^{\left( 2\right)
} $ \textit{given }$\lambda _{0}^{\left( 2\right) }$\textit{\ at time} $t=0$%
, \textit{is} \textit{given by}

\begin{eqnarray}
E\left( L_{t}^{\left( 2\right) }\mid \lambda _{0}^{\left( 2\right) }\right)
&=&L_{0}^{\left( 2\right) }+\mu _{1_{K}}\left\{ 
\begin{array}{c}
\left( \lambda _{0}^{\left( 2\right) }-\frac{\mu _{1_{F_{2}}}\text{ }\rho
+a^{\left( 2\right) }\delta ^{\left( 2\right) }}{\delta ^{\left( 2\right)
}-\mu _{1_{H}}}\right) \left( \frac{1-e^{-\left( \delta ^{\left( 2\right)
}-\mu _{1_{H}}\right) t}}{\delta ^{\left( 2\right) }-\mu _{1_{H}}}\right) \\ 
+\left( \frac{\mu _{1_{F_{2}}}\text{ }\rho +a^{\left( 2\right) }\delta
^{\left( 2\right) }}{\delta ^{\left( 2\right) }-\mu _{1_{H}}}\right) t%
\end{array}%
\right\} ,\text{ \textit{for} }\delta ^{\left( 2\right) }\neq \mu _{1_{H}}, 
\notag \\
&&  \TCItag{4.18} \\
E\left( L_{t}^{\left( 2\right) }\mid \lambda _{0}^{\left( 2\right) }\right)
&=&L_{0}^{\left( 2\right) }+\mu _{1_{K}}\left\{ \lambda _{0}^{\left(
2\right) }t+\left( \frac{\mu _{1_{F_{2}}}\text{ }\rho +a^{\left( 2\right)
}\delta ^{\left( 2\right) }}{2}\right) t^{2}\right\} ,\text{ \textit{for} }%
\delta ^{\left( 2\right) }=\mu _{1_{H}},  \TCItag{4.19}
\end{eqnarray}

\textit{where}%
\begin{equation*}
\mu _{1_{K}}=\int\limits_{0}^{\infty }\zeta ^{\left( 2\right) }dK(\zeta
^{\left( 2\right) }).
\end{equation*}

\begin{proof}
See Appendix A.
\end{proof}

\bigskip

\textbf{Corollary 4.1. }\textit{For the stationary distribution of the
process} $\lambda _{t}^{\left( 1\right) }$, \textit{given} $L_{0}^{\left(
1\right) }=0$, \textit{the expectation of the process} $L_{t}^{\left(
1\right) }$\textit{is given by}

\begin{equation}
E\left( L_{t}^{\left( 1\right) }\right) =\mu _{1_{J}}\left( \frac{\mu
_{1_{F_{1}}}\text{ }\rho +a^{\left( 1\right) }\delta ^{\left( 1\right) }}{%
\delta ^{\left( 1\right) }-\mu _{1_{G}}}\right) t,\text{ \ }\delta ^{\left(
1\right) }>\mu _{1_{G}}  \tag{4.20}
\end{equation}

\textit{and for the stationary distribution of the process} $\lambda
_{t}^{\left( 2\right) }$, \textit{given} $L_{0}^{\left( 2\right) }=0$, 
\textit{the expectation of the process} $L_{t}^{\left( 2\right) }$\textit{is
given by}

\begin{equation}
E\left( L_{t}^{\left( 2\right) }\right) =\mu _{1_{K}}\left( \frac{\mu
_{1_{F_{2}}}\text{ }\rho +a^{\left( 2\right) }\delta ^{\left( 2\right) }}{%
\delta ^{\left( 2\right) }-\mu _{1_{H}}}\right) t,\text{ \ }\delta ^{\left(
2\right) }>\mu _{1_{H}}.  \tag{4.21}
\end{equation}

\begin{proof}
See Appendix B.
\end{proof}

\bigskip

We now derive the joint expectation of $L_{t}^{\left( 1\right) }$ and $%
L_{t}^{\left( 2\right) }$ in Theorem 3.6, for which we start with a lemma to
show the joint expectation of $\lambda _{t}^{\left( 1\right) }L_{t}^{\left(
2\right) }$ and the joint expectation of $\lambda _{t}^{\left( 2\right)
}L_{t}^{\left( 1\right) }$, respectively. \ For simplicity, we use the case
for the stationary distribution of the process $\lambda _{t}^{\left(
i\right) }$ ($i=1,2$). \ It can serve a reasonable approximation for the
joint expectation of $L_{t}^{\left( 1\right) }$ and $L_{t}^{\left( 2\right)
} $ provided that the process the has been running for a relatively long
period and is close to the stationary (asymptotic) state.

\bigskip

\textbf{Lemma 4.1. \ }\textit{For the stationary distribution of the process}
$\lambda _{t}^{\left( 1\right) }$ \textit{and} $\lambda _{t}^{\left(
2\right) }$, \textit{given} $L_{0}^{\left( 2\right) }=0$, \textit{the joint
expectation of }$\lambda _{t}^{\left( 1\right) }$\textit{\ and }$%
L_{t}^{\left( 2\right) }$ \textit{is} \textit{given by}

\begin{equation*}
E\left( \lambda _{t}^{\left( 1\right) }L_{t}^{\left( 2\right) }\right) =\mu
_{1_{K}}\left( a^{\left( 1\right) }\delta ^{\left( 1\right) }+\mu
_{1_{F_{1}}}\rho \right) \left\{ \frac{\mu _{1_{F_{2}}}\text{ }\rho
+a^{\left( 2\right) }\delta ^{\left( 2\right) }}{\left( \delta ^{\left(
2\right) }-\mu _{1_{H}}\right) \left( \delta ^{\left( 1\right) }-\mu
_{1_{G}}\right) }\right\} t
\end{equation*}%
\begin{eqnarray*}
&&+\mu _{1_{K}}\left( \frac{1-e^{-\left( \delta ^{\left( 1\right) }-\mu
_{1_{G}}\right) t}}{\delta ^{\left( 1\right) }-\mu _{1_{G}}}\right) \\
&&\times \left[ 
\begin{array}{c}
\left[ \left( \frac{1}{\left( \delta ^{\left( 1\right) }-\mu _{1_{G}}\right)
+\left( \delta ^{\left( 2\right) }-\mu _{1_{H}}\right) }\right) \left\{ 
\begin{array}{c}
\left( a^{\left( 2\right) }\delta ^{\left( 2\right) }+\mu _{1_{F_{2}}}\text{ 
}\rho \right) \left( \frac{\mu _{1_{F_{1}}}\rho +a^{\left( 1\right) }\delta
^{\left( 1\right) }}{\delta ^{\left( 1\right) }-\mu _{1_{G}}}\right) \\ 
+\left( a^{\left( 1\right) }\delta ^{\left( 1\right) }+\mu _{1_{F_{1}}}\text{
}\rho \right) \left( \frac{\mu _{1_{F_{2}}}\rho +a^{\left( 2\right) }\delta
^{\left( 2\right) }}{\delta ^{\left( 2\right) }-\mu _{1_{H}}}\right) +\mu
_{1_{F_{1,2}}}\rho%
\end{array}%
\right\} \right] \\ 
-\left( a^{\left( 1\right) }\delta ^{\left( 1\right) }+\mu _{1_{F_{1}}}\rho
\right) \left\{ \frac{\mu _{1_{F_{2}}}\text{ }\rho +a^{\left( 2\right)
}\delta ^{\left( 2\right) }}{\left( \delta ^{\left( 2\right) }-\mu
_{1_{H}}\right) \left( \delta ^{\left( 1\right) }-\mu _{1_{G}}\right) }%
\right\}%
\end{array}%
\right]
\end{eqnarray*}

\begin{equation}
\text{\textit{for} }\delta ^{\left( 1\right) }>\mu _{1_{G}}\text{\ \textit{%
and} \ }\delta ^{\left( 2\right) }>\mu _{1_{H}}\text{,}  \tag{4.22}
\end{equation}

\textit{and} \textit{given} $L_{0}^{\left( 1\right) }=0$, \textit{the joint
expectation of }$\lambda _{t}^{\left( 2\right) }$\textit{\ and }$%
L_{t}^{\left( 1\right) }$ \textit{is} \textit{given by}

\begin{equation*}
E\left( \lambda _{t}^{\left( 2\right) }L_{t}^{\left( 1\right) }\right) =\mu
_{1_{J}}\left( a^{\left( 2\right) }\delta ^{\left( 2\right) }+\mu
_{1_{F_{2}}}\rho \right) \left\{ \frac{\mu _{1_{F_{1}}}\text{ }\rho
+a^{\left( 1\right) }\delta ^{\left( 1\right) }}{\left( \delta ^{\left(
2\right) }-\mu _{1_{H}}\right) \left( \delta ^{\left( 1\right) }-\mu
_{1_{G}}\right) }\right\} t
\end{equation*}%
\begin{eqnarray*}
&&+\mu _{1_{J}}\left( \frac{1-e^{-\left( \delta ^{\left( 2\right) }-\mu
_{1_{H}}\right) t}}{\delta ^{\left( 2\right) }-\mu _{1_{H}}}\right) \\
&&\times \left[ 
\begin{array}{c}
\left[ \left( \frac{1}{\left( \delta ^{\left( 1\right) }-\mu _{1_{G}}\right)
+\left( \delta ^{\left( 2\right) }-\mu _{1_{H}}\right) }\right) \left\{ 
\begin{array}{c}
\left( a^{\left( 2\right) }\delta ^{\left( 2\right) }+\mu _{1_{F_{2}}}\text{ 
}\rho \right) \left( \frac{\mu _{1_{F_{1}}}\rho +a^{\left( 1\right) }\delta
^{\left( 1\right) }}{\delta ^{\left( 1\right) }-\mu _{1_{G}}}\right) \\ 
+\left( a^{\left( 1\right) }\delta ^{\left( 1\right) }+\mu _{1_{F_{1}}}\text{
}\rho \right) \left( \frac{\mu _{1_{F_{2}}}\rho +a^{\left( 2\right) }\delta
^{\left( 2\right) }}{\delta ^{\left( 2\right) }-\mu _{1_{H}}}\right) +\mu
_{1_{F_{1,2}}}\rho%
\end{array}%
\right\} \right] \\ 
-\left( a^{\left( 2\right) }\delta ^{\left( 2\right) }+\mu _{1_{F_{2}}}\rho
\right) \left\{ \frac{\mu _{1_{F_{1}}}\text{ }\rho +a^{\left( 1\right)
}\delta ^{\left( 1\right) }}{\left( \delta ^{\left( 2\right) }-\mu
_{1_{H}}\right) \left( \delta ^{\left( 1\right) }-\mu _{1_{G}}\right) }%
\right\}%
\end{array}%
\right]
\end{eqnarray*}

\begin{equation}
\text{\textit{for} }\delta ^{\left( 1\right) }>\mu _{1_{G}}\text{\ \textit{%
and} \ }\delta ^{\left( 2\right) }>\mu _{1_{H}}\text{,}  \tag{4.23}
\end{equation}

\begin{proof}
See Appendix C.
\end{proof}

\bigskip

\textbf{Theorem 4.2. \ }\textit{For the stationary distribution of the
process} $\lambda _{t}^{\left( 1\right) }$ \textit{and} $\lambda
_{t}^{\left( 2\right) }$, \textit{given} $L_{0}^{\left( 1\right)
}=L_{0}^{\left( 2\right) }=0$, \textit{the joint expectation of }$%
L_{t}^{\left( 1\right) }$\textit{\ and }$L_{t}^{\left( 2\right) }$ \textit{is%
} \textit{given by}

\begin{equation*}
E\left( L_{t}^{\left( 1\right) }L_{t}^{\left( 2\right) }\right) =\mu
_{1_{J}}\mu _{1_{K}}\left( \frac{a^{\left( 1\right) }\delta ^{\left(
1\right) }+\mu _{1_{F_{1}}}\rho }{2}\right) \left\{ \frac{\mu _{1_{F_{2}}}%
\text{ }\rho +a^{\left( 2\right) }\delta ^{\left( 2\right) }}{\left( \delta
^{\left( 2\right) }-\mu _{1_{H}}\right) \left( \delta ^{\left( 1\right)
}-\mu _{1_{G}}\right) }\right\} t^{2}
\end{equation*}

\begin{equation*}
+\mu _{1_{J}}\mu _{1_{K}}\left( \frac{1}{\delta ^{\left( 1\right) }-\mu
_{1_{G}}}\right) \left\{ t-\left( \frac{1-e^{-\left( \delta ^{\left(
1\right) }-\mu _{1_{G}}\right) t}}{\delta ^{\left( 1\right) }-\mu _{1_{G}}}%
\right) \right\}
\end{equation*}

\begin{equation*}
\times \left[ 
\begin{array}{c}
\left[ \left( \frac{1}{\left( \delta ^{\left( 1\right) }-\mu _{1_{G}}\right)
+\left( \delta ^{\left( 2\right) }-\mu _{1_{H}}\right) }\right) \left\{ 
\begin{array}{c}
\left( a^{\left( 2\right) }\delta ^{\left( 2\right) }+\mu _{1_{F_{2}}}\text{ 
}\rho \right) \left( \frac{\mu _{1_{F_{1}}}\rho +a^{\left( 1\right) }\delta
^{\left( 1\right) }}{\delta ^{\left( 1\right) }-\mu _{1_{G}}}\right) \\ 
+\left( a^{\left( 1\right) }\delta ^{\left( 1\right) }+\mu _{1_{F_{1}}}\text{
}\rho \right) \left( \frac{\mu _{1_{F_{2}}}\rho +a^{\left( 2\right) }\delta
^{\left( 2\right) }}{\delta ^{\left( 2\right) }-\mu _{1_{H}}}\right) +\mu
_{1_{F_{1,2}}}\rho%
\end{array}%
\right\} \right] \\ 
-\left( a^{\left( 1\right) }\delta ^{\left( 1\right) }+\mu _{1_{F_{1}}}\rho
\right) \left\{ \frac{\mu _{1_{F_{2}}}\text{ }\rho +a^{\left( 2\right)
}\delta ^{\left( 2\right) }}{\left( \delta ^{\left( 2\right) }-\mu
_{1_{H}}\right) \left( \delta ^{\left( 1\right) }-\mu _{1_{G}}\right) }%
\right\}%
\end{array}%
\right]
\end{equation*}

\begin{equation*}
+\mu _{1_{K}}\mu _{1_{J}}\left( \frac{a^{\left( 2\right) }\delta ^{\left(
2\right) }+\mu _{1_{F_{2}}}\rho }{2}\right) \left\{ \frac{\mu _{1_{F_{1}}}%
\text{ }\rho +a^{\left( 1\right) }\delta ^{\left( 1\right) }}{\left( \delta
^{\left( 2\right) }-\mu _{1_{H}}\right) \left( \delta ^{\left( 1\right)
}-\mu _{1_{G}}\right) }\right\} t^{2}
\end{equation*}

\begin{equation*}
+\mu _{1_{K}}\mu _{1_{J}}\left( \frac{1}{\delta ^{\left( 2\right) }-\mu
_{1_{H}}}\right) \left\{ t-\left( \frac{1-e^{-\left( \delta ^{\left(
2\right) }-\mu _{1_{H}}\right) t}}{\delta ^{\left( 2\right) }-\mu _{1_{H}}}%
\right) \right\}
\end{equation*}

\begin{equation*}
\times \left[ 
\begin{array}{c}
\left[ \left( \frac{1}{\left( \delta ^{\left( 1\right) }-\mu _{1_{G}}\right)
+\left( \delta ^{\left( 2\right) }-\mu _{1_{H}}\right) }\right) \left\{ 
\begin{array}{c}
\left( a^{\left( 2\right) }\delta ^{\left( 2\right) }+\mu _{1_{F_{2}}}\text{ 
}\rho \right) \left( \frac{\mu _{1_{F_{1}}}\rho +a^{\left( 1\right) }\delta
^{\left( 1\right) }}{\delta ^{\left( 1\right) }-\mu _{1_{G}}}\right) \\ 
+\left( a^{\left( 1\right) }\delta ^{\left( 1\right) }+\mu _{1_{F_{1}}}\text{
}\rho \right) \left( \frac{\mu _{1_{F_{2}}}\rho +a^{\left( 2\right) }\delta
^{\left( 2\right) }}{\delta ^{\left( 2\right) }-\mu _{1_{H}}}\right) +\mu
_{1_{F_{1,2}}}\rho%
\end{array}%
\right\} \right] \\ 
-\left( a^{\left( 2\right) }\delta ^{\left( 2\right) }+\mu _{1_{F_{2}}}\rho
\right) \left\{ \frac{\mu _{1_{F_{1}}}\text{ }\rho +a^{\left( 1\right)
}\delta ^{\left( 1\right) }}{\left( \delta ^{\left( 2\right) }-\mu
_{1_{H}}\right) \left( \delta ^{\left( 1\right) }-\mu _{1_{G}}\right) }%
\right\}%
\end{array}%
\right] ,
\end{equation*}

\begin{equation}
\text{\textit{for} }\delta ^{\left( 1\right) }\neq \mu _{1_{G}}\text{\ 
\textit{and} \ }\delta ^{\left( 2\right) }\neq \mu _{1_{H}}.  \tag{4.24}
\end{equation}

\begin{proof}
See Appendix D.
\end{proof}

\bigskip

Based on Theorem 4.2 and Corollary 4.1, we can easily obtain the covariance
between $L_{t}^{\left( 1\right) }$\ and $L_{t}^{\left( 2\right) }$, i.e.

\begin{equation}
Cov\left( L_{t}^{\left( 1\right) },L_{t}^{\left( 2\right) }\right) =E\left(
L_{t}^{\left( 1\right) }L_{t}^{\left( 2\right) }\right) -E\left(
L_{t}^{\left( 1\right) }\right) E\left( L_{t}^{\left( 2\right) }\right) 
\tag{4.25}
\end{equation}%
and the linear correlation coefficient between $L_{t}^{\left( 1\right) }$\
and $L_{t}^{\left( 2\right) }$, i.e.

\begin{equation}
Corr\left( L_{t}^{\left( 1\right) },L_{t}^{\left( 2\right) }\right) =\frac{%
Cov\left( L_{t}^{\left( 1\right) },L_{t}^{\left( 2\right) }\right) }{\sqrt{%
Var\left( L_{t}^{\left( 1\right) }\right) }\sqrt{Var\left( L_{t}^{\left(
2\right) }\right) }}  \tag{4.26}
\end{equation}%
and hence we omit their corresponding expressions. \ We show their numerical
values in cyber insurance context in Section 5.

For the correlation coefficient calculation, we need variance of $%
L_{t}^{\left( 1\right) }$ and $L_{t}^{\left( 2\right) }$, respectively, for
which we start with a lemma to show the joint expectation of $\lambda
_{t}^{\left( 1\right) }L_{t}^{\left( 1\right) }$ and the joint expectation
of $\lambda _{t}^{\left( 2\right) }L_{t}^{\left( 2\right) }$, respectively.

\bigskip

\textbf{Lemma 4.2. \ }\textit{For the stationary distribution of the process}
$\lambda _{t}^{\left( 1\right) }$ \textit{and} $\lambda _{t}^{\left(
2\right) }$, \textit{given} $L_{0}^{\left( 1\right) }=0$, \textit{the joint
expectation of }$\lambda _{t}^{\left( 1\right) }$\textit{\ and }$%
L_{t}^{\left( 1\right) }$ \textit{is} \textit{given by}

\begin{equation*}
E\left( \lambda _{t}^{\left( 1\right) }L_{t}^{\left( 1\right) }\right)
=\left( a^{\left( 1\right) }\delta ^{\left( 1\right) }+\mu _{1_{F_{1}}}\rho
\right) \mu _{1_{J}}\left\{ \frac{\mu _{1_{F_{1}}}\text{ }\rho +a^{\left(
1\right) }\delta ^{\left( 1\right) }}{\left( \delta ^{\left( 1\right) }-\mu
_{1_{G}}\right) ^{2}}\right\} \left\{ t-\left( \frac{1-e^{-\left( \delta
^{\left( 1\right) }-\mu _{1_{G}}\right) t}}{\delta ^{\left( 1\right) }-\mu
_{1_{G}}}\right) \right\}
\end{equation*}%
\begin{eqnarray*}
&&+\mu _{1_{J}}\left( \frac{1-e^{-\left( \delta ^{\left( 1\right) }-\mu
_{1_{G}}\right) t}}{\delta ^{\left( 1\right) }-\mu _{1_{G}}}\right) \\
&&\times \left[ \frac{\left\{ 2\left( \mu _{1_{F_{1}}}\text{ }\rho
+a^{\left( 1\right) }\delta ^{\left( 1\right) }\right) +\mu _{2_{G}}\right\}
\left( \mu _{1_{F_{1}}}\text{ }\rho +a^{\left( 1\right) }\delta ^{\left(
1\right) }\right) }{2\left( \delta ^{\left( 1\right) }-\mu _{1_{G}}\right)
^{2}}+\frac{\mu _{2_{F_{1}}}\text{ }\rho }{2\left( \delta ^{\left( 1\right)
}-\mu _{1_{G}}\right) }\right] \\
&&+\mu _{1_{G}}\mu _{1_{J}}\left( \frac{1-e^{-\left( \delta ^{\left(
1\right) }-\mu _{1_{G}}\right) t}}{\delta ^{\left( 1\right) }-\mu _{1_{G}}}%
\right) \left( \frac{\mu _{1_{F_{1}}}\text{ }\rho +a^{\left( 1\right)
}\delta ^{\left( 1\right) }}{\delta ^{\left( 1\right) }-\mu _{1_{G}}}\right)
\end{eqnarray*}

\begin{equation}
\text{\textit{for} }\delta ^{\left( 1\right) }>\mu _{1_{G}}\text{\ ,} 
\tag{4.27}
\end{equation}

\textit{and} \textit{given} $L_{0}^{\left( 1\right) }=0$, \textit{the joint
expectation of }$\lambda _{t}^{\left( 2\right) }$\textit{\ and }$%
L_{t}^{\left( 2\right) }$ \textit{is} \textit{given by}

\begin{equation*}
E\left( \lambda _{t}^{\left( 2\right) }L_{t}^{\left( 2\right) }\right)
=\left( a^{\left( 2\right) }\delta ^{\left( 2\right) }+\mu _{1_{F_{2}}}\rho
\right) \mu _{1_{K}}\left\{ \frac{\mu _{1_{F_{2}}}\text{ }\rho +a^{\left(
2\right) }\delta ^{\left( 2\right) }}{\left( \delta ^{\left( 2\right) }-\mu
_{1_{H}}\right) ^{2}}\right\} \left\{ t-\left( \frac{1-e^{-\left( \delta
^{\left( 2\right) }-\mu _{1_{H}}\right) t}}{\delta ^{\left( 2\right) }-\mu
_{1_{H}}}\right) \right\}
\end{equation*}%
\begin{eqnarray*}
&&+\mu _{1_{K}}\left( \frac{1-e^{-\left( \delta ^{\left( 2\right) }-\mu
_{1_{H}}\right) t}}{\delta ^{\left( 2\right) }-\mu _{1_{H}}}\right) \\
&&\times \left[ \frac{\left\{ 2\left( \mu _{1_{F_{2}}}\text{ }\rho
+a^{\left( 2\right) }\delta ^{\left( 2\right) }\right) +\mu _{2_{H}}\right\}
\left( \mu _{1_{F_{2}}}\text{ }\rho +a^{\left( 2\right) }\delta ^{\left(
2\right) }\right) }{2\left( \delta ^{\left( 2\right) }-\mu _{1_{H}}\right)
^{2}}+\frac{\mu _{2_{F_{2}}}\text{ }\rho }{2\left( \delta ^{\left( 2\right)
}-\mu _{1_{H}}\right) }\right] \\
&&+\mu _{1_{H}}\mu _{1_{K}}\left( \frac{1-e^{-\left( \delta ^{\left(
2\right) }-\mu _{1_{H}}\right) t}}{\delta ^{\left( 2\right) }-\mu _{1_{H}}}%
\right) \left( \frac{\mu _{1_{F_{2}}}\text{ }\rho +a^{\left( 2\right)
}\delta ^{\left( 2\right) }}{\delta ^{\left( 2\right) }-\mu _{1_{H}}}\right)
\end{eqnarray*}

\begin{equation}
\text{\textit{for} }\delta ^{\left( 2\right) }>\mu _{1_{H}}\text{,} 
\tag{4.28}
\end{equation}

\begin{proof}
See Appendix E.
\end{proof}

\bigskip

\textbf{Theorem 4.3. \ }\textit{For the stationary distribution of the
process} $\lambda _{t}^{\left( 1\right) }$ \textit{and} $\lambda
_{t}^{\left( 2\right) }$, \textit{given} $L_{0}^{\left( 1\right)
}=L_{0}^{\left( 2\right) }=0$, \textit{the second moment of the process of }$%
L_{t}^{\left( 1\right) }$\textit{\ is} \textit{given by}

\begin{equation*}
E\left\{ \left( L_{t}^{\left( 1\right) }\right) ^{2}\right\}
\end{equation*}

\begin{eqnarray}
&=&2\mu _{1_{J}}\left[ 
\begin{array}{c}
\left( \frac{a^{\left( 1\right) }\delta ^{\left( 1\right) }+\mu
_{1_{F_{1}}}\rho }{2}\right) \mu _{1_{J}}\left\{ \frac{\mu _{1_{F_{1}}}\text{
}\rho +a^{\left( 1\right) }\delta ^{\left( 1\right) }}{\left( \delta
^{\left( 1\right) }-\mu _{1_{G}}\right) ^{2}}\right\} t^{2} \\ 
-\left( a^{\left( 1\right) }\delta ^{\left( 1\right) }+\mu _{1_{F_{1}}}\rho
\right) \mu _{1_{J}}\left\{ \frac{\mu _{1_{F_{1}}}\text{ }\rho +a^{\left(
1\right) }\delta ^{\left( 1\right) }}{\left( \delta ^{\left( 1\right) }-\mu
_{1_{G}}\right) ^{3}}\right\} \left\{ t-\left( \frac{1-e^{-\left( \delta
^{\left( 1\right) }-\mu _{1_{G}}\right) t}}{\delta ^{\left( 1\right) }-\mu
_{1_{G}}}\right) \right\} \\ 
+\mu _{1_{J}}\text{ }\left[ \frac{\left\{ 2\left( \mu _{1_{F_{1}}}\text{ }%
\rho +a^{\left( 1\right) }\delta ^{\left( 1\right) }\right) +\mu
_{2_{G}}\right\} \left( \mu _{1_{F_{1}}}\text{ }\rho +a^{\left( 1\right)
}\delta ^{\left( 1\right) }\right) }{2\left( \delta ^{\left( 1\right) }-\mu
_{1_{G}}\right) ^{3}}+\frac{\mu _{2_{F_{1}}}\text{ }\rho }{2\left( \delta
^{\left( 1\right) }-\mu _{1_{G}}\right) ^{2}}\right] \left\{ t-\left( \frac{%
1-e^{-\left( \delta ^{\left( 1\right) }-\mu _{1_{G}}\right) t}}{\delta
^{\left( 1\right) }-\mu _{1_{G}}}\right) \right\} \\ 
+\mu _{1_{G}}\mu _{1_{J}}\text{ }\left\{ \frac{\mu _{1_{F_{1}}}\text{ }\rho
+a^{\left( 1\right) }\delta ^{\left( 1\right) }}{\left( \delta ^{\left(
1\right) }-\mu _{1_{G}}\right) ^{2}}\right\} \left\{ t-\left( \frac{%
1-e^{-\left( \delta ^{\left( 1\right) }-\mu _{1_{G}}\right) t}}{\delta
^{\left( 1\right) }-\mu _{1_{G}}}\right) \right\}%
\end{array}%
\right]  \notag \\
&&+\mu _{2_{J}}\left( \frac{\mu _{1_{F_{1}}}\text{ }\rho +a^{\left( 1\right)
}\delta ^{\left( 1\right) }}{\delta ^{\left( 1\right) }-\mu _{1_{G}}}\right)
t  \notag \\
&&  \TCItag{4.29}
\end{eqnarray}

\textit{and the second moment of the process of }$L_{t}^{\left( 2\right) }$%
\textit{\ is} \textit{given by}

\begin{equation*}
E\left\{ \left( L_{t}^{\left( 2\right) }\right) ^{2}\right\}
\end{equation*}

\begin{eqnarray}
&=&2\mu _{1_{K}}\left[ 
\begin{array}{c}
\left( \frac{a^{\left( 2\right) }\delta ^{\left( 2\right) }+\mu
_{1_{F_{2}}}\rho }{2}\right) \mu _{1_{K}}\left\{ \frac{\mu _{1_{F_{2}}}\text{
}\rho +a^{\left( 2\right) }\delta ^{\left( 2\right) }}{\left( \delta
^{\left( 2\right) }-\mu _{1_{H}}\right) ^{2}}\right\} t^{2} \\ 
-\left( a^{\left( 2\right) }\delta ^{\left( 2\right) }+\mu _{1_{F_{2}}}\rho
\right) \mu _{1_{K}}\left\{ \frac{\mu _{1_{F_{2}}}\text{ }\rho +a^{\left(
2\right) }\delta ^{\left( 2\right) }}{\left( \delta ^{\left( 2\right) }-\mu
_{1_{H}}\right) ^{3}}\right\} \left\{ t-\left( \frac{1-e^{-\left( \delta
^{\left( 2\right) }-\mu _{1_{H}}\right) t}}{\delta ^{\left( 2\right) }-\mu
_{1_{H}}}\right) \right\} \\ 
+\mu _{1_{K}}\text{ }\left[ \frac{\left\{ 2\left( \mu _{1_{F_{2}}}\text{ }%
\rho +a^{\left( 2\right) }\delta ^{\left( 2\right) }\right) +\mu
_{2_{H}}\right\} \left( \mu _{1_{F_{2}}}\text{ }\rho +a^{\left( 2\right)
}\delta ^{\left( 2\right) }\right) }{2\left( \delta ^{\left( 2\right) }-\mu
_{1_{H}}\right) ^{3}}+\frac{\mu _{2_{F_{2}}}\text{ }\rho }{2\left( \delta
^{\left( 2\right) }-\mu _{1_{H}}\right) ^{2}}\right] \left\{ t-\left( \frac{%
1-e^{-\left( \delta ^{\left( 2\right) }-\mu _{1_{H}}\right) t}}{\delta
^{\left( 2\right) }-\mu _{1_{H}}}\right) \right\} \\ 
+\mu _{1_{H}}\mu _{1_{K}}\text{ }\left\{ \frac{\mu _{1_{F_{2}}}\text{ }\rho
+a^{\left( 2\right) }\delta ^{\left( 2\right) }}{\left( \delta ^{\left(
2\right) }-\mu _{1_{H}}\right) ^{2}}\right\} \left\{ t-\left( \frac{%
1-e^{-\left( \delta ^{\left( 2\right) }-\mu _{1_{H}}\right) t}}{\delta
^{\left( 2\right) }-\mu _{1_{H}}}\right) \right\}%
\end{array}%
\right]  \notag \\
&&+\mu _{2_{K}}\left( \frac{\mu _{1_{F_{2}}}\text{ }\rho +a^{\left( 2\right)
}\delta ^{\left( 2\right) }}{\delta ^{\left( 2\right) }-\mu _{1_{H}}}\right)
t.  \notag \\
&&  \TCItag{4.30}
\end{eqnarray}

\begin{proof}
See Appendix F.
\end{proof}

\bigskip

\textbf{Corollary 4.2. \ }\textit{For the stationary distribution of the
process} $\lambda _{t}^{\left( 1\right) }$ \textit{and} $\lambda
_{t}^{\left( 2\right) }$, \textit{given} $L_{0}^{\left( 1\right)
}=L_{0}^{\left( 2\right) }=0$, \textit{the variance of the process of }$%
L_{t}^{\left( 1\right) }$\textit{\ is} \textit{given by}

\begin{equation*}
Var\left( L_{t}^{\left( 1\right) }\right)
\end{equation*}

\begin{eqnarray}
&=&2\mu _{1_{J}}\left[ 
\begin{array}{c}
\mu _{1_{J}}\text{ }\left[ \frac{\left\{ 2\left( \mu _{1_{F_{1}}}\text{ }%
\rho +a^{\left( 1\right) }\delta ^{\left( 1\right) }\right) +\mu
_{2_{G}}\right\} \left( \mu _{1_{F_{1}}}\text{ }\rho +a^{\left( 1\right)
}\delta ^{\left( 1\right) }\right) }{2\left( \delta ^{\left( 1\right) }-\mu
_{1_{G}}\right) ^{3}}+\frac{\mu _{2_{F_{1}}}\text{ }\rho }{2\left( \delta
^{\left( 1\right) }-\mu _{1_{G}}\right) ^{2}}\right] \left\{ t-\left( \frac{%
1-e^{-\left( \delta ^{\left( 1\right) }-\mu _{1_{G}}\right) t}}{\delta
^{\left( 1\right) }-\mu _{1_{G}}}\right) \right\} \\ 
+\mu _{1_{G}}\mu _{1_{J}}\text{ }\left\{ \frac{\mu _{1_{F_{1}}}\text{ }\rho
+a^{\left( 1\right) }\delta ^{\left( 1\right) }}{\left( \delta ^{\left(
1\right) }-\mu _{1_{G}}\right) ^{2}}\right\} \left\{ t-\left( \frac{%
1-e^{-\left( \delta ^{\left( 1\right) }-\mu _{1_{G}}\right) t}}{\delta
^{\left( 1\right) }-\mu _{1_{G}}}\right) \right\} \\ 
-\left( a^{\left( 1\right) }\delta ^{\left( 1\right) }+\mu _{1_{F_{1}}}\rho
\right) \mu _{1_{J}}\left\{ \frac{\mu _{1_{F_{1}}}\text{ }\rho +a^{\left(
1\right) }\delta ^{\left( 1\right) }}{\left( \delta ^{\left( 1\right) }-\mu
_{1_{G}}\right) ^{3}}\right\} \left\{ t-\left( \frac{1-e^{-\left( \delta
^{\left( 1\right) }-\mu _{1_{G}}\right) t}}{\delta ^{\left( 1\right) }-\mu
_{1_{G}}}\right) \right\}%
\end{array}%
\right]  \notag \\
&&+\mu _{2_{J}}\left( \frac{\mu _{1_{F_{1}}}\text{ }\rho +a^{\left( 1\right)
}\delta ^{\left( 1\right) }}{\delta ^{\left( 1\right) }-\mu _{1_{G}}}\right)
t  \notag \\
&&  \TCItag{4.31}
\end{eqnarray}

\bigskip

\textit{and the variance of the process of }$L_{t}^{\left( 2\right) }$%
\textit{\ is} \textit{given by}

\begin{equation*}
Var\left( L_{t}^{\left( 2\right) }\right)
\end{equation*}

\begin{eqnarray}
&=&2\mu _{1_{K}}\left[ 
\begin{array}{c}
\mu _{1_{K}}\text{ }\left[ \frac{\left\{ 2\left( \mu _{1_{F_{2}}}\text{ }%
\rho +a^{\left( 2\right) }\delta ^{\left( 2\right) }\right) +\mu
_{2_{H}}\right\} \left( \mu _{1_{F_{2}}}\text{ }\rho +a^{\left( 2\right)
}\delta ^{\left( 2\right) }\right) }{2\left( \delta ^{\left( 2\right) }-\mu
_{1_{H}}\right) ^{3}}+\frac{\mu _{2_{F_{2}}}\text{ }\rho }{2\left( \delta
^{\left( 2\right) }-\mu _{1_{H}}\right) ^{2}}\right] \left\{ t-\left( \frac{%
1-e^{-\left( \delta ^{\left( 2\right) }-\mu _{1_{H}}\right) t}}{\delta
^{\left( 2\right) }-\mu _{1_{H}}}\right) \right\} \\ 
+\mu _{1_{H}}\mu _{1_{K}}\text{ }\left\{ \frac{\mu _{1_{F_{2}}}\text{ }\rho
+a^{\left( 2\right) }\delta ^{\left( 2\right) }}{\left( \delta ^{\left(
2\right) }-\mu _{1_{H}}\right) ^{2}}\right\} \left\{ t-\left( \frac{%
1-e^{-\left( \delta ^{\left( 2\right) }-\mu _{1_{H}}\right) t}}{\delta
^{\left( 2\right) }-\mu _{1_{H}}}\right) \right\} \\ 
-\left( a^{\left( 2\right) }\delta ^{\left( 2\right) }+\mu _{1_{F_{2}}}\rho
\right) \mu _{1_{K}}\left\{ \frac{\mu _{1_{F_{2}}}\text{ }\rho +a^{\left(
2\right) }\delta ^{\left( 2\right) }}{\left( \delta ^{\left( 2\right) }-\mu
_{1_{H}}\right) ^{3}}\right\} \left\{ t-\left( \frac{1-e^{-\left( \delta
^{\left( 2\right) }-\mu _{1_{H}}\right) t}}{\delta ^{\left( 2\right) }-\mu
_{1_{H}}}\right) \right\}%
\end{array}%
\right]  \notag \\
&&+\mu _{2_{K}}\left( \frac{\mu _{1_{F_{2}}}\text{ }\rho +a^{\left( 2\right)
}\delta ^{\left( 2\right) }}{\delta ^{\left( 2\right) }-\mu _{1_{H}}}\right)
t.  \notag \\
&&  \TCItag{4.32}
\end{eqnarray}

\begin{proof}
See Appendix G.
\end{proof}

\bigskip

The corresponding results for Lemma 4.1-4.2, Theorem 4.2-4.3 and Corollary
4.2 can be obtained without using the case for the stationary distribution
of the process $\lambda _{t}^{\left( i\right) }$ ($i=1,2$). \ However their
expressions would be very lengthy formulas with various exponential
functions.

\bigskip

\textbf{5. Insurance application}

\bigskip

The proposed bivariate compound dynamic contagion process may be interpreted
in the context of cyber insurance. \ An initial cyber attack/incident/shock
(e.g. a computer virus) may be the magnitude of joint contribution to
intensities for two different business risks/lines at the same time. \ In
the bivariate compound dynamic contagion process, they are positive
externally-excited \textit{joint} jumps with its distribution $F(x^{\left(
1\right) },x^{\left( 2\right) }),$ $x^{\left( 1\right) }>0,$ $x^{\left(
2\right) }>0$, where margins are $F_{X^{\left( 1\right) }}$ and $%
F_{X^{\left( 2\right) }}$ at the corresponding random times $\left\{
T_{1,i}\right\} _{i=1,2,\cdots }$ following a Poisson process $M_{t}$ with
constant rate $\rho >0$.

After-cyber attacks/incidents/shocks (e.g. infections) may be the magnitudes
of contribution to intensity for each business risk/line at the different
time. \ In the bivariate compound dynamic contagion process, they are
positive self-excited jumps with distribution function $G(y)$, $y>0$, at the
corresponding random times $\left\{ T_{2,j}\right\} _{j=1,2,\cdots }$ and
another positive self-excited jumps with distribution function $H(z)$, $z>0$%
, at the corresponding random times $\left\{ T_{2,k}\right\} _{k=1,2,\cdots
} $. \ The impact of each attack/incident/shock decays exponentially with
constant rate $\delta $.

The number of losses/claims released from the first business risk/line, $%
N_{t}^{\left( 1\right) }$ is driven by a series of after-cyber
attacks/incidents/shocks $\left\{ Y_{j}\right\} _{j=1,2,\cdots }$ and
initial cyber attacks/incidents/shocks $\left\{ X_{i}^{\left( 1\right)
}\right\} _{i=1,2,\cdots }$via its intensity $\lambda _{t}^{\left( 1\right)
} $, \ and the number losses/claims released from the second business
risk/line, $N_{t}^{\left( 2\right) }$ is driven by a series of after-cyber
attacks/incidents/shocks $\left\{ Z_{k}\right\} _{k=1,2,\cdots }$ and
initial cyber attacks/incidents/shocks $\left\{ X_{i}^{\left( 2\right)
}\right\} _{i=1,2,\cdots }$via its intensity $\lambda _{t}^{\left( 2\right)
} $, where initial cyber attacks/incidents/shocks $\left\{ X_{i}^{\left(
1\right) },X_{i}^{\left( 2\right) }\right\} _{i=1,2,\cdots }$ occur to two
different business risks/lines simultaneously/collaterally with constant
intensity $\rho $.

$L_{t}^{\left( 1\right) }$ is the aggregate loss from the first business
risk/line, where loss/claim distribution function is given by $J(\xi
^{\left( 1\right) })$, $\xi ^{\left( 1\right) }>0$, \ and $L_{t}^{\left(
2\right) }$ is the aggregate loss from the second business risk/line, where
loss/claim distribution function is given by $K(\xi ^{\left( 2\right) })$, $%
\xi ^{\left( 2\right) }>0$.

\bigskip

\textbf{5.1. \ Univariate case}

\bigskip

Set $a^{\left( 1\right) }=0$ in (2.1), then from (4.20) the expectation of
the process $L_{t}^{\left( 1\right) }$is given by

\begin{equation}
E\left( L_{t}^{\left( 1\right) }\right) =\mu _{1_{J}}\left( \frac{\mu
_{1_{F_{1}}}\text{ }\rho }{\delta ^{\left( 1\right) }-\mu _{1_{G}}}\right) t,%
\text{ \ }\delta ^{\left( 1\right) }>\mu _{1_{G}}  \tag{5.1}
\end{equation}%
and from (4.31) its variance is given by

\begin{equation*}
Var\left( L_{t}^{\left( 1\right) }\right)
\end{equation*}

\begin{eqnarray}
&=&2\mu _{1_{J}}\left[ 
\begin{array}{c}
\mu _{1_{J}}\text{ }\left\{ \frac{\left( 2\mu _{1_{F_{1}}}\text{ }\rho +\mu
_{2_{G}}\right) \times \mu _{1_{F_{1}}}\text{ }\rho }{2\left( \delta
^{\left( 1\right) }-\mu _{1_{G}}\right) ^{3}}+\frac{\mu _{2_{F_{1}}}\text{ }%
\rho }{2\left( \delta ^{\left( 1\right) }-\mu _{1_{G}}\right) ^{2}}\right\}
\left\{ t-\left( \frac{1-e^{-\left( \delta ^{\left( 1\right) }-\mu
_{1_{G}}\right) t}}{\delta ^{\left( 1\right) }-\mu _{1_{G}}}\right) \right\}
\\ 
+\mu _{1_{G}}\mu _{1_{J}}\text{ }\left\{ \frac{\mu _{1_{F_{1}}}\text{ }\rho 
}{\left( \delta ^{\left( 1\right) }-\mu _{1_{G}}\right) ^{2}}\right\}
\left\{ t-\left( \frac{1-e^{-\left( \delta ^{\left( 1\right) }-\mu
_{1_{G}}\right) t}}{\delta ^{\left( 1\right) }-\mu _{1_{G}}}\right) \right\}
\\ 
-\mu _{1_{J}}\mu _{1_{F_{1}}}\rho \left\{ \frac{\mu _{1_{F_{1}}}\text{ }\rho 
}{\left( \delta ^{\left( 1\right) }-\mu _{1_{G}}\right) ^{3}}\right\}
\left\{ t-\left( \frac{1-e^{-\left( \delta ^{\left( 1\right) }-\mu
_{1_{G}}\right) t}}{\delta ^{\left( 1\right) }-\mu _{1_{G}}}\right) \right\}%
\end{array}%
\right]  \notag \\
&&+\mu _{2_{J}}\left( \frac{\mu _{1_{F_{1}}}\text{ }\rho }{\delta ^{\left(
1\right) }-\mu _{1_{G}}}\right) t  \notag \\
&&  \TCItag{5.2}
\end{eqnarray}

If there are no self-excited jumps, from (5.1) we have%
\begin{equation}
\mu _{1_{J}}\left( \frac{\mu _{1_{F_{1}}}\text{ }\rho }{\delta ^{\left(
1\right) }}\right) t\text{,}  \tag{5.3}
\end{equation}%
which is the expectation of compound shot-noise Cox process, and can also be
found in Dassios and Jang (2003) and Jang and Fu (2012). \ From (5.2), the
corresponding variance is given by

\begin{equation}
\frac{\left( \mu _{1_{J}}\right) ^{2}\mu _{2_{F_{1}}}\text{ }\rho }{\left(
\delta ^{\left( 1\right) }\right) ^{2}}\left\{ t-\left( \frac{1-e^{-\delta
^{\left( 1\right) }t}}{\delta ^{\left( 1\right) }}\right) \right\} +\mu
_{2_{J}}\left( \frac{\mu _{1_{F_{1}}}\text{ }\rho }{\delta ^{\left( 1\right)
}}\right) t.  \tag{5.4}
\end{equation}

Let us now illustrate the calculations of above expectations as cyber
insurance premiums. \ For $F\left( x^{\left( 1\right) }\right) $, we use an
exponential distribution, i.e. 
\begin{equation*}
1-e^{-\alpha x^{\left( 1\right) }}\text{, \ }\alpha >0
\end{equation*}%
and for $G(y)$, we use a Loggamma distribution with probability density, i.e.

\begin{equation*}
\frac{\varsigma ^{c}}{\psi \Gamma \left( c\right) }\left\{ \ln \left( \frac{y%
}{\psi }+1\right) \right\} ^{c-1}\left( \frac{y}{\psi }+1\right)
^{-\varsigma -1},\psi >0,\ \varsigma >0\ \ \text{and}\ \ c>0
\end{equation*}%
to capture the effect of sudden increases of the intensity, i.e. after-cyber
attacks/incidents/shocks driven by initial cyber attacks/incidents/shocks. \
For $J(\xi ^{\left( 1\right) }),$ we use a Pareto distribution with
probability density, i.e.

\begin{equation*}
\frac{\Gamma \left( \omega ^{\left( 1\right) }+k^{\left( 1\right) }\right) 
\text{ }\left\{ \zeta ^{\left( 1\right) }\right\} ^{\omega ^{\left( 1\right)
}}\text{ }\left\{ \xi ^{\left( 1\right) }\right\} ^{k^{\left( 1\right) }-1}}{%
\Gamma \left( \omega ^{\left( 1\right) }\right) \text{ }\Gamma \left(
k^{\left( 1\right) }\right) \text{ }\left( \zeta ^{\left( 1\right) }+\xi
^{\left( 1\right) }\right) ^{\omega ^{\left( 1\right) }+k^{\left( 1\right) }}%
},\ \omega ^{\left( 1\right) }>0,\ \zeta ^{\left( 1\right) }>0\ \ \text{and}%
\ \ k^{\left( 1\right) }>0
\end{equation*}%
to accommodate catastrophic losses/claims generated from the first business
risk/line due to initial and after cyber attacks/incidents/shocks. \ We
assume interest rates to be constant.

\bigskip

\underline{\textbf{Example 5.1}}

We assume that the frequency of initial cyber attack/incident/shock (e.g. a
computer virus) to single business risk/line is $3$ per unit time period
(say, per year) with the average of contribution to intensity, $10$. \ Once
the virus is executed, it replicates itself by modifying other computer
programs causing a series of infection to this business risk/line IT system.
\ The mean of contribution to intensity by after-cyber
attacks/incidents/shocks (e.g. infections), which are unknown at the arrival
times of initial cyber attacks/incidents/shocks, is assumed to be $2.8805$.
\ We assume that the mean of catastrophic losses/claims due to initial and
after cyber attacks/incidents/shocks is $12$.

Hence the parameter values to calculate the expectations are

\begin{eqnarray*}
\delta ^{\left( 1\right) } &=&3,\text{ \ }\rho =3,\text{ }{\alpha =0.1},%
\text{ \ }\psi =1,\text{ \ }\varsigma =2.75\text{, \ }c=3, \\
\omega ^{\left( 1\right) } &=&3,\ \ \zeta ^{\left( 1\right) }=4,\text{ \ }%
k^{\left( 1\right) }=6\text{ \ and \ }t=1.
\end{eqnarray*}%
and from (5.1)-(5.4), their calculations are shown in Table 5.1.

\begin{center}
\begin{tabular}{|l|l|l|}
\multicolumn{3}{l}{\ \ \ \ \ \ \ \ \ \ \ \ \ \ \ \ \ \ \ \ \ \ \ \ \ \ \ \ \
\ \ \ \ \ \ \ \ \ \ \ \ \ \ \ \ \ \ \ \ \ \ \ \ \ \ \ \ \ \ \ \ \ \ \textbf{%
Table 5.1}} \\ \hline
& 
\begin{tabular}{l}
Univariate compound \\ 
dynamic contagion process%
\end{tabular}
& 
\begin{tabular}{l}
Univariate compound \\ 
shot-noise Cox process%
\end{tabular}
\\ \hline
\ \ \ \ \ \ \ \ \ \ \ \ \ Mean & $\ \ \ \ \ \ \ \ \ \ \ \ \ \ \ \ \ 3,011.71$
& $\ \ \ \ \ \ \ \ \ \ \ \ \ \ \ \ \ 120$ \\ \hline
\ \ \ \ \ \ \ \ \ \ \ Variance & $\ \ \ \ \ \ \ \ \ \ \ \ \ \ 6,713,295.5$ & 
$\ \ \ \ \ \ \ \ \ \ \ \ \ \ \ 9,919$ \\ \hline
\begin{tabular}{l}
Mean-standard deviation \\ 
principle premium%
\end{tabular}
& $\ \ \ \ \ \ \ \ \ \ \ \ \ \ \ \ \ 5,602.7$ & $\ \ \ \ \ \ \ \ \ \ \ \ \ \
219.59$ \\ \hline
\end{tabular}
\end{center}

\bigskip

\textbf{Remark 5}: \ Table 5.1 shows that mean-standard deviation principle
premium, $5,602.7$ calculated based on (5.1)-(5.2) is extremely higher than
its counterpart $219.59$ calculated based on (5.3)-(5.4). \ It is because
after-cyber attacks/incidents/shocks (e.g. infections) driven by initial
cyber attacks/incidents/shocks (e.g. a computer virus). \ In other words, $%
\mu _{1_{G}}$, which is the mean of after-cyber attacks/incidents/shocks, is
the main driver to raise the premium extremely higher than its counterpart.
\ Hence the significance of after-cyber attacks/incidents/shocks driven from
an initial attack/incident/shock depends on its measure $G(y)$.

Due to the digitalisation of business and economic activities, all types of
risk are touched by cyber nowadays. \ To deal with new challenge insurers
face - risks arising from cyber space, they need new tools to measure these
risks. \ The mean-standard deviation principle premium value calculated
based on (5.1)-(5.2) clearly justifies that the univariate compound dynamic
contagion process can be used for modelling aggregate losses/claims from
cyber attacks/incidents.

\bigskip

\textbf{5.2. \ Bivariate case}

Set $a^{\left( 1\right) }=0,$ $a^{\left( 2\right) }=0$ and $\delta =\delta
^{\left( 1\right) }=\delta ^{\left( 2\right) }$, then from (4.20) and
(4.21), the expectation of the process $L_{t}^{\left( 1\right) }$ is given by

\begin{equation}
E\left( L_{t}^{\left( 1\right) }\right) =\mu _{1_{J}}\left( \frac{\mu
_{1_{F_{1}}}\text{ }\rho }{\delta -\mu _{1_{G}}}\right) t,\text{ \ }\delta
>\mu _{1_{G}}  \tag{5.5}
\end{equation}%
and the expectation of the process $L_{t}^{\left( 2\right) }$ is given by

\begin{equation}
E\left( L_{t}^{\left( 2\right) }\right) =\mu _{1_{K}}\left( \frac{\mu
_{1_{F_{2}}}\text{ }\rho }{\delta -\mu _{1_{H}}}\right) t,\text{ \ }\delta
>\mu _{1_{H}}.  \tag{5.6}
\end{equation}%
Let us assume that an insurance company charges cyber insurance premium as
follows:

\begin{eqnarray}
&&E\left( L_{t}^{\left( 1\right) }+L_{t}^{\left( 2\right) }\right) +\phi 
\sqrt{Var\left( L_{t}^{\left( 1\right) }+L_{t}^{\left( 2\right) }\right) } 
\notag \\
&=&E\left( L_{t}^{\left( 1\right) }\right) +E\left( L_{t}^{\left( 2\right)
}\right) +\phi \sqrt{Var\left( L_{t}^{\left( 1\right) }\right) +Var\left(
L_{t}^{\left( 2\right) }\right) +2Cov\left( L_{t}^{\left( 1\right)
},L_{t}^{\left( 2\right) }\right) },  \notag \\
&&  \TCItag{5.7}
\end{eqnarray}%
where $0\leq \phi \leq 1$ and $\phi \sqrt{Var\left( L_{t}^{\left( 1\right)
}+L_{t}^{\left( 2\right) }\right) }$ can be considered as a security loading.

To calculate the covariance, we need to specify externally-excited joint
jump distribution $F(x^{\left( 1\right) },x^{\left( 2\right) })$, for which
we offer four choices of copulas: (1) the Farlie-Gumbel-Morgenstern (FGM)
copula, (2) the Gaussian copula, (3) the $t$ copula and (4) the Gumbel
copula. \ The Farlie-Gumbel-Morgenstern (FGM) family copula is given by%
\begin{equation}
C_{\theta }(u_{1},u_{2})=u_{1}u_{2}+\theta u_{1}u_{2}(1-u_{1})(1-u_{2}), 
\tag{5.8}
\end{equation}%
where $u_{1}\in \left[ 0,1\right] $, $u_{2}\in \left[ 0,1\right] $ and $%
\theta \in \left[ -1,1\right] $. \ The Gaussian family copula is given by%
\begin{equation}
C_{\theta }(u_{1},u_{2})=\Phi _{\Sigma }\left( \Phi ^{-1}\left( u_{1}\right)
,\Phi ^{-1}\left( u_{2}\right) \right) ,  \tag{5.9}
\end{equation}%
where $\Phi ^{-1}$ is the inverse cumulative distribution function (c.d.f.)
of a standard univariate normal, $\Phi _{\Sigma }$ denotes the c.d.f. for a
bivariate normal distribution with mean vector zero and covariance matrix $%
\Sigma $, where $\Sigma $ the $2\times 2$ matrix with $1$ on the diagonal
and correlation coefficient $\theta $ otherwise, $u_{1}\in \left[ 0,1\right] 
$, $u_{2}\in \left[ 0,1\right] $ and $\theta \in \left[ -1,1\right] $. \ The 
\textit{t} copula is given by

\begin{equation}
C_{\theta }(u_{1},u_{2})=t_{\varepsilon ,\Sigma }\left( t_{\nu }^{-1}\left(
u_{1}\right) ,t_{\nu }^{-1}\left( u_{2}\right) \right) ,  \tag{5.10}
\end{equation}%
where $t_{\nu }^{-1}$ is the inverse cumulative distribution function
(c.d.f.) of a standard univariate $t$, $t_{\varepsilon ,\Sigma }$ denotes
the c.d.f. for a bivariate $t$ distribution with mean vector zero and
covariance matrix $\Sigma $, where $\Sigma $ the $2\times 2$ matrix with $1$
on the diagonal and correlation coefficient $\theta $ otherwise, $%
\varepsilon $ is the degrees of freedom, $u_{1}\in \left[ 0,1\right] $, $%
u_{2}\in \left[ 0,1\right] $ and $\theta \in \left[ -1,1\right] $. \ The
Gumbel copulas are given by

\begin{equation}
C_{\theta }(u_{1},u_{2})=\exp \left[ \left\{ \left( -\ln \left( u_{1}\right)
\right) ^{-\theta }+\left( -\ln \left( u_{2}\right) \right) ^{-\theta }-1%
\text{ }\right\} ^{-\frac{1}{\theta }}\right] ,  \tag{5.11}
\end{equation}%
where $u_{1}\in \left[ 0,1\right] $, $u_{2}\in \left[ 0,1\right] $ and $%
\theta \in \lbrack 1,\infty )$.

For $F\left( x^{\left( 2\right) }\right) $, we also use an exponential
distribution, i.e.%
\begin{equation*}
F\left( x^{\left( 2\right) }\right) =1-e^{-\beta x^{\left( 2\right) }}\left(
\beta >0\right)
\end{equation*}%
and for $H(z)$, we use a Fr\'{e}chet distribution with probability density,
i.e.

\begin{equation*}
\frac{\epsilon }{\varphi }\left( \frac{z}{\varphi }\right) ^{-\epsilon
-1}e^{-\left( \frac{z}{\varphi }\right) ^{-\epsilon }},\text{ }\varphi >0%
\text{ \ and \ }\epsilon >0
\end{equation*}%
to capture the effect of sudden increases of the intensity, i.e. after-cyber
attacks/incidents/shocks driven by initial cyber attacks/incidents/shocks. \
For $K(\xi ^{\left( 2\right) }),$ we use another Pareto distribution with
probability density, i.e.

\begin{equation*}
\frac{\Gamma \left( \omega ^{\left( 2\right) }+k^{\left( 2\right) }\right) 
\text{ }\left\{ \zeta ^{\left( 2\right) }\right\} ^{\omega ^{\left( 2\right)
}}\text{ }\left\{ \xi ^{\left( 2\right) }\right\} ^{k^{\left( 2\right) }-1}}{%
\Gamma \left( \omega ^{\left( 2\right) }\right) \text{ }\Gamma \left(
k^{\left( 2\right) }\right) \text{ }\left( \zeta ^{\left( 2\right) }+\xi
^{\left( 2\right) }\right) ^{\omega ^{\left( 2\right) }+k^{\left( 2\right) }}%
},\ \omega ^{\left( 2\right) }>0,\ \zeta ^{\left( 2\right) }>0\ \ \text{and}%
\ \ k^{\left( 2\right) }>0
\end{equation*}%
to accommodate catastrophic losses/claims generated from the second business
risk/line due to initial and after cyber attacks/incidents/shocks.

For the next four examples, we assume that the frequency of initial \textit{%
joint} cyber attack/incident/shock (e.g. a computer virus) to two business
risks/lines is $3$ per unit time period (say, per year) with the same
average of contributions to both intensities, $10$. \ Once the virus is
executed, it replicates itself by modifying other computer programs causing
a series of infection to two business risks/lines IT systems, separately. \
The mean of contribution to the first \& second business risk/line intensity
by after-cyber attacks/incidents/shocks (e.g. infections), which are unknown
at the arrival times of initial cyber attacks/incidents/shocks, is assumed
to be $2.8805$ and $2.7082$, respectively. \ We assume that the mean of
catastrophic losses/claims from two business risks/lines due to initial and
after cyber attacks/incidents/shocks is $12$ and $8$, respectively. \ As the
security loading factor, this insurance company uses 1.

Hence the parameter values used to calculate cyber loss insurance premiums
are%
\begin{eqnarray*}
\beta &=&{0.1},\text{ \ }\varphi =2,\ \ \epsilon =3\text{,\ \ }\phi =1, \\
\omega ^{\left( 2\right) } &=&4,\ \ \zeta ^{\left( 2\right) }=4\ \ \text{and}%
\ \ k^{\left( 2\right) }=6
\end{eqnarray*}%
and using the parameter values in Example 5.1, let us now illustrate the
calculations of cyber loss insurance premiums at different value of $\theta
, $ comparing their counterparts when there are no after cyber
attacks/incidents/shocks.

\bigskip

\underline{\textbf{Example 5.2}} (FGM copula)

Due to the Farlie-Gumbel-Morgenstern (FGM) copulas simplicity and analytical
tractability, we have

\begin{equation}
\mu _{1_{F_{1,2}}}=\int\limits_{0}^{\infty }\int\limits_{0}^{\infty
}x^{\left( 1\right) }x^{\left( 2\right) }dF\left( x^{\left( 1\right)
},x^{\left( 2\right) }\right) =\frac{1}{\alpha \beta }\left( 1+\frac{\theta 
}{4}\right)  \tag{5.12}
\end{equation}%
to calculate $E\left( L_{t}^{\left( 1\right) }L_{t}^{\left( 2\right)
}\right) $ in $Cov\left( L_{t}^{\left( 1\right) },L_{t}^{\left( 2\right)
}\right) $. \ Cyber loss insurance premium calculations are shown in Table
5.2,

\begin{center}
\begin{tabular}{|l|l|l|}
\hline
\multicolumn{3}{|l|}{\textbf{Table 5.2 \ \ \ \ \ \ \ \ \ \ \ \ }Cyber loss
insurance premium} \\ \hline
\multicolumn{3}{|l|}{} \\ \hline
$\theta $ & 
\begin{tabular}{l}
Bivariate compound \\ 
dynamic contagion process%
\end{tabular}
& 
\begin{tabular}{l}
Bivariate compound \\ 
shot-noise Cox process%
\end{tabular}
\\ \hline
$-1$ & $\ \ \ \ \ \ \ \ \ \ \ \ \ 6481.74$ & $\ \ \ \ \ \ \ \ \ \ \ \ \ \ \
331.28$ \\ \hline
$-0.5$ & $\ \ \ \ \ \ \ \ \ \ \ \ \ 6484.83$ & $\ \ \ \ \ \ \ \ \ \ \ \ \ \
\ 333.34$ \\ \hline
$0$ & $\ \ \ \ \ \ \ \ \ \ \ \ \ 6487.92$ & $\ \ \ \ \ \ \ \ \ \ \ \ \ \ \
335.38$ \\ \hline
$0.5$ & $\ \ \ \ \ \ \ \ \ \ \ \ \ 6491.01$ & $\ \ \ \ \ \ \ \ \ \ \ \ \ \ \
337.38$ \\ \hline
$1$ & $\ \ \ \ \ \ \ \ \ \ \ \ \ 6494.09$ & $\ \ \ \ \ \ \ \ \ \ \ \ \ \ \
339.36$ \\ \hline
\end{tabular}
\end{center}

where for bivariate compound dynamic contagion case, we have{\ 
\begin{eqnarray*}
E\left( L_{t}^{\left( 1\right) }\right) &=&3011.71\text{ \ and \ }Var\left(
L_{t}^{\left( 1\right) }\right) =6,713,296\text{,} \\
E\left( L_{t}^{\left( 2\right) }\right) &=&822.582\text{ \ and \ }Var\left(
L_{t}^{\left( 2\right) }\right) =197,473
\end{eqnarray*}%
} and for bivariate compound shot-noise case, we have{\ 
\begin{eqnarray*}
E\left( L_{t}^{\left( 1\right) }\right) &=&120\text{ \ and \ }Var\left(
L_{t}^{\left( 1\right) }\right) =9,919.32\text{,} \\
E\left( L_{t}^{\left( 2\right) }\right) &=&80\text{ \ \ and \ }Var\left(
L_{t}^{\left( 2\right) }\right) =4,035.25.
\end{eqnarray*}%
}

The covariances between $L_{t}^{\left( 1\right) }$\ and $L_{t}^{\left(
2\right) }$ and their corresponding linear correlation coefficients at
different value of $\theta $, compared to their counterparts when there are
no self-excited jumps are shown in Table 5.3 and Table 5.4, respectively.

\begin{center}
\begin{tabular}{|l|l|l|}
\hline
\multicolumn{3}{|l|}{\textbf{Table 5.3 \ \ \ \ \ \ \ \ \ \ \ \ \ \ \ \ \ \ \
\ }$Cov\left( L_{t}^{\left( 1\right) },L_{t}^{\left( 2\right) }\right) $} \\ 
\hline
\multicolumn{3}{|l|}{} \\ \hline
$\theta $ & 
\begin{tabular}{l}
Bivariate compound \\ 
dynamic contagion process%
\end{tabular}
& 
\begin{tabular}{l}
Bivariate compound \\ 
shot-noise Cox process%
\end{tabular}
\\ \hline
$-1$ & $\ \ \ \ \ \ \ \ \ \ \ \ \ \ 49123.16$ & $\ \ \ \ \ \ \ \ \ \ \ \ \ \
\ 1639.83$ \\ \hline
$-0.5$ & $\ \ \ \ \ \ \ \ \ \ \ \ \ \ 57310.35$ & $\ \ \ \ \ \ \ \ \ \ \ \ \
\ \ 1913.13$ \\ \hline
$0$ & $\ \ \ \ \ \ \ \ \ \ \ \ \ \ 65497.54$ & $\ \ \ \ \ \ \ \ \ \ \ \ \ \
\ 2186.44$ \\ \hline
$0.5$ & $\ \ \ \ \ \ \ \ \ \ \ \ \ \ 73684.73$ & $\ \ \ \ \ \ \ \ \ \ \ \ \
\ \ 2459.74$ \\ \hline
$1$ & $\ \ \ \ \ \ \ \ \ \ \ \ \ \ 81871.93$ & $\ \ \ \ \ \ \ \ \ \ \ \ \ \
\ 2733.05$ \\ \hline
\end{tabular}

\bigskip

\begin{tabular}{|l|l|l|}
\hline
\multicolumn{3}{|l|}{\textbf{Table 5.4 \ \ \ \ \ \ \ \ \ \ \ \ \ \ \ \ \ \ \
\ \ }$Corr\left( L_{t}^{\left( 1\right) },L_{t}^{\left( 2\right) }\right) $}
\\ \hline
\multicolumn{3}{|l|}{} \\ \hline
$\theta $ & 
\begin{tabular}{l}
Bivariate compound \\ 
dynamic contagion process%
\end{tabular}
& 
\begin{tabular}{l}
Bivariate compound \\ 
shot-noise Cox process%
\end{tabular}
\\ \hline
$-1$ & $\ \ \ \ \ \ \ \ \ \ \ \ \ 0.04266$ & $\ \ \ \ \ \ \ \ \ \ \ \ \
0.25919$ \\ \hline
$-0.5$ & $\ \ \ \ \ \ \ \ \ \ \ \ \ 0.04977$ & $\ \ \ \ \ \ \ \ \ \ \ \ \
0.30239$ \\ \hline
$0$ & $\ \ \ \ \ \ \ \ \ \ \ \ \ 0.05689$ & $\ \ \ \ \ \ \ \ \ \ \ \ \
0.34559$ \\ \hline
$0.5$ & $\ \ \ \ \ \ \ \ \ \ \ \ \ 0.06400$ & $\ \ \ \ \ \ \ \ \ \ \ \ \
0.38879$ \\ \hline
$1$ & $\ \ \ \ \ \ \ \ \ \ \ \ \ 0.07111$ & $\ \ \ \ \ \ \ \ \ \ \ \ \
0.43199$ \\ \hline
\end{tabular}

\bigskip
\end{center}

\underline{\textbf{Example 5.2}} (Gaussian copula)

For the Gaussian copulas, using the programming language R cyber loss
insurance premium calculations are shown in Table 5.5,

\begin{center}
\begin{tabular}{|l|l|l|}
\hline
\multicolumn{3}{|l|}{\textbf{Table 5.5 \ \ \ \ \ \ \ \ \ \ \ \ }Cyber loss
insurance premium} \\ \hline
\multicolumn{3}{|l|}{} \\ \hline
$\theta $ & 
\begin{tabular}{l}
Bivariate compound \\ 
dynamic contagion process%
\end{tabular}
& 
\begin{tabular}{l}
Bivariate compound \\ 
shot-noise Cox process%
\end{tabular}
\\ \hline
$-0.99$ & $\ \ \ \ \ \ \ \ \ \ \ \ 6,472.08$ & $\ \ \ \ \ \ \ \ \ \ \ \
324.61$ \\ \hline
$-0.5$ & $\ \ \ \ \ \ \ \ \ \ \ \ 6,478.91$ & $\ \ \ \ \ \ \ \ \ \ \ \
329.36 $ \\ \hline
$0$ & $\ \ \ \ \ \ \ \ \ \ \ \ 6,487.92$ & $\ \ \ \ \ \ \ \ \ \ \ \ 335.38$
\\ \hline
$0.5$ & $\ \ \ \ \ \ \ \ \ \ \ \ 6,499.08$ & $\ \ \ \ \ \ \ \ \ \ \ \ 342.51$
\\ \hline
$0.99$ & $\ \ \ \ \ \ \ \ \ \ \ \ 6,512.20$ & $\ \ \ \ \ \ \ \ \ \ \ \
350.49 $ \\ \hline
\end{tabular}
\end{center}

The covariances between $L_{t}^{\left( 1\right) }$\ and $L_{t}^{\left(
2\right) }$ and their corresponding linear correlation coefficients at
different value of $\theta $, compared to their counterparts when there are
no self-excited jumps are shown in Table 5.6 and Table 5.7, respectively.

\begin{center}
\begin{tabular}{|l|l|l|}
\hline
\multicolumn{3}{|l|}{\textbf{Table 5.6 \ \ \ \ \ \ \ \ \ \ \ \ \ \ \ \ \ \ \
\ }$Cov\left( L_{t}^{\left( 1\right) },L_{t}^{\left( 2\right) }\right) $} \\ 
\hline
\multicolumn{3}{|l|}{} \\ \hline
$\theta $ & 
\begin{tabular}{l}
Bivariate compound \\ 
dynamic contagion process%
\end{tabular}
& 
\begin{tabular}{l}
Bivariate compound \\ 
shot-noise Cox process%
\end{tabular}
\\ \hline
$-0.99$ & $\ \ \ \ \ \ \ \ \ \ 23,571.72$ & $\ \ \ \ \ \ \ \ \ \ \ \ 786.87 $
\\ \hline
$-0.5$ & $\ \ \ \ \ \ \ \ \ \ 41,632.69$ & $\ \ \ \ \ \ \ \ \ 1,389.78$ \\ 
\hline
$0$ & $\ \ \ \ \ \ \ \ \ \ 65,497.54$ & $\ \ \ \ \ \ \ \ \ 2,186.44$ \\ 
\hline
$0.5$ & $\ \ \ \ \ \ \ \ \ \ 95,172.84$ & $\ \ \ \ \ \ \ \ \ 3,177.06$ \\ 
\hline
$0.99$ & $\ \ \ \ \ \ \ \ 130,216.13$ & $\ \ \ \ \ \ \ \ \ 4,346.88$ \\ 
\hline
\end{tabular}

\bigskip

\begin{tabular}{|l|l|l|}
\hline
\multicolumn{3}{|l|}{\textbf{Table 5.7 \ \ \ \ \ \ \ \ \ \ \ \ \ \ \ \ \ \ \
\ \ }$Corr\left( L_{t}^{\left( 1\right) },L_{t}^{\left( 2\right) }\right) $}
\\ \hline
\multicolumn{3}{|l|}{} \\ \hline
$\theta $ & 
\begin{tabular}{l}
Bivariate compound \\ 
dynamic contagion process%
\end{tabular}
& 
\begin{tabular}{l}
Bivariate compound \\ 
shot-noise Cox process%
\end{tabular}
\\ \hline
$-0.99$ & $\ \ \ \ \ \ \ \ \ \ \ 0.02047$ & $\ \ \ \ \ \ \ \ \ \ 0.12437$ \\ 
\hline
$-0.5$ & $\ \ \ \ \ \ \ \ \ \ \ 0.03616$ & $\ \ \ \ \ \ \ \ \ \ 0.21967$ \\ 
\hline
$0$ & $\ \ \ \ \ \ \ \ \ \ \ 0.05689$ & $\ \ \ \ \ \ \ \ \ \ 0.34559$ \\ 
\hline
$0.5$ & $\ \ \ \ \ \ \ \ \ \ \ 0.08266$ & $\ \ \ \ \ \ \ \ \ \ 0.50217$ \\ 
\hline
$0.99$ & $\ \ \ \ \ \ \ \ \ \ \ 0.11309$ & $\ \ \ \ \ \ \ \ \ \ 0.68707$ \\ 
\hline
\end{tabular}
\end{center}

\bigskip

\underline{\textbf{Example 5.3}} (\textit{t} copula with $\varepsilon=5$)

For the \textit{t} copulas, using the programming language R cyber loss
insurance premium calculations are shown in Table 5.8

\begin{center}
\begin{tabular}{|l|l|l|}
\hline
\multicolumn{3}{|l|}{\textbf{Table 5.8 \ \ \ \ \ \ \ \ \ \ \ \ }Cyber loss
insurance premium} \\ \hline
\multicolumn{3}{|l|}{} \\ \hline
$\theta $ & 
\begin{tabular}{l}
Bivariate compound \\ 
dynamic contagion process%
\end{tabular}
& 
\begin{tabular}{l}
Bivariate compound \\ 
shot-noise Cox process%
\end{tabular}
\\ \hline
$-0.99$ & $\ \ \ \ \ \ \ \ \ \ \ \ 6,472.08$ & $\ \ \ \ \ \ \ \ \ \ \ 324.62$
\\ \hline
$-0.5$ & $\ \ \ \ \ \ \ \ \ \ \ \ 6,479.53$ & $\ \ \ \ \ \ \ \ \ \ \ 329.78 $
\\ \hline
$0$ & $\ \ \ \ \ \ \ \ \ \ \ \ 6,488.87$ & $\ \ \ \ \ \ \ \ \ \ \ 336.00$ \\ 
\hline
$0.5$ & $\ \ \ \ \ \ \ \ \ \ \ \ 6,499.76$ & $\ \ \ \ \ \ \ \ \ \ \ 342.93$
\\ \hline
$0.99$ & $\ \ \ \ \ \ \ \ \ \ \ \ 6,512.21$ & $\ \ \ \ \ \ \ \ \ \ \ 350.50 $
\\ \hline
\end{tabular}
\end{center}

The covariances between $L_{t}^{\left( 1\right) }$\ and $L_{t}^{\left(
2\right) }$ and their corresponding linear correlation coefficients at
different value of $\theta $, comparing their counterparts when there are no
self-excited jumps are shown in Table 5.9 and Table 5.10, respectively.

\begin{center}
\begin{tabular}{|l|l|l|}
\hline
\multicolumn{3}{|l|}{\textbf{Table 5.9 \ \ \ \ \ \ \ \ \ \ \ \ \ \ \ \ \ \ \
\ }$Cov\left( L_{t}^{\left( 1\right) },L_{t}^{\left( 2\right) }\right) $} \\ 
\hline
\multicolumn{3}{|l|}{} \\ \hline
$\theta $ & 
\begin{tabular}{l}
Bivariate compound \\ 
dynamic contagion process%
\end{tabular}
& 
\begin{tabular}{l}
Bivariate compound \\ 
shot-noise Cox process%
\end{tabular}
\\ \hline
$-0.99$ & $\ \ \ \ \ \ \ \ \ \ \ \ 23,595.21$ & $\ \ \ \ \ \ \ \ \ \ \ \
787.66$ \\ \hline
$-0.5$ & $\ \ \ \ \ \ \ \ \ \ \ \ 43,268.73$ & $\ \ \ \ \ \ \ \ \ 1,444.40$
\\ \hline
$0$ & $\ \ \ \ \ \ \ \ \ \ \ \ 68,008.96$ & $\ \ \ \ \ \ \ \ \ 2,270.28$ \\ 
\hline
$0.5$ & $\ \ \ \ \ \ \ \ \ \ \ \ 96,986.52$ & $\ \ \ \ \ \ \ \ \ 3,237.61$
\\ \hline
$0.99$ & $\ \ \ \ \ \ \ \ \ \ 130,248.18$ & $\ \ \ \ \ \ \ \ \ 4,347.95$ \\ 
\hline
\end{tabular}

\bigskip

\begin{tabular}{|l|l|l|}
\hline
\multicolumn{3}{|l|}{\textbf{Table 5.10 \ \ \ \ \ \ \ \ \ \ \ \ \ \ \ \ \ \
\ \ \ }$Corr\left( L_{t}^{\left( 1\right) },L_{t}^{\left( 2\right) }\right) $%
} \\ \hline
\multicolumn{3}{|l|}{} \\ \hline
$\theta $ & 
\begin{tabular}{l}
Bivariate compound \\ 
dynamic contagion process%
\end{tabular}
& 
\begin{tabular}{l}
Bivariate compound \\ 
shot-noise Cox process%
\end{tabular}
\\ \hline
$-0.99$ & $\ \ \ \ \ \ \ \ \ \ \ \ \ 0.02049$ & $\ \ \ \ \ \ \ \ \ \ \
0.12450$ \\ \hline
$-0.5$ & $\ \ \ \ \ \ \ \ \ \ \ \ \ 0.03758$ & $\ \ \ \ \ \ \ \ \ \ \
0.22830 $ \\ \hline
$0$ & $\ \ \ \ \ \ \ \ \ \ \ \ \ 0.05907$ & $\ \ \ \ \ \ \ \ \ \ \ 0.35884$
\\ \hline
$0.5$ & $\ \ \ \ \ \ \ \ \ \ \ \ \ 0.08423$ & $\ \ \ \ \ \ \ \ \ \ \ 0.51174$
\\ \hline
$0.99$ & $\ \ \ \ \ \ \ \ \ \ \ \ \ 0.11312$ & $\ \ \ \ \ \ \ \ \ \ \
0.68724 $ \\ \hline
\end{tabular}

\bigskip
\end{center}

\underline{\textbf{Example 5.4}} (Gumbel copula)

For the Gaussian copulas, using the programming language R cyber loss
insurance premium calculations are shown in Table 5.11.

\begin{center}
\begin{tabular}{|l|l|l|}
\hline
\multicolumn{3}{|l|}{\textbf{Table 5.11 \ \ \ \ \ \ \ \ \ \ \ \ }Cyber loss
insurance premium} \\ \hline
\multicolumn{3}{|l|}{} \\ \hline
$\theta $ & 
\begin{tabular}{l}
Bivariate compound \\ 
dynamic contagion process%
\end{tabular}
& 
\begin{tabular}{l}
Bivariate compound \\ 
shot-noise Cox process%
\end{tabular}
\\ \hline
$1.001$ & $\ \ \ \ \ \ \ \ \ 6,487.97$ & $\ \ \ \ \ \ \ \ \ \ \ 335.41$ \\ 
\hline
$2$ & $\ \ \ \ \ \ \ \ \ 6,506.88$ & $\ \ \ \ \ \ \ \ \ \ \ 347.30$ \\ \hline
$5$ & $\ \ \ \ \ \ \ \ \ 6,511.66$ & $\ \ \ \ \ \ \ \ \ \ \ 350.17$ \\ \hline
$10$ & $\ \ \ \ \ \ \ \ \ 6,512.29$ & $\ \ \ \ \ \ \ \ \ \ \ 350.55$ \\ 
\hline
$100$ & $\ \ \ \ \ \ \ \ \ 6,512.49$ & $\ \ \ \ \ \ \ \ \ \ \ 350.67$ \\ 
\hline
\end{tabular}
\end{center}

The covariances between $L_{t}^{\left( 1\right) }$\ and $L_{t}^{\left(
2\right) }$ and their corresponding linear correlation coefficients at
different value of $\theta $, compared to their counterparts when there are
no self-excited jumps are shown in Table 5.12 and Table 5.13, respectively.{%
\ }

\begin{center}
\begin{tabular}{|l|l|l|}
\hline
\multicolumn{3}{|l|}{\textbf{Table 5.12 \ \ \ \ \ \ \ \ \ \ \ \ \ \ \ \ \ \
\ \ }$Cov\left( L_{t}^{\left( 1\right) },L_{t}^{\left( 2\right) }\right) $}
\\ \hline
\multicolumn{3}{|l|}{} \\ \hline
$\theta $ & 
\begin{tabular}{l}
Bivariate compound \\ 
dynamic contagion process%
\end{tabular}
& 
\begin{tabular}{l}
Bivariate compound \\ 
shot-noise Cox process%
\end{tabular}
\\ \hline
$1.001$ & $\ \ \ \ \ \ \ \ \ \ \ 65,637.60$ & $\ \ \ \ \ \ \ \ \ 2,191.12$
\\ \hline
$2$ & $\ \ \ \ \ \ \ \ \ 115,986.56$ & $\ \ \ \ \ \ \ \ \ 3,871.86$ \\ \hline
$5$ & $\ \ \ \ \ \ \ \ \ 128,771.15$ & $\ \ \ \ \ \ \ \ \ 4,298.64$ \\ \hline
$10$ & $\ \ \ \ \ \ \ \ \ 130,456.78$ & $\ \ \ \ \ \ \ \ \ 4,354.91$ \\ 
\hline
$100$ & $\ \ \ \ \ \ \ \ \ 130,990.50$ & $\ \ \ \ \ \ \ \ \ 4,372.73$ \\ 
\hline
\end{tabular}

\begin{tabular}{|l|l|l|}
\hline
\multicolumn{3}{|l|}{\textbf{Table 5.13 \ \ \ \ \ \ \ \ \ \ \ \ \ \ \ \ \ \
\ \ \ }$Corr\left( L_{t}^{\left( 1\right) },L_{t}^{\left( 2\right) }\right) $%
} \\ \hline
\multicolumn{3}{|l|}{} \\ \hline
$\theta $ & 
\begin{tabular}{l}
Bivariate compound \\ 
dynamic contagion process%
\end{tabular}
& 
\begin{tabular}{l}
Bivariate compound \\ 
shot-noise Cox process%
\end{tabular}
\\ \hline
$1.001$ & $\ \ \ \ \ \ \ \ \ \ \ \ \ 0.05701$ & $\ \ \ \ \ \ \ \ \ \ 0.34633$
\\ \hline
$2$ & $\ \ \ \ \ \ \ \ \ \ \ \ \ 0.10074$ & $\ \ \ \ \ \ \ \ \ \ 0.61199$ \\ 
\hline
$5$ & $\ \ \ \ \ \ \ \ \ \ \ \ \ 0.11184$ & $\ \ \ \ \ \ \ \ \ \ 0.67945$ \\ 
\hline
$10$ & $\ \ \ \ \ \ \ \ \ \ \ \ \ 0.11330$ & $\ \ \ \ \ \ \ \ \ \ 0.68834$
\\ \hline
$100$ & $\ \ \ \ \ \ \ \ \ \ \ \ \ 0.11377$ & $\ \ \ \ \ \ \ \ \ \ 0.69116$
\\ \hline
\end{tabular}
\end{center}

\bigskip

\textbf{Remark 6}: \ Table 5.2, 5.5, 5.8 and 5.11 show that cyber loss
insurance premium values calculated using the bivariate compound dynamic
contagion process are significantly higher than their counterparts
calculated using the bivariate compound shot-noise Cox process at a
different value of $\theta $. \ The covariances in Table 5.3, 5.6, 5.9 and
5,12 also support this. \ It is because two means for after-cyber
attacks/incidents/shocks, i.e. $\mu _{1_{G}}$ and $\mu _{1_{H}}$, and $\mu
_{1_{F_{1,2}}}$ are involved in calculating cyber loss insurance premium
values using (5.7). \ Hence the significance of two separate after-cyber
attacks/incidents/shocks impacts driven from initial joint cyber
attack/incident/shock depends on two measures $G(y)$ and $H(z)$. \ It will
be of interest to examine cyber loss insurance premium values using other
joint measures for initial cyber attack/incident/shock as well as other
measures for after-cyber attacks/incidents/shocks.

\bigskip

\textbf{Remark 7}: \ Table 5.4, 5.7, 5.10 and 5.13 show that the linearities
between $L_{t}^{\left( 1\right) }$\ and $L_{t}^{\left( 2\right) }$
calculated using the bivariate compound dynamic contagion process and the
bivariate compound shot-noise Cox process at a different value of $\theta .$
\ They show the former linearities between $L_{t}^{\left( 1\right) }$\ and $%
L_{t}^{\left( 2\right) }$ significantly lower than the latter linearities
between $L_{t}^{\left( 1\right) }$\ and $L_{t}^{\left( 2\right) }$. \ It is
because two separate after-cyber attacks/incidents/shocks weaken the
linearity between $L_{t}^{\left( 1\right) }$\ and $L_{t}^{\left( 2\right) }$%
. \ Therefore it will be also of interest to compare bivariate distribution
for compound dynamic contagion case with its counterpart, in particular
seeing their two tail corners inverting bivariate Fast Fourier transform
using bivariate Laplace transform of the process $(L_{t}^{\left( 1\right) },$
$L_{t}^{\left( 2\right) })$ shown in Section 3.

\bigskip

To make easier for statistical analysis, further business applications and
research, we close this section providing the simulation algorithm for one
sample path of the bivariate compound dynamic contagion process $\left(
\left( {%
\begin{tabular}{l}
$L_{t}^{\left( 1\right) }$ \\ 
$L_{t}^{\left( 2\right) }$%
\end{tabular}%
}\right) \left( 
\begin{tabular}{l}
$N_{t}^{\left( 1\right) }$ \\ 
$N_{t}^{\left( 2\right) }$%
\end{tabular}%
\right) ,\left( 
\begin{tabular}{l}
$\lambda _{t}^{\left( 1\right) }$ \\ 
$\lambda _{t}^{\left( 2\right) }$%
\end{tabular}%
\right) \right) $, with $m$ jump times $\left[ 
\begin{tabular}{l}
$\{T_{1}^{\ast \left( 1\right) },T_{2}^{\ast \left( 1\right) },\cdots
,T_{m}^{\ast \left( 1\right) }\}$ \\ 
$\{T_{1}^{\ast \left( 2\right) },T_{2}^{\ast \left( 2\right) },\cdots
,T_{m}^{\ast \left( 2\right) }\}$%
\end{tabular}%
\right] $ in the process $\left( 
\begin{tabular}{l}
$\lambda _{t}^{\left( 1\right) }$ \\ 
$\lambda _{t}^{\left( 2\right) }$%
\end{tabular}%
\right) $ (see Figure 1). \ This algorithm has been extended from Dassios
and Zhao (2011) Section 5 algorithm, where they have shown how to simulate
the univariate dynamic contagion process.

\bigskip

\textbf{Algorithm 5.1}. (The bivariate compound dynamic contagion process
simulation algorithm)

\bigskip

\begin{quote}
1. Set the initial conditions $T_{0}^{\ast \left( 1\right) }=T_{0}^{\ast
\left( 2\right) }=0,$ $\lambda _{T_{0}^{\ast \left( 1\right) +}}^{\left(
1\right) }=\lambda _{0}^{\left( 1\right) }>a^{\left( 1\right) }$, $\lambda
_{T_{0}^{\ast \left( 2\right) +}}^{\left( 2\right) }=\lambda _{0}^{\left(
2\right) }>a^{\left( 2\right) }$ and\textit{\ }$i\in \{0,1,2,\ldots ,m-1\}$.

\bigskip

2. Simulate the $(i+1)^{\text{th}}$ externally excited joint jump waiting
time $E_{i+1}^{\ast }$ by%
\begin{equation*}
E_{i+1}^{\ast }=-\frac{1}{\rho }\ln U,\text{ \ \ \ \ }U\sim \text{U}[0,1]%
\text{.}
\end{equation*}

\bigskip

3. (i) Simulate the $(i+1)^{\text{th}}$ self-excited jump waiting time $%
S_{i+1}^{\ast \left( 1\right) }$ by

\begin{equation*}
S_{i+1}^{\ast \left( 1\right) }=%
\begin{cases}
S_{1,i+1}^{\ast \left( 1\right) }\wedge S_{2,i+1}^{\ast \left( 1\right) } & 
\left( d_{i+1}^{\left( 1\right) }>0\right) \\ 
S_{2,i+1}^{\ast \left( 1\right) } & \left( d_{i+1}^{\left( 1\right)
}<0\right)%
\end{cases}%
,
\end{equation*}

where 
\begin{equation*}
d_{i+1}^{\left( 1\right) }=1+\frac{\delta ^{\left( 1\right) }\ln U_{11}}{%
\lambda _{T_{i}^{\ast \left( 1\right) +}}^{\left( 1\right) }-a^{\left(
1\right) }}\text{, \ \ \ \ \ \ \ \ }U_{11}\sim \text{U}[0,1]
\end{equation*}

and

\begin{equation*}
S_{1,i+1}^{\ast \left( 1\right) }=-\frac{1}{\delta ^{\left( 1\right) }}\ln
d_{i+1}^{\left( 1\right) }\text{; \ \ \ \ }S_{2,i+1}^{\ast \left( 1\right)
}=-\frac{1}{a^{\left( 1\right) }}\ln U_{12}\text{, \ \ \ \ \ \ \ }U_{12}\sim 
\text{U}[0,1].
\end{equation*}

\bigskip

(ii) Similarly, simulate the $(i+1)^{\text{th}}$ self-excited jump waiting
time $S_{i+1}^{\ast \left( 2\right) }$ by

\begin{equation*}
S_{i+1}^{\ast \left( 2\right) }=%
\begin{cases}
S_{1,i+1}^{\ast \left( 2\right) }\wedge S_{2,i+1}^{\ast \left( 2\right) } & 
\left( d_{i+1}^{\left( 2\right) }>0\right) \\ 
S_{2,i+1}^{\ast \left( 2\right) } & \left( d_{i+1}^{\left( 2\right)
}<0\right)%
\end{cases}%
,
\end{equation*}

where 
\begin{equation*}
d_{i+1}^{\left( 2\right) }=1+\frac{\delta ^{\left( 2\right) }\ln U_{21}}{%
\lambda _{T_{i}^{\ast \left( 2\right) +}}^{\left( 2\right) }-a^{\left(
2\right) }}\text{, \ \ \ \ \ \ \ \ }U_{21}\sim \text{U}[0,1]
\end{equation*}

and

\begin{equation*}
S_{1,i+1}^{\ast \left( 2\right) }=-\frac{1}{\delta ^{\left( 2\right) }}\ln
d_{i+1}^{\left( 2\right) }\text{; \ \ \ \ }S_{2,i+1}^{\ast \left( 2\right)
}=-\frac{1}{a^{\left( 2\right) }}\ln U_{22},\text{ \ \ \ \ \ \ \ }U_{22}\sim 
\text{U}[0,1].\text{\ }
\end{equation*}

\bigskip

4. Simulate the $(i+1)^{\text{th}}$ overall jump time $T_{i+1}^{\ast }$ by 
\begin{equation*}
T_{i+1}^{\ast }=T_{i}^{\ast }+S_{i+1}^{\ast \left( 1\right) }\wedge
S_{i+1}^{\ast \left( 2\right) }\wedge E_{i+1}^{\ast },\text{ where \ }%
T_{0}^{\ast }=T_{0}^{\ast \left( 1\right) }=T_{0}^{\ast \left( 2\right) }=0
\end{equation*}

\bigskip

5. (i) The $(i+1)^{\text{th}}$ jump time for the process $\lambda
_{t}^{\left( 1\right) }$ is given by the overall jump time $T_{i+1}^{\ast }$
in Step 4, i.e.{\ 
\begin{equation*}
T_{i+1}^{\ast \left( 1\right) }=T_{i+1}^{\ast }=%
\begin{cases}
T_{i}^{\ast }+S_{i+1}^{\ast \left( 1\right) }\text{ \ \ }\left(
S_{i+1}^{\ast \left( 1\right) }\wedge S_{i+1}^{\ast \left( 2\right) }\wedge
E_{i+1}^{\ast }=S_{i+1}^{\ast \left( 1\right) }\right) \\ 
T_{i}^{\ast }+E_{i+1}^{\ast }\text{\ \ \ }\left( S_{i+1}^{\ast \left(
1\right) }\wedge S_{i+1}^{\ast \left( 2\right) }\wedge E_{i+1}^{\ast
}=E_{i+1}^{\ast }\right)%
\end{cases}%
,
\end{equation*}%
} where $S_{i+1}^{\ast \left( 1\right) }\wedge S_{i+1}^{\ast \left( 2\right)
}\wedge E_{i+1}^{\ast }=S_{i+1}^{\ast \left( 2\right) }$ is irrelevant to
the $(i+1)^{\text{th}}$ jump time for the process $\lambda _{t}^{\left(
1\right) }$.

\bigskip

(ii) Similarly, the $(i+1)^{\text{th}}$ jump time for the process $\lambda
_{t}^{\left( 2\right) }$ is given by the overall jump time $T_{i+1}^{\ast }$
in Step 4, i.e. {\ 
\begin{equation*}
T_{i+1}^{\ast \left( 2\right) }=T_{i+1}^{\ast }=%
\begin{cases}
T_{i}^{\ast }+S_{i+1}^{\ast \left( 2\right) }\text{ \ }\left( S_{i+1}^{\ast
\left( 1\right) }\wedge S_{i+1}^{\ast \left( 2\right) }\wedge E_{i+1}^{\ast
}=S_{i+1}^{\ast \left( 2\right) }\right) \\ 
T_{i}^{\ast }+E_{i+1}^{\ast }\text{ \ \ }\left( S_{i+1}^{\ast \left(
1\right) }\wedge S_{i+1}^{\ast \left( 2\right) }\wedge E_{i+1}^{\ast
}=E_{i+1}^{\ast }\right)%
\end{cases}%
,
\end{equation*}%
} where $S_{i+1}^{\ast \left( 1\right) }\wedge S_{i+1}^{\ast \left( 2\right)
}\wedge E_{i+1}^{\ast }=S_{i+1}^{\ast \left( 1\right) }$ is irrelevant to
the $(i+1)^{\text{th}}$ jump time for the process $\lambda _{t}^{\left(
2\right) }$.

\bigskip

6. The changes at jump time $T_{i+1}^{\ast \left( 1\right) }$ in the
intensity process $\lambda _{t}^{\left( 1\right) }$ is given by%
\begin{equation*}
\lambda _{T_{i+1}^{\ast \left( 1\right) +}}^{\left( 1\right) }=%
\begin{cases}
\lambda _{T_{i+1}^{\ast \left( 1\right) -}}^{\left( 1\right) }+Y_{i+1},\text{
\ }Y_{i+1}\sim G(y)\text{ \ }\left( S_{i+1}^{\ast \left( 1\right) }\wedge
S_{i+1}^{\ast \left( 2\right) }\wedge E_{i+1}^{\ast }=S_{i+1}^{\ast \left(
1\right) }\right) \\ 
\lambda _{T_{i+1}^{\ast \left( 1\right) -}}^{\left( 1\right)
}+X_{i+1}^{\left( 1\right) },\text{ \ \ \ \ \ \ \ \ \ \ \ \ \ \ \ \ \ \ }%
\left( S_{i+1}^{\ast \left( 1\right) }\wedge S_{i+1}^{\ast \left( 2\right)
}\wedge E_{i+1}^{\ast }=E_{i+1}^{\ast }\right)%
\end{cases}%
\end{equation*}

and the changes at jump time $T_{i+1}^{\ast \left( 2\right) }$ in the
intensity process $\lambda _{t}^{\left( 2\right) }$ is given by 
\begin{equation*}
\lambda _{T_{i+1}^{\ast \left( 2\right) +}}^{\left( 2\right) }=%
\begin{cases}
\lambda _{T_{i+1}^{\ast \left( 2\right) -}}^{\left( 2\right) }+Z_{i+1},\text{
\ }Z_{i+1}\sim H(z)\text{ \ }\left( S_{i+1}^{\ast \left( 1\right) }\wedge
S_{i+1}^{\ast \left( 2\right) }\wedge E_{i+1}^{\ast }=S_{i+1}^{\ast \left(
2\right) }\right) \\ 
\lambda _{T_{i+1}^{\ast \left( 2\right) -}}^{\left( 2\right)
}+X_{i+1}^{\left( 2\right) },\text{ \ \ \ \ \ \ \ \ \ \ \ \ \ \ \ \ \ \ \ }%
\left( S_{i+1}^{\ast \left( 1\right) }\wedge S_{i+1}^{\ast \left( 2\right)
}\wedge E_{i+1}^{\ast }=E_{i+1}^{\ast }\right)%
\end{cases}%
,
\end{equation*}

where
\end{quote}

\begin{equation*}
\lambda _{T_{i+1}^{\ast \left( 1\right) -}}^{\left( 1\right) }=(\lambda
_{T_{i}^{\ast \left( 1\right) +}}^{\left( 1\right) }-a^{\left( 1\right) })e^{%
{-\delta ^{\left( 1\right) }(T_{i+1}^{\ast \left( 1\right) }-T_{i}^{\ast
\left( 1\right) })}}+a^{\left( 1\right) },
\end{equation*}

\begin{quote}
\bigskip
\end{quote}

\begin{equation*}
\lambda _{T_{i+1}^{\ast \left( 2\right) -}}^{\left( 2\right) }=(\lambda
_{T_{i}^{\ast \left( 2\right) +}}^{\left( 2\right) }-a^{\left( 2\right) })e^{%
{-\delta ^{\left( 2\right) }(T_{i+1}^{\ast \left( 2\right) }-T_{i}^{\ast
\left( 2\right) })}}+a^{\left( 2\right) },
\end{equation*}

\begin{equation*}
\left( X_{i+1}^{\left( 1\right) },\text{ }X_{i+1}^{\left( 2\right) }\right)
\sim F(x^{\left( 1\right) },x^{\left( 2\right) }),
\end{equation*}

\begin{quote}
where the joint distribution of the vector $\left( X_{i+1}^{\left( 1\right)
},\text{ }X_{i+1}^{\left( 2\right) }\right) $ is assumed to be of the form $%
C_{\theta }(F\left( x^{\left( 1\right) }\right) ,F\left( x^{\left( 2\right)
}\right) )$ with $C_{\theta }$ being a given copula.

\bigskip

7. The change at jump time $T_{i+1}^{\ast \left( 1\right) }$ in the point
process $N_{t}^{\left( 1\right) }$ is given by 
\begin{equation*}
N_{T_{i+1}^{\ast \left( 1\right) +}}^{\left( 1\right) }=%
\begin{cases}
N_{T_{i+1}^{\ast \left( 1\right) -}}^{\left( 1\right) }+1\text{ \ \ }\left(
S_{i+1}^{\ast \left( 1\right) }\wedge S_{i+1}^{\ast \left( 2\right) }\wedge
E_{i+1}^{\ast }=S_{i+1}^{\ast \left( 1\right) }\right) \\ 
N_{T_{i+1}^{\ast \left( 1\right) -}}^{\left( 1\right) }\text{ \ \ \ \ \ \ \ }%
\left( S_{i+1}^{\ast \left( 1\right) }\wedge S_{i+1}^{\ast \left( 2\right)
}\wedge E_{i+1}^{\ast }=E_{i+1}^{\ast }\right)%
\end{cases}%
\end{equation*}

and the change at jump time $T_{i+1}^{\ast \left( 2\right) }$ in the point
process $N_{t}^{\left( 2\right) }$ is given by 
\begin{equation*}
N_{T_{i+1}^{\ast \left( 2\right) +}}^{\left( 2\right) }=%
\begin{cases}
N_{T_{i+1}^{\ast \left( 2\right) -}}^{\left( 2\right) }+1\text{ \ \ }\left(
S_{i+1}^{\ast \left( 1\right) }\wedge S_{i+1}^{\ast \left( 2\right) }\wedge
E_{i+1}^{\ast }=S_{i+1}^{\ast \left( 2\right) }\right) \\ 
N_{T_{i+1}^{\ast \left( 2\right) -}}^{\left( 2\right) }\text{ \ \ \ \ \ \ \ }%
\left( S_{i+1}^{\ast \left( 1\right) }\wedge S_{i+1}^{\ast \left( 2\right)
}\wedge E_{i+1}^{\ast }=E_{i+1}^{\ast }\right)%
\end{cases}%
\end{equation*}

{8. The change at jump time $T_{i+1}^{\ast \left( 1\right) }$ in the
compound point process $L_{t}^{\left( 1\right) }$ is given by 
\begin{equation*}
L_{T_{i+1}^{\ast \left( 1\right) +}}^{\left( 1\right) }=%
\begin{cases}
L_{T_{i+1}^{\ast \left( 1\right) -}}^{\left( 1\right) }+\xi _{i+1}^{\left(
1\right) },\text{ \ }\xi _{i+1}^{\left( 1\right) }\sim J(\xi ^{(1)})\text{ \
\ }\left( S_{i+1}^{\ast \left( 1\right) }\wedge S_{i+1}^{\ast \left(
2\right) }\wedge E_{i+1}^{\ast }=S_{i+1}^{\ast \left( 1\right) }\right) \\ 
L_{T_{i+1}^{\ast \left( 1\right) -}}^{\left( 1\right) }\text{ \ \ \ \ \ \ \
\ \ \ \ \ \ \ \ \ \ \ \ \ \ \ \ \ \ \ \ \ \ \ \ }\left( S_{i+1}^{\ast \left(
1\right) }\wedge S_{i+1}^{\ast \left( 2\right) }\wedge E_{i+1}^{\ast
}=E_{i+1}^{\ast }\right)%
\end{cases}%
\end{equation*}%
}
\end{quote}

and the change at jump time $T_{i+1}^{\ast \left( 2\right) }$ in the
compound point process $L_{t}^{\left( 2\right) }$ is given by 
\begin{equation*}
L_{T_{i+1}^{\ast \left( 2\right) +}}^{\left( 2\right) }=%
\begin{cases}
L_{T_{i+1}^{\ast \left( 2\right) -}}^{\left( 2\right) }+\xi _{i+1}^{\left(
2\right) },\text{ \ }\xi _{i+1}^{\left( 2\right) }\sim K(\xi ^{(2)})\text{ \
\ }\left( S_{i+1}^{\ast \left( 1\right) }\wedge S_{i+1}^{\ast \left(
2\right) }\wedge E_{i+1}^{\ast }=S_{i+1}^{\ast \left( 2\right) }\right) \\ 
L_{T_{i+1}^{\ast \left( 2\right) -}}^{\left( 2\right) }\text{ \ \ \ \ \ \ \
\ \ \ \ \ \ \ \ \ \ \ \ \ \ \ \ \ \ \ \ \ \ \ \ \ }\left( S_{i+1}^{\ast
\left( 1\right) }\wedge S_{i+1}^{\ast \left( 2\right) }\wedge E_{i+1}^{\ast
}=E_{i+1}^{\ast }\right)%
\end{cases}%
\end{equation*}

\bigskip

% ------------------------------------------------------------------------------------------------------
%% Figure 

\begin{figure}[p]
    \centering
        \subfloat{\includegraphics[width=0.95\textwidth]{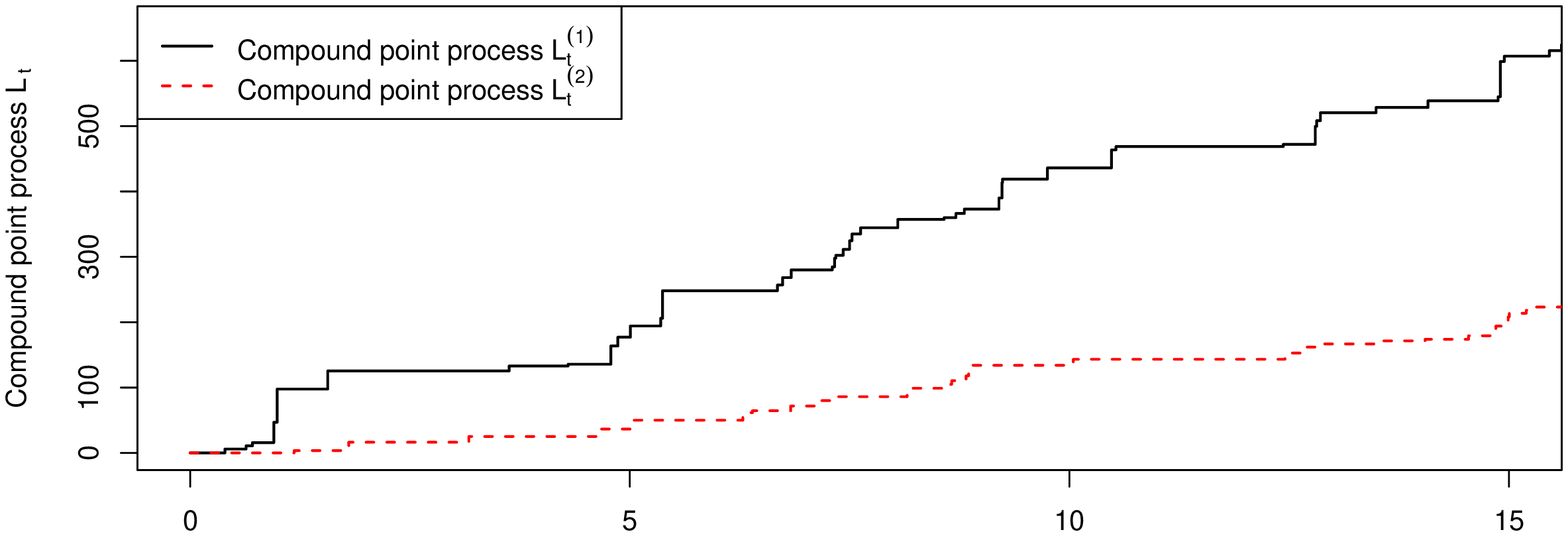}}
      
        \subfloat{\includegraphics[width=0.95\textwidth]{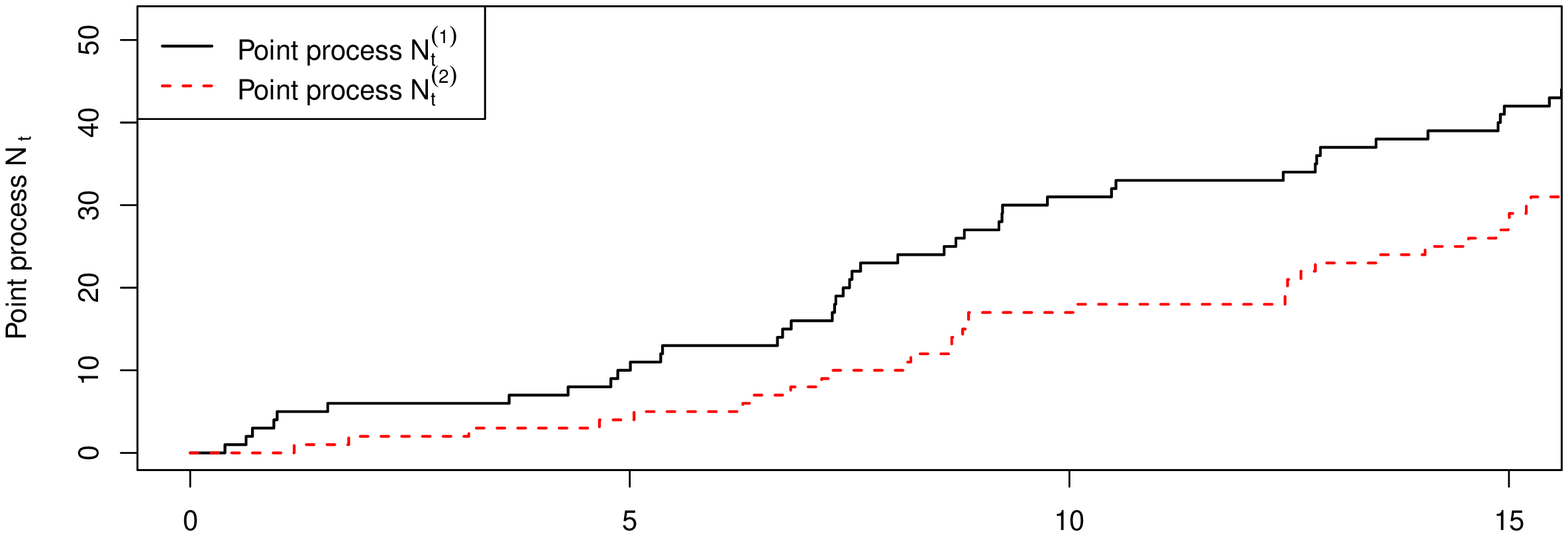}}

        \subfloat{\includegraphics[width=0.95\textwidth]{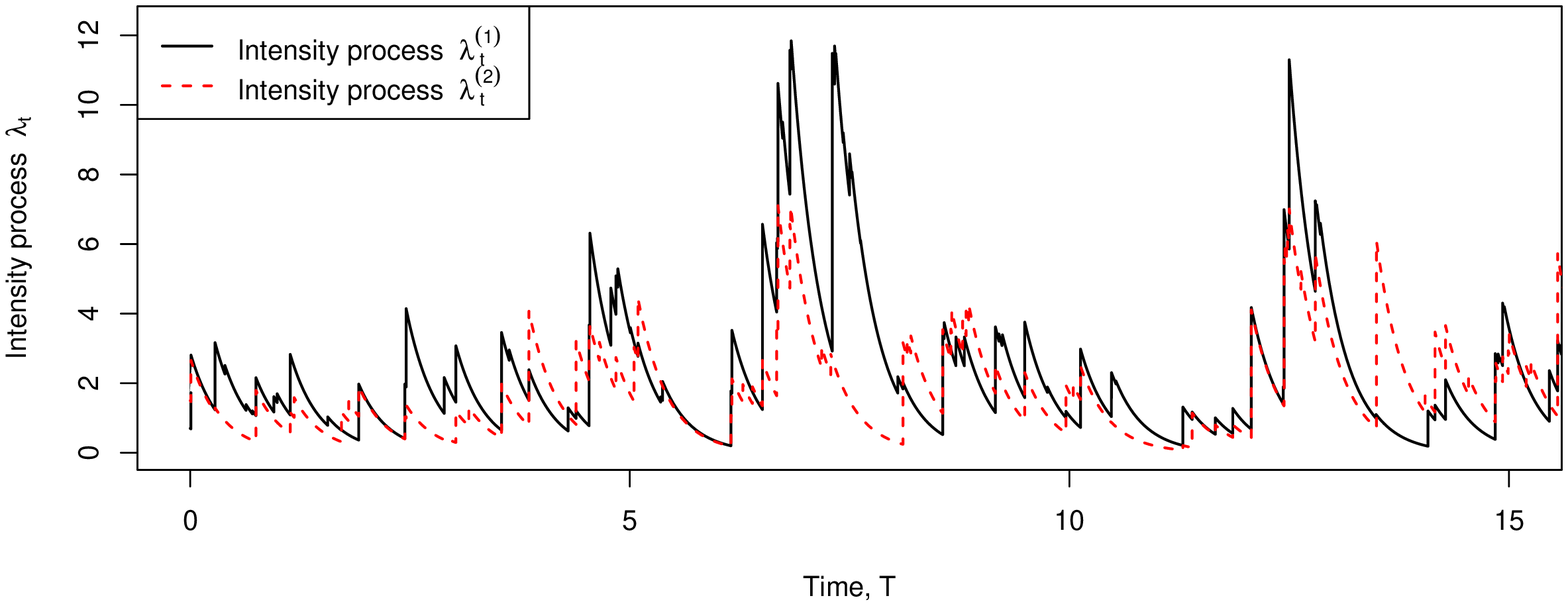}}

\caption{ Simulated sample path of the bivariate compound dynamic contagion process: Intensity process $\left(\lambda_t^{(1)}, \lambda_t^{(2)}\right)$, point processes $\left(N_t^{(1)}, N_t^{(2)}\right)$, 
and compound point processes $\left(L_t^{(1)}, L_t^{(2)}\right)$. FGM copula is considered with parameter $\theta$. 
The parameters for the process 1 and  process 2  are 
$\Big(a^{(1)},a^{(2)}, \rho, \delta^{(1)},\delta^{(2)};
\alpha, \psi, \varsigma, c;
\beta, \varphi, \epsilon;
\theta;
\omega^{(1)},\omega^{(2)},$ $ \zeta^{(1)},\zeta^{(2)}, k^{(1)},  k^{(2)};
\lambda_0^{(1)}, \lambda_0^{(2)}\Big)$
$= \left(0,0,3,3,3; 0.5,1,11,3; 1,0.5,15; 0.5; 3,4, 4,4, 6,6; 0.7, 1.5  \right) $.}
\end{figure}

% ------------------------------------------------------------------------------------------------------

\bigskip

\textbf{6. Conclusion}

\bigskip

Digitalisation of business and economic activities have changed the risk
landscape to cyber space. A cyber attack can trigger multiple, catastrophic
and contagious losses to corporates and governments due to IT system
interdependence. It is a real threat to all organisations as the number of
cyber attacks and its complex way of doing so are rising.

Cyber insurance can be purchased to cover economic and financial losses
occurring from cyber incidents. \ However, due to the complexity of cyber
risks, i.e. multiple, catastrophic and contagious losses, it is difficult
for insurers to price cyber insurance products. To provide insurers with a
tool to deal with the ongoing challenge of new risks, we introduce a
bivariate \textit{compound} dynamic contagion process, which accommodate the
interdependence of IT system and the frequency and impact of cyber events.

Our numerical results confirm that cyber loss insurance premiums calculated
using the bivariate compound dynamic contagion process are significantly
higher than their counterparts calculated using a bivariate compound
shot-noise Cox process. \ For that purpose, we provided moment-based
insurance premium calculations using a log gamma distribution and a Fr\'{e}%
chet distribution for two separate \textit{self-excited} jumps (i.e.
after-cyber attacks/incidents/shocks), two different exponential
distributions and four different copulas (i.e. the Farlie-Gumbel-Morgenstern
(FGM) family copula, the Gaussian family copula, the \textit{t} copula and
the Gumbel copula) for \textit{externally excited joint} jumps (i.e. initial
cyber attacks/incidents/shocks). \ Two Pareto distributions were used to
represent catastrophic cyber losses from contagious cyber attacks. This
suggests that the bivariate compound dynamic contagion process can be
considered for modelling two aggregate cyber losses to calculate cyber loss
insurance premiums accommodating waves of events and with the critical
aspects of the interdependence of IT system and the impact of cyber events
taken into account. \ For further research, we may consider the extension of
dimension, other copulas, other measures for initial and after cyber
attacks/incidents/shocks, and other measures for cyber losses. \ As
loss/claim size and after-cyber attacks/incidents/shocks (self-exciting
jumps) could be correlated, considering the dependency $\left\{
Y_{j}\right\} _{j=1,2,\cdots }$ and $\left\{ \Xi _{j}^{\left( 1\right)
}\right\} _{j=1,2,\cdots }$and $\left\{ Z_{k}\right\} _{k=1,2,\cdots }$ and $%
\left\{ \Xi _{k}^{\left( 1\right) }\right\} _{k=1,2,\cdots }$, respectively
could be another object of further research.

Cyber attacks would occur more often as all types of risk are touched by
cyber space due to digitalisation of business and economic activities, so
the proposed bivariate compound dynamic contagion process can be an improved
model for insurance companies to quantify cyber losses. \ The bivariate
compound dynamic contagion process is also very much applicable to credit,
insurance, market and other operational risks. \ We hope that what we
presented in this paper provides practitioners with feasible models to
quantify cyber losses, and to deal with a variety of problems in economics,
finance and insurance.

\bigskip

\textbf{Acknowledgements}
\begin{quote}
Rosy Oh's research was supported by Basic Science Research Program through the National Research Foundation of Korea(NRF) funded by the Ministry of Education (Grant No. 2019R1A6A1A11051177 and 2020R1I1A1A01067376).
\end{quote}

\bigskip

\textbf{References}

\begin{quote}
Allianz (2016), Allianz Risk Barometer Top Business Risks.

A\"{\i}t-Sahalia, Y., Cacho-Diaz, J. A. and Laeven, R. J. (2015) : Modeling
financial contagion using mutually exciting jump processes, Journal of
FinancialEconomics 117 (3), 585-606.

A\"{\i}t-Sahalia, Y., Laeven, R. J. and Pelizzon, L. (2014) : Mutual
excitation in Eurozone sovereign CDS, Journal of Econometrics, Journal of
FinancialEconomics 183(2), 151--167.

Bauwens, L. and Hautsch, N. (2009). Modelling Financial High Frequency Data
Using Point Processes, In: Handbook of Financial Time Series, T.G. Andersen,
R.A. Davis, J.-P. Kreiss and T. Mikosch (eds), Springer.

Biener, C., Eling, M. and Wirfs, J. H. (2015): Insurability of Cyber Risk:
An Empirical Analysis, The Geneva Papers on Risk and Insurance - Issues and
Practice, 40(1), 131--158.

Bowsher, C. G. (2007) : Modelling security market events in continuous time:
Intensity based, multivariate point process models, Journal of Econometrics,
141(2), 876-912.

Chavez-Demoulin, V., Davison, A. C. and McNeil, A. J. (2005): Estimating
Value-at-Risk: A point process approach, Quantitative Finance, 5(2), 227-234.

Dassios, A. and Jang, J. (2003) : Pricing of catastrophe reinsurance \&
derivatives using the Cox process with shot noise intensity, Finance \&
Stochastics, 7/1, 73-95.

Dassios, A. and Zhao, H. (2011) : A dynamic contagion process, Advances in
Applied Probability, 43, 814-846.

Dassios, A. and Zhao, H. (2012) : Ruin by Dynamic Contagion Claims,
Insurance: Mathematics \& Economics, 51/1, 93-106.

Dassios, A. and Zhao, H. (2017a) : A generalized contagions process with an
application to credit risk, International Journal of Theoretical and Applied
Finance, 20(1), 1750003 (33 pages).

Dassios, A. and Zhao, H. (2017b) : Efficient simulation of clustering jumps
with CIR intensity, Operations Research, 65(6), 1494-1515.

Davis, M. H. A.(1984) : Piecewise deterministic Markov processes: A general
class of non diffusion stochastic models. J. R. Stat. Soc. B 46, 353--388.

Dong, X. (2014): Compensators and diffusion approximation of point processes
and applications, Ph.D Thesis. Imperial College London.

Embrechts, P., Liniger, T. and Lin, L. (2011) : Multivariate Hawkes
Processes: an Application to Financial Data, Journal of Applied Probability,
pecial Volume 48(A), 367-378

Errais, E., Giesecke, K. and Goldberg, L. R. (2010) : Affine Point Processes
and Portfolio Credit Risk, SIAM Journal on Financial Mathematics, 1, 642-665.

Gao, X, Zhou, X and Zhu, L. (2018) : Transform analysis for Hawkes processes
with applications in dark pool trading, \ Quantitative Finance, 18(2),
265--282.

Giesecke, K. and Kim, B. (2011) : Risk Analysis of Collateralized Debt
Obligations, Operations Research, 59(1), 32--49.

Hawkes, A. G. (1971a) : Point spectra of some mutually exciting point
processes, Journal of the Royal Statistical Society. Series B
(Methodological ) 33 (3), 438--443.

Hawkes, A. G. (1971b) : Spectra of some self-exciting and mutually exciting
point processes, Biometrika, 58(1), 83-90.

Hawkes, A. G. and Oakes, D. (1974) : A cluster process representation of a
self-exciting process, Journal of Applied Probability, 11, 493-503.

Herath, H. S. B. and Herath, T. S. (2011) : Copula Based Actuarial Model for
Pricing Cyber-Insurance Policies. Insurance Markets and Companies: Analyses
and Actuarial Computations, 2 (1). 7-20.

Jang, J. and Dassios, A. (2013) : A Bivariate Shot Noise Self-Exciting
Process for Insurance, Insurance: Mathematics \& Economics, 53/3, 524--532.

Lu, X. and Abergel, F. (2018): High-dimensional Hawkes processes for limit
order books: modelling, empirical analysis and numerical calibration, \
Quantitative Finance, 18(20), 249--264.

McNeil, A. J., Frey, R. and Embrechts, P. (2005): Quantitative Risk
Management: Concepts, Techniques and Tools, Princeton University Press, USA.

Mukhopadhyay, A., Chatterjee, S., Saha, D., Mahanti, A. and Sadhukhan, S. K.
(2006) : e-Risk management with insurance: A framework using copula aided
Bayesian belief networks. In Proceedings of the 39th Annual Hawaii
International Conference on System Sciences (HICSS'06), vol. 6,
126.1--126.6. Hoboken, NJ: IEEE.

Rambaldi, M., Bacry, E. and Lillo, F. (2017) : The role of volume in order
book dynamics: a multivariate Hawkes process analysis, Quantitative Finance,
17(7), 999--1020.

Stabile, G. and Torrisi, G. L. (2010) : Risk processes with non-stationary
Hawkes claims arrivals, Methodology and Computing in Applied Probability,
12(3), 415-429.

Xu, M. and Hua, L. (2017) : Cybersecurity Insurance: Modeling and Pricing,
Society of Actuaries. Schaumburg, Illinois.

Yang, S. Y., Liu, A., Chen, J. and Hawkes, A. (2018) :\ Applications of a
multivariate Hawkes process to joint modeling of sentiment and market return
events, \ Quantitative Finance, 18(2), 295--310.
\end{quote}

\bigskip

\textbf{Appendix}

\bigskip

\textbf{A} \ \textbf{Proof of Theorem 4.1}

\bigskip

Setting $\mathcal{A}$ $f\left( \lambda ^{\left( 1\right) },n^{\left(
1\right) },l^{(1)},\lambda ^{\left( 2\right) },n^{\left( 2\right)
},l^{(2)},t\right) =l^{\left( 1\right) }$ in (2.3), we have 
\begin{equation*}
\mathcal{A}\text{ }l^{\left( 1\right) }=\mu _{1_{J}}\lambda ^{\left(
1\right) }.
\end{equation*}

As $L_{t}^{\left( 1\right) }-L_{0}^{\left( 1\right) }-\int\limits_{0}^{t}%
\mathcal{A}$ $l_{s}^{\left( 1\right) }ds$ is a $\Im $-martingale, we have%
\begin{equation*}
E\left\{ L_{t}^{\left( 1\right) }-\int\limits_{0}^{t}\mathcal{A}\text{ }%
l_{s}^{\left( 1\right) }ds\mid \lambda _{0}^{\left( 1\right) }\right\}
=L_{0}^{\left( 1\right) }.
\end{equation*}

Hence%
\begin{equation*}
E\left( L_{t}^{\left( 1\right) }\mid \lambda _{0}^{\left( 1\right) }\right)
=L_{0}^{\left( 1\right) }+E\left\{ \int\limits_{0}^{t}\mathcal{A}\text{ }%
l_{s}^{\left( 1\right) }ds\mid \lambda _{0}^{\left( 1\right) }\right\}
=L_{0}^{\left( 1\right) }+\mu _{1_{J}}\int\limits_{0}^{t}E\left( \lambda
_{s}^{\left( 1\right) }\mid \lambda _{0}^{\left( 1\right) }\right) ds
\end{equation*}

and (4.16) and (4.17) follow using (4.1) and (4.2) in Proposition 4.1. \
Similarly, (4.18) and (4.19) can be obtained.

\bigskip

\textbf{B} \ \textbf{Proof of Corollary 4.1}

\bigskip

From the proof in Theorem 4.1, we have%
\begin{equation*}
E\left( L_{t}^{\left( 1\right) }-L_{0}^{\left( 1\right) }\mid \lambda
_{0}^{\left( 1\right) }\right) =\mu _{1_{J}}\int\limits_{0}^{t}E\left(
\lambda _{s}^{\left( 1\right) }\mid \lambda _{0}^{\left( 1\right) }\right) ds
\end{equation*}

and also we know $E\left( \lambda _{t}^{\left( 1\right) }\right) $ from
(4.5), then by assuming that $L_{0}^{\left( 1\right) }=0$, we have

\begin{equation*}
E\left( L_{t}^{\left( 1\right) }\right) =E\left( L_{t}^{\left( 1\right)
}-L_{0}^{\left( 1\right) }\right) =\mu _{1_{J}}\int\limits_{0}^{t}E\left(
\lambda _{s}^{\left( 1\right) }\right) ds=\mu _{1_{J}}\left( \frac{\mu
_{1_{F_{1}}}\text{ }\rho +a^{\left( 1\right) }\delta ^{\left( 1\right) }}{%
\delta ^{\left( 1\right) }-\mu _{1_{G}}}\right) t.
\end{equation*}

Similarly, we have%
\begin{equation*}
E\left( L_{t}^{\left( 2\right) }\right) =E\left( L_{t}^{\left( 2\right)
}-L_{0}^{\left( 2\right) }\right) =\mu _{1_{K}}\int\limits_{0}^{t}E\left(
\lambda _{s}^{\left( 2\right) }\right) ds=\mu _{1_{K}}\left( \frac{\mu
_{1_{F_{2}}}\text{ }\rho +a^{\left( 2\right) }\delta ^{\left( 2\right) }}{%
\delta ^{\left( 2\right) }-\mu _{1_{H}}}\right) t.
\end{equation*}

\bigskip

\textbf{C} \ \textbf{Proof of Lemma 4.1}

\bigskip

Setting $\mathcal{A}$ $f\left( \Lambda ^{\left( 1\right) },\lambda ^{\left(
1\right) },n^{\left( 1\right) },l^{(1)},\Lambda ^{\left( 2\right) },\lambda
^{\left( 2\right) },n^{\left( 2\right) },l^{(2)},t\right) =\lambda ^{\left(
1\right) }l^{\left( 2\right) }$ in (2.3), we have%
\begin{equation*}
\mathcal{A}\text{ }\left( \lambda ^{\left( 1\right) }l^{\left( 2\right)
}\right) =-\left( \delta ^{\left( 1\right) }-\mu _{1_{G}}\right) \lambda
^{\left( 1\right) }l^{\left( 2\right) }+\left( a^{\left( 1\right) }\delta
^{\left( 1\right) }+\mu _{1_{F_{1}}}\rho \right) l^{\left( 2\right) }+\mu
_{1_{K}}\lambda ^{\left( 1\right) }\lambda ^{\left( 2\right) }.
\end{equation*}

As $\lambda _{t}^{\left( 1\right) }L_{t}^{\left( 2\right) }-\lambda
_{0}^{\left( 1\right) }L_{0}^{\left( 2\right) }-\int\limits_{0}^{t}\mathcal{A%
}$ $\left( \lambda _{s}^{\left( 1\right) }L_{s}^{\left( 2\right) }\right) ds$
is a $\Im $-martingale, given $L_{0}^{\left( 2\right) }=0$ we have the ODE%
\begin{equation*}
\frac{dE\left( \lambda _{t}^{\left( 1\right) }L_{t}^{\left( 2\right)
}\right) }{dt}=-\left( \delta ^{\left( 1\right) }-\mu _{1_{G}}\right)
E\left( \lambda _{t}^{\left( 1\right) }L_{t}^{\left( 2\right) }\right)
+\left( a^{\left( 1\right) }\delta ^{\left( 1\right) }+\mu _{1_{F_{1}}}\rho
\right) E\left( L_{t}^{\left( 2\right) }\right) +\mu _{1_{K}}E\left( \lambda
_{t}^{\left( 1\right) }\lambda _{t}^{\left( 2\right) }\right)
\end{equation*}

with the initial condition $E\left( \lambda _{0}^{\left( 1\right)
}L_{0}^{\left( 2\right) }\right) =0$. \ The solution of this ODE using
(4.21) and (4.9) is given by (4.22). \ Similarly, we have (4.23).

\bigskip

\textbf{D} \ \textbf{Proof of Theorem 4.2}

\bigskip

Setting $\mathcal{A}$ $f\left( \Lambda ^{\left( 1\right) },\lambda ^{\left(
1\right) },n^{\left( 1\right) },l^{(1)},\Lambda ^{\left( 2\right) },\lambda
^{\left( 2\right) },n^{\left( 2\right) },l^{(2)},t\right) =l^{\left(
1\right) }l^{\left( 2\right) }$ in (2.3), we have%
\begin{equation*}
\mathcal{A}\text{ }\left( l^{\left( 1\right) }l^{\left( 2\right) }\right)
=\mu _{1_{J}}\lambda ^{\left( 1\right) }l^{\left( 2\right) }+\mu
_{1_{K}}\lambda ^{\left( 2\right) }l^{\left( 1\right) }.
\end{equation*}

As $L_{t}^{\left( 1\right) }L_{t}^{\left( 2\right) }-L_{0}^{\left( 1\right)
}L_{0}^{\left( 2\right) }-\int\limits_{0}^{t}\mathcal{A}$ $\left(
L_{s}^{\left( 1\right) }L_{s}^{\left( 2\right) }\right) ds$ is a $\Im $%
-martingale, we have%
\begin{equation*}
E\left\{ L_{t}^{\left( 1\right) }L_{t}^{\left( 2\right) }-\int\limits_{0}^{t}%
\mathcal{A}\left( L_{s}^{\left( 1\right) }L_{s}^{\left( 2\right) }\right)
ds\mid \lambda _{0}^{\left( 1\right) },\lambda _{0}^{\left( 2\right)
}\right\} =L_{0}^{\left( 1\right) }L_{0}^{\left( 2\right) }.
\end{equation*}

Hence%
\begin{equation*}
E\left( L_{t}^{\left( 1\right) }L_{t}^{\left( 2\right) }\mid \lambda
_{0}^{\left( 1\right) },\lambda _{0}^{\left( 2\right) }\right)
=L_{0}^{\left( 1\right) }L_{0}^{\left( 2\right) }+E\left\{
\int\limits_{0}^{t}\mathcal{A}\left( L_{s}^{\left( 1\right) }L_{s}^{\left(
2\right) }\right) ds\mid \lambda _{0}^{\left( 1\right) },\lambda
_{0}^{\left( 2\right) }\right\}
\end{equation*}%
\begin{equation*}
=L_{0}^{\left( 1\right) }L_{0}^{\left( 2\right) }+\mu
_{1_{J}}\int\limits_{0}^{t}E\left( \lambda _{s}^{\left( 1\right)
}L_{s}^{\left( 2\right) }\mid \lambda _{0}^{\left( 1\right) },\lambda
_{0}^{\left( 2\right) }\right) ds+\mu _{1_{K}}\int\limits_{0}^{t}E\left(
\lambda _{s}^{\left( 2\right) }L_{s}^{\left( 1\right) }\mid \lambda
_{0}^{\left( 1\right) },\lambda _{0}^{\left( 2\right) }\right) ds.
\end{equation*}

Using (4.22) and (4.23), with $L_{0}^{\left( 1\right) }=L_{0}^{\left(
2\right) }=0$, we have 
\begin{equation*}
E\left( L_{t}^{\left( 1\right) }L_{t}^{\left( 2\right) }\right) =\mu
_{1_{J}}\int\limits_{0}^{t}E\left( \lambda _{s}^{\left( 1\right)
}L_{s}^{\left( 2\right) }\right) ds+\mu _{1_{K}}\int\limits_{0}^{t}E\left(
\lambda _{s}^{\left( 2\right) }L_{s}^{\left( 1\right) }\right) ds
\end{equation*}

and the result follows.

\bigskip

\textbf{E} \ \textbf{Proof of Lemma 4.2}

\bigskip

Setting $\mathcal{A}$ $f\left( \Lambda ^{\left( 1\right) },\lambda ^{\left(
1\right) },n^{\left( 1\right) },l^{(1)},\Lambda ^{\left( 2\right) },\lambda
^{\left( 2\right) },n^{\left( 2\right) },l^{(2)},t\right) =\lambda ^{\left(
1\right) }l^{\left( 1\right) }$ in (2.3), we have%
\begin{equation*}
\mathcal{A}\text{ }\left( \lambda ^{\left( 1\right) }l^{\left( 1\right)
}\right) =-\left( \delta ^{\left( 1\right) }-\mu _{1_{G}}\right) \lambda
^{\left( 1\right) }l^{\left( 1\right) }+\left( a^{\left( 1\right) }\delta
^{\left( 1\right) }+\mu _{1_{F_{1}}}\rho \right) l^{\left( 1\right) }+\mu
_{1_{J}}\left\{ \lambda ^{\left( 1\right) }\right\} ^{2}+\mu _{1_{G}}\mu
_{1_{J}}\lambda ^{\left( 1\right) }.
\end{equation*}

As $\lambda _{t}^{\left( 1\right) }L_{t}^{\left( 1\right) }-\lambda
_{0}^{\left( 1\right) }L_{0}^{\left( 1\right) }-\int\limits_{0}^{t}\mathcal{A%
}$ $\left( \lambda _{s}^{\left( 1\right) }L_{s}^{\left( 1\right) }\right) ds$
is a $\Im $-martingale, given $L_{0}^{\left( 1\right) }=0$ we have the ODE,%
\begin{eqnarray*}
&&\frac{dE\left( \lambda _{t}^{\left( 1\right) }L_{t}^{\left( 1\right)
}\right) }{dt}=-\left( \delta ^{\left( 1\right) }-\mu _{1_{G}}\right)
E\left( \lambda _{t}^{\left( 1\right) }L_{t}^{\left( 1\right) }\right)
+\left( a^{\left( 1\right) }\delta ^{\left( 1\right) }+\mu _{1_{F_{1}}}\rho
\right) E\left( L_{t}^{\left( 1\right) }\right) \\
&&+\mu _{1_{J}}E\left[ \left\{ \lambda _{t}^{\left( 1\right) }\right\} ^{2}%
\right] +\mu _{1_{G}}\mu _{1_{J}}E\left( \lambda _{t}^{\left( 1\right)
}\right) .
\end{eqnarray*}

with the initial condition $E\left( \lambda _{0}^{\left( 1\right)
}L_{0}^{\left( 1\right) }\right) =0$. \ \ The solution of this ODE using
(4.20), (4.14) and (4.5), is given by (4.27). \ Similarly, we have (4.28).

\bigskip

\textbf{F} \ \textbf{Proof of Theorem 4.3}

\bigskip

Setting $\mathcal{A}$ $f\left( \Lambda ^{\left( 1\right) },\lambda ^{\left(
1\right) },n^{\left( 1\right) },l^{(1)},\Lambda ^{\left( 2\right) },\lambda
^{\left( 2\right) },n^{\left( 2\right) },l^{(2)},t\right) =\left\{ l^{\left(
1\right) }\right\} ^{2}$ in (2.3), we have%
\begin{equation*}
\mathcal{A}\text{ }\left\{ l^{\left( 1\right) }\right\} ^{2}=2\mu
_{1_{J}}\lambda ^{\left( 1\right) }l^{\left( 1\right) }+\mu _{2_{J}}\lambda
^{\left( 1\right) }.
\end{equation*}

As $\left\{ L_{t}^{\left( 1\right) }\right\} ^{2}-\left\{ L_{0}^{\left(
1\right) }\right\} ^{2}-\int\limits_{0}^{t}\mathcal{A}$ $\left\{
L_{s}^{\left( 1\right) }\right\} ^{2}ds$ is a $\Im $-martingale, given $%
L_{0}^{\left( 1\right) }=0$ we have 
\begin{equation*}
E\left[ \left\{ L_{t}^{\left( 1\right) }\right\} ^{2}\right] =2\mu
_{1_{J}}\int\limits_{0}^{t}E\left( \lambda _{s}^{\left( 1\right)
}L_{s}^{\left( 1\right) }\right) ds+\mu _{2_{J}}\int\limits_{0}^{t}E\left(
\lambda _{s}^{\left( 1\right) }\right) ds
\end{equation*}

and (4.29) follows using (4.27) and (4.5). Similarly, we have (4.30).

\bigskip

\textbf{G} \ \textbf{Proof of Corollary 4.2}

\bigskip

By $Var\left\{ L_{t}^{\left( 1\right) }\right\} =E\left[ \left\{
L_{t}^{\left( 1\right) }\right\} ^{2}\right] -\left\{ E\left( L_{t}^{\left(
1\right) }\right) \right\} ^{2}$ and using (4.29) and (4.20), (4.31)
follows. \ Similarly, we have (4.32).

\end{document}